\documentclass[review]{elsarticle}
\usepackage{lineno,hyperref}
\usepackage{epsfig}
\usepackage{natbib}
\usepackage{graphicx}
\usepackage{slashbox}
\usepackage{multirow}
\usepackage{lscape}
\usepackage{mathrsfs,amssymb}
\usepackage{amsmath}
\usepackage{amssymb}
\usepackage{multirow}
\usepackage{tabularx}
\usepackage{amsmath}
\usepackage[flushleft]{threeparttable}
\newcommand       \Angstrom     {\,{\rm \AA}}

\newcommand       \cm           {\,{\rm cm}}

\newcommand       \nm           {\,{\rm nm}}

\newcommand       \K            {\,{\rm K}}

\newcommand       \s            {\,{\rm s}}

\newcommand       \simlt        {\lesssim}
\newcommand       \simgt        {\gtrsim}
\newcommand       \gtsim        {\gtrsim}

\newcommand       \mum          {\,{\rm \mu m}}

\newcommand       \simali       {\sim\,}

\newcommand       \Iaro         {I_{3.3}}
\newcommand       \Iali         {I_{3.4}}
\newcommand       \Iratio         {I_{3.4}/I_{3.3}}
\newcommand       \Aratio        {A_{3.4}/A_{3.3}}
\newcommand       \Aali           {A_{3.4}}
\newcommand       \Aaro          {A_{3.3}}
\newcommand       \nuMea         {\nu_{\rm Me,1}}
\newcommand       \nuMeb         {\nu_{\rm Me,2}}
\newcommand       \nuMec         {\nu_{\rm Me,3}}
\newcommand       \km        {\,{\rm km}}
\newcommand       \mol       {\,{\rm mol}}
\newcommand       \cals       {\,{\rm cal}}
\newcommand       \kcal       {\,{\rm kcal}}
\newcommand       \Me        {{\rm Me}}
\newcommand       \Ali        {3.4}
\newcommand       \Aro        {3.3}
\newcommand       \Arel       {A_{\rm ali}/A_{\rm aro}}
\newcommand       \Etot      {E_{\rm tot}}
\newcommand       \fali      {f_{\rm ali}}
\newcommand       \faro      {f_{\rm aro}}
\newcommand       \Li        {L_i}
\newcommand       \Lj        {L_j}
\journal{Journal of \LaTeX\ Templates}

\begin{document}

\begin{frontmatter}

\title{The Carriers of the Unidentified Infrared Emission Features:
Clues from Polycyclic Aromatic Hydrocarbons
with Aliphatic Sidegroups\tnoteref{mytitlenote}}




\author[mymainaddress,mysecondaryaddress]{X.J. Yang}
\ead{xjyang@xtu.edu.cn}

\author[mythirdaddress]{R. Glaser}
\ead{glaserr@missouri.edu}

\author[mysecondaryaddress]{Aigen Li}
\ead{lia@missouri.edu}

\author[mymainaddress,mysecondaryaddress]{J.X. Zhong}
\ead{jxzhong@xtu.edu.cn}

\address[mymainaddress]{Department of Physics, 
                                          Xiangtan University, 
                                          411105 Xiangtan, 
                                          Hunan Province, China}
\address[mysecondaryaddress]{Department of Physics and Astronomy,
                  University of Missouri,
                  Columbia, MO 65211, USA}

\address[mythirdaddress]{Department of Chemistry,
                  University of Missouri,
                  Columbia, MO 65211, USA}

\begin{abstract}
The unidentified infrared emission (UIE) features
at 3.3, 6.2, 7.7, 8.6, 11.3 and 12.7$\mum$ are
ubiquitously seen in a wide variety of astrophysical
regions in the Milky Way and nearby galaxies as well as
distant galaxies at redshifts $z\simgt4$.
The UIE features are characteristic of
the stretching and bending vibrations of aromatic
hydrocarbon materials. The 3.3$\mum$ feature which
results from the C--H stretching vibration in aromatic
species is often accompanied by a weaker feature
at 3.4$\mum$. The 3.4$\mum$ feature is often thought to
result from the C--H stretch of aliphatic groups attached
to the aromatic systems.
The ratio of the observed intensity of
the 3.3$\mum$ aromatic C--H feature ($\Iaro$)
to that of the 3.4$\mum$ aliphatic C--H feature ($\Iali$)
allows one to estimate the aliphatic fraction
(e.g., $N_{\rm C,aliph}/N_{\rm C,arom}$,
the number of C atoms in aliphatic units
to that in aromatic rings)
of the carriers of the UIE features,
provided that the intrinsic oscillator strengths
(per chemical bond) of the 3.3$\mum$ aromatic
C--H stretch ($\Aaro$) and the 3.4$\mum$ aliphatic
C--H stretch ($\Aali$) are known.
In this review we summarize the computational
results on $\Aaro$ and $\Aali$ 
and their implications for the aromaticity 
and aliphaticity of the UIE carriers. 
We use density functional theory and 
second-order perturbation theory
to derive $\Aaro$ and $\Aali$ 
from the infrared vibrational spectra of 
seven polycyclic aromatic hydrocarbon (PAH)
molecules with various aliphatic substituents
(e.g., methyl-, dimethyl-, ethyl-, propyl-,
butyl-PAHs, and PAHs with unsaturated alkyl chains).
The mean band strengths of the aromatic ($\Aaro$)
and aliphatic ($\Aali$) C--H stretches are derived
and then employed to estimate the aliphatic fraction
of the carriers of the UIE features by comparing
the ratio of the intrinsic band strength of the
two stretches ($\Aratio$) with
the ratio of the observed intensities ($\Iratio$).
We conclude that the UIE emitters are predominantly
aromatic, as revealed by the observationally-derived
mean ratio of $\langle\Iratio\rangle\approx 0.12$ 
and the computationally-derived mean ratio of
$\langle\Aali/\Aaro\rangle\approx 1.76$ 
which suggest an upper limit
of $N_{\rm C,aliph}/N_{\rm C,arom}\approx0.02$
for the aliphatic fraction of the UIE carriers.
\end{abstract}


\end{frontmatter}


\section{Introduction\label{sec:intro}}
A series of strong and relatively broad infrared (IR)
emission features at 3.3, 6.2, 7.7, 8.6, 11.3, and
12.7$\mum$ are ubiquitously seen in almost all astronomical
objects with associated gas and dust,
including protoplanetary nebulae (PPNe),
planetary nebulae (PNe), young stellar objects,
HII regions, reflection nebulae, the Galactic IR cirrus,
and starburst galaxies (see Tielens 2008).
These features are a common characteristic of
the interstellar medium (ISM) of the Milky Way
and nearby galaxies as well as distant galaxies
out to redshifts of $z\simgt4$
(e.g., see Riechers et al.\ 2014).
Since their first detection four decades ago in
two planetary nebulae (NGC\,7027 and BD$+$30$^{\rm o}$3639,
Gillett et al.\ 1973), the carriers of these IR emission
features have remained unidentified. Because of this,
they are collectively known as
the  ``unidentified infrared emission'' (UIE or UIR) bands.
Nevertheless, it is now generally accepted that these features
are characteristic of the stretching
and bending vibrations of some sorts of
aromatic hydrocarbon materials and therefore,
the UIE features are sometimes also referred to as
the  ``aromatic infrared bands'' (AIB).

The identification of the exact carriers of the UIE bands
is crucial for modern astrophysics:
(1) the UIE bands account for $>$\,10--20\% of
    the total IR power of the Milky Way
    and star-forming galaxies
    (Tielens 2008; Smith et al.\ 2007),
    and therefore by implication,
    their carriers must be an important
    absorber of starlight in the ultraviolet
    (UV) wavelength region
    (e.g., see Joblin et al.\ 1992,
     Cecchi-Pestellini et al.\ 2008,
     Mulas et al.\ 2013);
(2) their carriers dominate the heating of
    the gas in the diffuse ISM by providing
    photoelectrons (Lepp \& Dalgarno 1988,
    Verstraete et al.\ 1990, Bakes \& Tielens 1994,
    Weingartner \& Draine 2001);
(3) as an important sink for electrons,
    their carriers dominate the ionization balance
    in molecular clouds and hence they influence
    the ion-molecule chemistry and the ambipolar
    diffusion process that sets the stage
    for star formation (see Verstraete 2011); and
(4) because of their ubiquitous presence
    in the ISM of the local and distant universe,
    the UIE bands are often taken as a reliable
    indicator of the star-formation rates
    across cosmic time up to high redshifts
    (see Calzetti 2011, but also see Peeters et al.\ 2004).


\subsection{The UIE Carriers: Aromatic or Aliphatic?
            \label{sec:introAroAli}}
A large number of candidate materials have been proposed
as carriers of the UIE bands. All of these materials
contain aromatic structures of fused benzene rings.
The major debate lies in the exact structure of
the UIE carriers:
are they free-flying,
predominantly {\it aromatic} gas-phase molecules,
or amorphous solids
with a {\it mixed aromatic/aliphatic} composition?
In this context, the proposed carriers can be divided
into two broad categories:
\vspace{-1mm}
\begin{itemize}
\item Gas-phase, free-flying polycyclic aromatic hydrocarbon (PAH)
      molecules (see Figure~\ref{fig:uie_carriers}a).
      The PAH model attributes the UIE bands
      to the vibrational modes of PAHs
      (L\'{e}ger \& Puget 1984; Allamandola et al.\ 1985, 1989),
      with the 3.3$\mum$ feature assigned to
      C--H stretching modes,
      the 6.2$\mum$ and 7.7$\mum$ features to
      C--C stretching modes,
      the 8.6$\mum$ feature to
      C--H in-plane bending modes,
      and the 11.3$\mum$ feature to
      C--H out-of-plane bending modes.
      The relative strengths of these bands depend on
      the chemical nature of the PAH molecule
      (i.e., size, structure, and charge)
      and their physical environments
      (e.g., the intensity and hardness of
       the starlight illuminating the PAHs,
      the electron density, and the gas temperature;
      see Bakes \& Tielens 1994, Weingartner \& Draine 2001).
      Ionized PAHs emit strongly at 6.2, 7.7, and 8.6$\mum$
      while neutral PAHs emit strongly at 3.3 and 11.3$\mum$.
      The PAH model readily explains the UIE band patterns
      observed in various regions in terms of
      a mixture of neutral and charged PAHs of
      different sizes (e.g., see Allamandola et al.\ 1999,
      Li \& Draine 2001).
      PAHs have also been proposed as a possible carrier
      of several other unidentified interstellar spectral
      features, including the diffuse interstellar bands
      (DIBs; see Salama et al.\ 2011),
      and the 2175$\Angstrom$ extinction bump
      (Joblin et al.\ 1992, Li \& Draine 2001,
       Cecchi-Pestellini et al.\ 2008,
       Steglich et al.\ 2010).
\item Amorphous solids with a mixed aromatic/aliphatic composition
      (see Figure~\ref{fig:uie_carriers} for illustration).
      This category includes hydrogenated amorphous carbon
      (HAC; Jones et al.\ 1990),
      quenched carbonaceous composites (QCC; Sakata et al.\ 1990),
      and coal or kerogen (Papoular et al.\ 1989).
      As originally suggested by Duley \& Williams (1981),
      all of these materials share
      the basic molecular structure of PAHs
      by containing arenes.\footnote{%
        A benzene ring is C$_6$H$_6$.
        If the H atoms are gone, then it is not really
        ``benzene'' anymore. It is an aromatic ring system
         which can be called ``arene''.
        Arene is a hydrocarbon with alternating double
        and single bonds between carbon atoms forming rings.
        }
      They also contain aliphatic C--H bonds as well as
      other molecular structures often with other elements
      besides C and H.
%
\end{itemize}



%


\subsection{HAC\label{sec:introHAC}}
HAC can be considered to be a collection of
molecular clusters loosely aggregated to form
an extended, three-dimensional bulk solid,
with discrete aromatic ``islands'' embedded
in a three-dimensional matrix held together
unsaturated (sp and sp$^{2}$)
and saturated (sp$^{3}$) spacers
(see Figure~\ref{fig:uie_carriers}b).
The aromatic units typically contain
$\simali$1--8 benzene rings.\footnote{%
  Compared to HAC, amorphous carbon (AC)
  has a smaller H content and a larger
  aromatic cluster size of typically
  $\simali$20--40 rings and even up to
  several hundred rings
  (see Robertson 1986).
  }
Bulk HAC dust was proposed as a major constituent
of interstellar grains (see Duley et al.\ 1989).
Jones et al.\ (1990) argued that the probable
deposition conditions of the dust in the ISM
will lead to the formation of interstellar HAC.
The HAC material coated on amorphous silicate dust
was invoked to account for the UV and visual (UV-vis)
extinction, the extended red emission (ERE; see Witt 2014).
and the blue luminescence (Vijh et al.\ 2005).
The HAC hypothesis for the UIE features postulates
that the energy of the absorbed photons is localized
in a small region of $\simali$1$\nm$ of the bulk,
submicrometer-sized dust and therefore the aromatic
islands are transiently heated to temperatures in
excess of those expected for the bulk material
(Duley \& Williams 1988). The UIE features are thought to
arise from the aromatic islands at their temperature spikes.





\subsection{QCC\label{sec:introQCC}}
QCC is an experimentally-synthesized condensate
of low molecular-weight hydrocarbons
generated from a hydrocarbon plasma.
It is synthesized by quenching the plasma
of low pressure methane gas
excited to high temperatures with a microwave generator
(Sakata et al.\ 1990).
The experimental conditions for making QCC are similar
to what one would expect in the atmospheres of cool
evolved stars where the rapid condensation of hydrocarbon
dust occurs.



QCC is composed of aromatic and aliphatic molecules
as well as radicals assembled in a random manner to
form an amorphous solid 
(see Figure~\ref{fig:uie_carriers}c).
QCC contains four kinds of organic components
such as
arenes, alkynes and polyynes (CC triple bonds),
olefins (CC double bonds),
and saturated hydrocarbon spacers
and substituents (e.g., alkyl groups).
The aromatic component typically contains
$\simali$1--4 rings and mostly only one or
two rings like benzene and naphthalene.
These rings are connected by aliphatic chains
(bridging) and can be randomly
cross-linked together
in a three-dimentional structure
(see Sakata et al.\ 1990).






The QCC model interprets the 7.7 and 8.6$\mum$ UIE bands
as arising from the ketone (C=O) bond of
a ``cross-conjugated ketone'' (CCK) molecular structure
within oxidized QCC.
The 6.2$\mum$ UIE band is attributed to
the C=O stretching
as well as the skeletal in-plane vibration
of C=C (Sakata et al.\ 1990).
The ``solo'' H atoms on carbon are thought to be
responsible for the 3.3 and 11.3$\mum$ UIE features.
While the oxidation of QCC may occur in the ISM,
it is less likely to occur in carbon stars because
of the lack of O atoms which are presumably all
locked up in CO in the atmospheres of carbon stars.


\subsection{Soot\label{sec:introSoot}}
\begin{figure*}
\centerline{
\includegraphics[scale=0.75,clip]{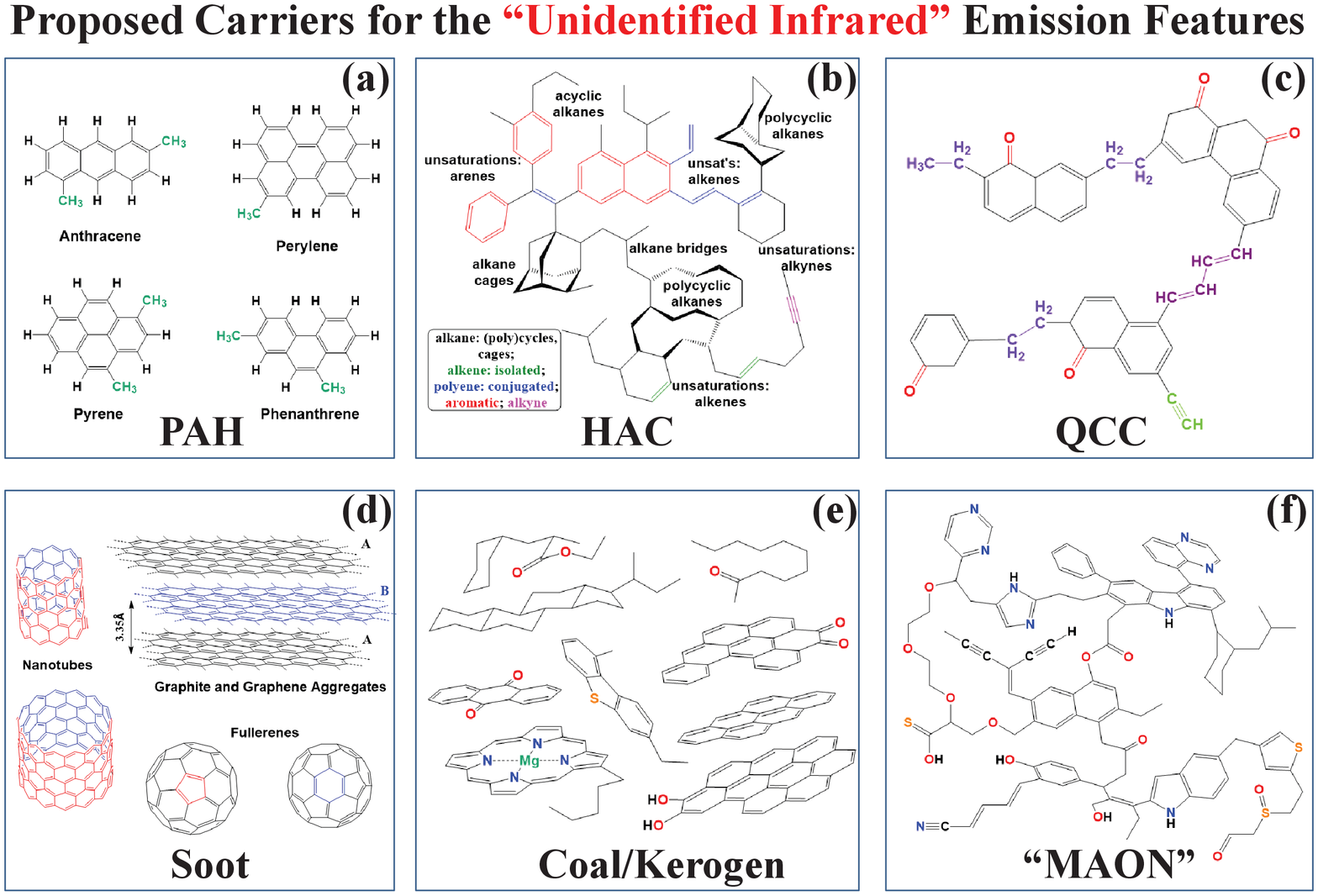}
}
\caption{\footnotesize
         \label{fig:uie_carriers}
        Schematic structures for
        the possible UIE carriers:
        (a) free-flying PAHs,
        (b) bulk HAC,
        (c) bulk QCC,
        (d) bulk soot,
        (e) bulk coal or kerogen, and
        (f) nano MAONs
            (``mixed aromatic/aliphatic
               organic nanoparticles'').
        }
\end{figure*}

Tielens (1990) argued that the condensation of
carbon dust from acetylene (C$_2$H$_2$) molecules
in the outflow from carbon-rich red giants
is probably very similar to that occurring
during the gas phase pyrolysis
of hydrocarbon molecules which leads to
formation of soot.
Soot is a general side product of the combustion
and pyrolysis of hydrocarbons
such as methane, acetylene, or benzene.
Soot consists of large planar PAH molecules
stacked together to form platelets
which are the building blocks of soot particles
(see Figure~\ref{fig:uie_carriers}d).
These platelets as well as the layers within
them are generally cross-linked by tetrahedrally
bonded carbon atoms and chains
(not shown in Figure~\ref{fig:uie_carriers}d).
The formation of soot starts from the conversion
of acetylene to small PAH molecules which rapidly
grow to large aromatic species.
The aromatic platelets are randomly stacked to form
three-dimensional structures of sizes of
$\simali$1--10$\nm$ which further grow into
soot particles by clustering and agglomerization
(see Tielens 1990). Other components of soot particles
include carbon nanotubes and fullerenes
(see Figure~\ref{fig:uie_carriers}d).
C$_{60}$ has recently been detected in
a wide range of astrophyical regions
(see Cami et al.\ 2010, Sellgren et al.\ 2010).

Balm \& Kroto (1990) assigned the 11.3$\mum$ UIE
feature to soot-like microparticles with internal H atoms.
Allamandola et al.\ (1985) compared the 5--10$\mum$
Raman spectrum of auto soot with the Orion UIE bands.
They found that soot has IR spectral features
in close correspondence to that seen in space.








\subsection{Coal and Kerogen\label{sec:introCoal}}
Papoular et al.\ (1989)
were the first ones to draw attention
to coal as a possible model for
understanding the UIE bands.
They showed that the absorption spectra
of vitrinite, the major organic component
of demineralized coal, resemble the observed
UIE bands.
%
Coal is mainly composed of C, H, and O
and differs from HAC partly by its higher oxygen content
($\simali$2 to 20\% by mass).
A large fraction of the carbon in coal is in the form of
condensed aromatic units
arranged in graphite-like ``bricks''.
These ``basic structural units''
(of $\simali$1.5$\nm$ in size)
are randomly oriented
and made of stacks of a few layers of
planar arene systems packed together
to form an irregular carbon skeleton
(see Figure~\ref{fig:uie_carriers}e).
Most of the H atoms are bonded to this carbon skeleton,
while O atoms bridge the gaps between the ``bricks''.
H and O form, together with C, simple functional
groups attached to the inter-connected ``bricks''
that are responsible for the vibrational bands
that mimic the UIE bands.

The IR spectra of coal differ considerably
according to the coal history
which is quantified by its ``rank'', ``order''
or carbon content (or, equivalently, its age,
or the mining depth of the seam in which it originated).
As coal ages, the concentration of H and O
in it decreases and the C content increases.
With increasing C content,
H/C decreases, O/C decreases,
aromaticity increases,
the degree of substitution in aromatic rings
decreases and the rank or order of coal improves.
For high ranking coals (with a carbon-content of $>$90\%),
the intensity of the aromatic C--H stretching band
relative to that of the aliphatic band increases
very steeply as a result of
a decrease of the number of aliphatic H atoms and
an increase of the aromatic H atoms.
According to Papoular et al.\ (1989),
the average UIE carrier is best mimicked by
anthracite, the highest ranking coal
(i.e., most graphite-like).


Papoular (2001) analyzed the absorption spectra
of terrestrial kerogen materials and argued
that kerogen could explain the observed UIE features.
Kerogen is a family of highly disordered
macromolecular organic materials made of
C, H and O, and traces of N and S
(see Figure~\ref{fig:uie_carriers}e).
It is a solid sedimentary, insoluble organic material
found in the upper crust of the Earth in dispersed form.
The main difference between kerogen
and coal is that the latter is found in
the form of bulk rocks and the former in
dispersed form (sand-like).
The term kerogen is also often used to designate
the insoluble, three-dimensional,
organic macromolecular skeleton
which is the main constituent of coal.
Upon aging, kerogen changes its composition
with oxygen expelled in the form of CO, CO$_2$ and H$_2$O
and hydrogen expelled in the form of methane (CH$_4$).
This will break the aliphatic chains and
allow aromatic rings to form and coalesce in clusters
in kerogen and therefore increase its aromaticity.
Papoular (2001) argued that the great diversity
of the astronomical UIE spectra could be explained
by kerogen of different evolutionary stages
characterized by different ratios
of O to C concentrations,
and of H to C concentrations.

Cataldo et al.\ (2013) compared the UIE features
observed in some PPNe with the experimentally-measured
absorption spectra of heavy petroleum fractions
and asphaltenes. A series of heavy petroleum fractions
(e.g., ``distillate aromatic extract'',
``residual aromatic extract'',
heavy aromatic fraction (BQ-1)
and asphaltenes derived from BQ-1)
were considered.
They found that the band pattern
of the UIE features
(particularly that of the aromatic-aliphatic
C--H stretching bands)
of certain PPNe is closely matched by
the BQ-1 heavy aromatic oil fraction
and by its asphaltene fraction.
Like coal or kerogen,
the heavy petroleum fractions
contain a mix of aromatic
and aliphatic structures.
They are composed of aromatic cores containing three
to four condensed aromatic rings surrounded
by cycloaliphatic (naphthenic) and aliphatic alkyl chains.
In comparison with coal, the heavy petroleum fractions
are viscous liquids at room temperature
and become glassy solids below $\simali$235\,K.

\subsection{Excitation Mechanism:
            Equilibrium Temperatures
            or Stochastic Heating?
            \label{sec:introExMech}}
The HAC, QCC, and coal/kerogen hypotheses
all assume that the UIE bands
arise following photon absorption in
small thermally-isolated aromatic units
within or attached to these bulk materials.
However, it does not appear possible to confine
the absorbed stellar photon energy within these aromatic
``islands'' for the time $\gtsim10^{-3}\s$ required for
the thermal emission process (see Li \& Draine 2002).
Bulk materials like HAC, QCC and coal
have a huge number of vibrational degrees of freedom and
therefore their heat capacities are so large that they will
attain an equilibrium temperature of $T$\,$\simali$20$\K$
in the diffuse ISM (see Li 2004).
With $T$\,$\simali$20$\K$,
they will not emit efficiently in the UIE bands
at $\lambda$\,$\simali$3--12$\mum$ (Draine \& Li 2007).

It has been observationally demonstrated that
the UIE profiles remain constant even the exciting
starlight intensities vary by five orders of magnitude
(e.g., see Boulanger et al.\ 1999).
The equilibrium temperature $T$ depends on
the starlight intensity $U$
[e.g., $T$\,$\propto$\,$U^{1/\left(4+\alpha\right)}$
if the far-IR emissivity of the dust is
proportional to $\lambda^{-\alpha}$].
Therefore, if the UIE bands arise from
bulk materials like HAC, QCC, or coal,
one would expect the UIE profiles to vary
with the starlight intensity.
Furthermore, Sellgren et al.\ (1983) also
showed that in some reflection nebulae
the UIE profiles and
the color temperatures
of the smooth continuum emission
underneath the 3.3$\mum$ UIE feature
show very little variation
from source to source and
within a given source
with distance $r$ from the central star.
Sellgren (1984) argued that the UIE features
and the associated continuum emission
could not be explained by thermal emission
from bulk dust in radiative equilibrium with
the central star.
Otherwise one would expect them to vary with $r$
as the equilibrium temperature $T$
of bulk dust is expected to decline
with $r$: $T\propto r^{-2/\left(4+\alpha\right)}$.

Recognizing the challenge of
bulk materials like HAC, QCC and coal
in being heated to emit the UIE bands,
Kwok \& Zhang (2011, 2013) recently proposed
the so-called MAON model: they argued that the UIE bands
arise from coal- or kerogen-like organic nanoparticles,
consisting of chain-like aliphatic hydrocarbon material
linking small units of aromatic rings,
where MAON stands for ``{\it mixed aromatic/aliphatic
organic nanoparticle}''
(see Figure~\ref{fig:uie_carriers}f).
The major improvement of the MAON model
over the earlier HAC, QCC and coal hypotheses is that
the MAON model hypothesizes
that the coal-like UIE carriers are {\it nanometer} in size
so that their heat capacities are smaller than or comparable
to the energy of the starlight photons that excite them.
Upon absorption of a single stellar photon,
they will be stochastically heated to high temperatures
to emit the UIE bands (see Draine \& Li 2001).
The stochastic heating nature of PAHs
guarantees that the UIE spectra
(scaled by the starlight intensity)
to remain the same
for different starlight intensities.\footnote{%
  Single-photon heating implies that the shape of the
  high-$T$ end of the temperature ($T$) probability
  distribution function for a nanoparticle is the same
  for different levels of starlight intensity,
  and what really matters is the mean photon energy
  (which determines to what peak temperature
  a nanoparticle will reach upon
  absorption of such a photon).
  }
This is true for both hard radiation fields
and soft radiation fields
(see Draine \& Li 2001, Li \& Draine 2002).
As demonstrated in Figure~1f of Draine \& Li (2007),
the UIE spectra predicted from the PAH model are
essentially the same even if the illuminating
starlight intensities differ by a factor of 10$^5$.


To summarize, it is fair to conclude that,
based on the brief descriptions of
the proposed carriers presented above,
the current views about the UIE carriers
generally agree that (1) the UIE features arise from
the {\it aromatic} C--C and C--H vibrational modes,
and (2) the carriers must be {\it nanometer} in size
or smaller (e.g., large molecules).
The dispute is mainly on the structure
of the UIE carriers:
are they predominantly {\it aromatic} (like PAHs),
or {\it largely aliphatic}
but mixed with small aromatic units (like MAONs)?

\subsection{Are the UIE Carriers Aromatic or Aliphatic?
            Constraints from the C--H Stretching Features
            \label{sec:introConstr}}
\begin{figure*}
\centerline{
\includegraphics[scale=0.75,clip]{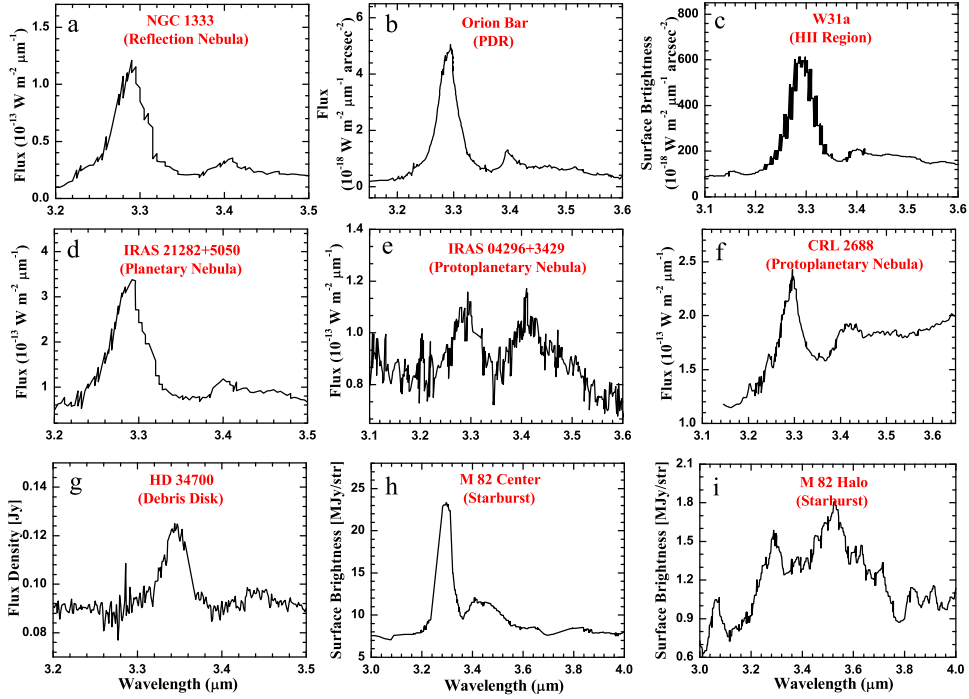}
}
\caption{\footnotesize
              \label{fig:astrospec}
              Aromatic and aliphatic C--H stretching 
              emission features seen in representative 
              astrophysical regions:
              (a) NGC~1333 (reflection nebula, Joblin et al.\ 1996),
              (b) Orion bar (photodissociation region [PDR], 
                   Sloan et al.\ 1997), 
              (c) W31a (HII region, Mori et al.\ 2014), 
              (d) IRAS~21282+5050 (planetary nebula, 
                   Nagata et al.\ 1988), 
              (e) IRAS~04296+3429 (protoplanetary nebula, 
                   Geballe et al.\ 1992), 
              (f) CRL~2688 (protoplanetary nebula, 
                  Geballe et al. 1992), 
              (g) HD~34700 (debris disk, Smith et al.\ 2004), 
              (h) M82 center (starburst galaxy,
                   Yamagishi et al.\ 2012), and 
              (i) M82 halo  (Yamagishi et al.\ 2012).       
               }
\end{figure*}

Are the UIE carriers aromatic or aliphatic?
A straightforward way to address this question
is to examine the {\it aliphatic fraction}
of the UIE carriers
(i.e., the fraction of carbon atoms in aliphatic chains).
Aliphatic hydrocarbons have a vibrational
band at 3.4$\mum$ due to the C--H stretching mode
(Pendleton \& Allamandola 2002).
In many interstellar and circumstellar environments
the 3.3$\mum$ emission feature is indeed often
accompanied by a weaker feature at 3.4$\mum$
(see Figure~\ref{fig:astrospec} for illustration).
As demonstrated by Li \& Draine (2012)
and Yang et al.\ (2013), one can place an upper limit
on the aliphatic fraction of the emitters
of the UIE features by assigning the 3.4$\mum$ emission
{\it exclusively} to aliphatic C--H
(also see Rouill{\'e} et al.\ 2012,
Steglich et al.\ 2013).\footnote{%
  This is indeed an {\it upper limit}
  as the 3.4$\mum$ emission feature could also be
  due to {\it anharmonicity} of the aromatic C--H
  stretch (Barker et al.\ 1987)
  and ``{\it superhydrogenated}'' PAHs
  whose edges contain excess H atoms
  (Bernstein et al.\ 1996, Sandford et al.\ 2013).
  }
This requires the knowledge of
the intrinsic oscillator strengths of
the 3.3$\mum$ aromatic C--H stretch ($\Aaro$)
and the 3.4$\mum$ aliphatic C--H stretch ($\Aali$),
where $\Aaro$ and $\Aali$ are on a per unit C--H bond basis.

In this review we summarize our recent
work on computing the IR vibrational
spectra of a range of PAH molecules
with various aliphatic sidegroups
(e.g., methyl-, dimethyl-, ethyl-, propyl-,
butyl-PAHs, and PAHs with unsaturated alkyl chains),
based on density functional theory 
and second-order perturbation theory.
The mean band strengths of the aromatic
and aliphatic C--H stretches are derived
and then employed to estimate the aliphatic fraction
of the carriers of the UIE features by comparing
the ratio of the intrinsic band strength of the
two stretches with
the ratio of the observed intensities.

In \S\ref{sec:Method} we describe
the computational methods
and the parent molecules
based on which we derive the band strengths
of the aromatic and aliphatic C--H stretches.
The structures and stabilities of methylated PAHs
are discussed in \S\ref{sec:structure}.
We report in \S\ref{sec:ParentPAHs} the computed frequencies
and intensities of the C--H stretching modes of
the parent PAHs
as well as that of their methylated derivatives
in \S\ref{sec:MethylPAHs}.
Theoretical level dependencies of
the computed band intensities
and approaches to intensity scaling
are discussed in detail in \S\ref{sec:I_scale}.
In \S\ref{sec:recomA} we present the recommended
mean band intensities for the aromatic and aliphatic
C--H stretches.
We estimate in \S\ref{sec:astro} the aliphatic fraction
of the UIE carriers from the mean ratio of the observed
intensities of the 3.3$\mum$ aromatic and
3.4$\mum$ aliphatic C--H features.
We summarize our major results in \S\ref{sec:summary}.

A considerable fraction of this review 
is concerned with the computational techniques 
and the resulting frequencies and intensities for 
the aromatic and aliphatic C--H stretching modes 
of a range of PAHs and their methylated derivatives.
For more details we refer the interested readers
to Yang et al.\ (2016a,b, 2017).
Readers who are interested only in
the mean ratio of the band strength of
the 3.4$\mum$ aliphatic C--H stretch
to that of the 3.3$\mum$ aromatic C--H stretch
($\Aali/\Aaro$)
and their implications for the aliphatic fraction
of the UIE carriers
may wish to proceed directly to \S\ref{sec:recomA}.

\section{Computational Methods and Target Molecules
         \label{sec:Method}
         }
We use the Gaussian09 software (Frisch et al.\ 2009)
to calculate the IR vibrational spectra for a range of
aromatic molecules with a methyl side chain
(see Figure~\ref{fig:MonoMethylPAHs}).
We have considered benzene (C$_6$H$_6$),
naphthalene (C$_{10}$H$_8$),
anthracene (C$_{14}$H$_{10}$),
phenanthrene (C$_{14}$H$_{10}$),
pyrene (C$_{16}$H$_{10}$),
perylene (C$_{20}$H$_{12}$),
and coronene (C$_{24}$H$_{12}$),
as well as all of their methyl derivatives
(see Figure ~\ref{fig:MonoMethylPAHs}).
%

We employ the hybrid density functional
theoretical method (B3LYP)
in conjunction with a variety of basis sets:
{\rm 6-31G$^{\ast}$},
{\rm 6-31+G$^{\ast}$},
{\rm 6-311+G$^{\ast}$},
{\rm 6-311G$^{\ast\ast}$},
{\rm 6-31+G$^{\ast\ast}$},
{\rm 6-31++G$^{\ast\ast}$},
{\rm 6-311+G$^{\ast\ast}$},
{\rm 6-311++G$^{\ast\ast}$},
{\rm 6-311+G(3df,3pd)}, and
{\rm 6-311++G(3df,3pd)}. 
Here the Slater-type atomic orbitals (AOs)
   are described by one or more 
   ``basis functions (BFs)''
   and each basis function usually is described by
   a sum of several Gaussian functions with various
   radial distributions (reflected in the exponents).
   For example, the ``6'' in 6-31G$^{\ast}$
   indicates that every core AO is described by
   one basis function which is expressed as
   a sum of 6 ``primitive'' Gaussian functions.
   The ``31'' part describes a ``split-valence'' basis set,
   that is, every valence AO is described by
   two independently varied basis functions,
   and it is primarily this feature that allows
   the electron density to adopt the best radial
   distribution for any given bonding situation.
   Similarly, ``-311'' refers to a ``triply-split valence''
   basis set (3 basis functions for every valence AO)
   and allows for even more freedom to describe
   the electronic wave function.
   While atomic orbitals have perfect
   s- and p-shapes, the electron density distributions
   within molecules are polarized (oriented)
   and so-called ``polarization functions''
   are added to the basis set to allow for
   an improved description of this polarization around atoms.
   Small admixtures of p-type basis functions
   polarize s-type AOs, small admixtures of d-type BFs
   polarize p-AOs, small admixtures of f-type BFs
   polarize d-AOs, etc.  Information about the types
   and number of polarization functions is provided
   after the ``G'' in the basis set descriptor,
   first for non-H atoms and then for H-atoms.
   For example, that large basis set 6-311+G(3df,3pd)
   contains three sets of d-functions and one set of
   f-functions on every carbon atom and it contains
   three sets of p-functions and one set of d-functions
   on every hydrogen. The basis set 6-31G$^{\ast\ast}$
   denotes 6-31G(d,p).
   ``Diffuse functions'' are spatially rather extended
   basis functions and they are important to reproduce
   electric multipoles with high accuracy.
   The augmentation of a basis set with diffuse functions
   is indicated by ``+'' signs, where the first ``+'' refers
   to heavy atoms (C in our cases)
   and the second ``+'' refers to H-atoms.
   In general, the quality of the wave function
   improves with the number of basis functions,
   with the number of primitives per basis function,
   with the number of polarization functions,
   and with the extent of diffuse-function augmentation.
   Since computer-time needs scale exponentially
   with the number of basis functions N,
   the art consists in understanding
   the essential requirements.
%

We also employ second-order
M$\o$ller-Plesset perturbation theory
(hereafter abbreviated as MP2)
in conjunction with the basis sets
{\rm 6-311+G$^{\ast\ast}$}
and {\rm 6-311++G(3df,3pd)}
for some of the molecules.
The MP2 computations were performed
either with the full active space of
all core and valence electrons
considered in the correlation energy computation,
denoted MP2(full),
or with the frozen core approximation
and the consideration of just the valence electrons
in the correlation treatment, denoted MP2(fc).

In this work all molecules are optimized 
and calculated at {\rm B3LYP/6-31G$^{\ast}$}.
Benzene, toluene and naphthalene
and methylnaphthalenes are studied
using several theoretical levels:
{\rm B3LYP/6-311+G$^{\ast\ast}$},
{\rm B3LYP/6-311+G(3df,3pd)},
{\rm MP2/6-311+G$^{\ast\ast}$},
and {\rm MP2/6-311+G(3df,3pd)}.
Toluene and isomers of methylpyrene are studied
at B3LYP using basis sets from {\rm 6-31G$^{\ast}$}
all the way up to {\rm 6-311++G(3df,3pd)}.
Scaling will be applied to frequencies and intensities.
We employ the frequency scale factors recommended for
the various theoretical levels (see Borowski 2012,
Andersson et al.\ 2005, Merrick et al.\ 2007)
and their values are listed in Table~\ref{Table:Freq_scale}.
For the intensity scaling factors, we will discuss
in detail in \S\ref{sec:I_scale} regarding
the theoretical level dependency of
the computed band intensities.

\section{Structures and Stabilities of Methylated PAHs
         \label{sec:structure}
         }
\begin{figure}[ht]
 \vspace{-2mm}
  \begin{center}
  \epsfig{file=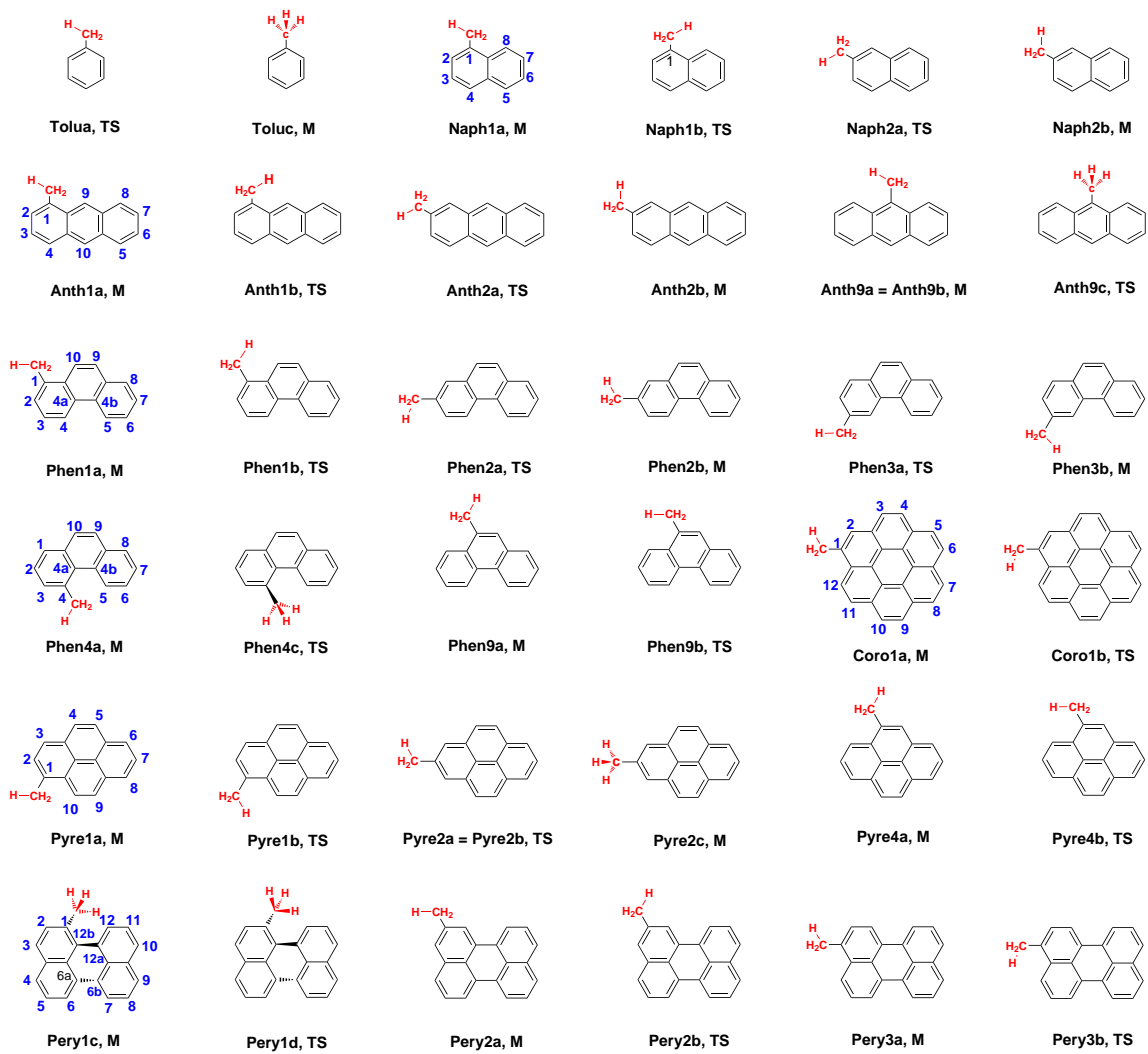,width=15.2cm}
  \end{center}
\vspace{-6mm}
\caption{\label{fig:MonoMethylPAHs} \footnotesize
         Structures of the mono-methyl (${\rm -CH_3}$) derivatives
         of seven aromatic molecules together with the standard IUPAC
         numbering:
         benzene (C$_6$H$_6$),
         naphthalene (C$_{10}$H$_8$),
         anthracene (C$_{14}$H$_{10}$),
         phenanthrene (C$_{14}$H$_{10}$),
         pyrene (C$_{16}$H$_{10}$),
         perylene (C$_{20}$H$_{12}$), and
         coronene (C$_{24}$H$_{12}$).
         We use the first four letters of the parent molecules to
         refer to them and attach the position number of
         the location of the methyl group
         (e.g., Naph1 for 1-methylnaphthalene).
         The mono-methyl derivative of benzene
         is known as toluene (i.e., ``Tolu'', C$_7$H$_8$).
         Depending on where the methyl side-group
         is attached, a molecule will have several
         isomers (e.g., monomethyl-pyrene has three isomers
         in which the -CH$_3$ group is attached
         to carbon 1, 2, or 4, respectively).
         We also indicate whether the structure
         is a minimum (M) or a transition state (TS)
         structure for the methyl rotation.
	 }
\vspace{-3mm}
\end{figure}

The molecules studied are shown in
Figure~\ref{fig:MonoMethylPAHs}
together with the standard
{\it International Union of Pure and Applied Chemistry}
(IUPAC) numbering scheme.\footnote{%
   http://www.iupac.org
   \label{ftnt:iupac}
   }
We use the first four letters of the molecules
to refer to them and attach the position number
of the location of the methyl group.
For example, 1-methylnaphthalene is referred to as Naph1.
The methyl conformations are indicated
in Figure~\ref{fig:MonoMethylPAHs}
and there are several possibilities.

Depending on the symmetry of the molecule,
there are one or two stereoisomers
in which one of the methyl-CH bonds lies
in the plane of the arene.
We differentiate between these stereoisomers
by addition of ``$a$'' or ``$b$'' to the name of
the structure isomer,
and the in-plane C--H bond points into
the less (more) crowded hemisphere
in the $a$-conformation
($b$-conformation).\footnote{%
   Take Naph1a and Naph1b as examples.
   In Naph1a, the in-plane methyl-H is four bonds
   away from the closest H-atom, H at C2.
   In Naph1b, the in-plane methyl-H is five bonds
   away from the closest H-atom, H at C8.
   This leaves more space between the in-plane
   methyl-H and H(C2) in Naph1a than
   between in-plane methyl-H and H(C8) in Naph1b,
   and Naph1a is less crowded than Naph1b.
   }
In most cases, either the $a$-conformation
or the $b$-conformation corresponds
to the minimum
while the other conformation
corresponds to the transition state structure
for methyl rotation.\footnote{%
    A structure on the potential energy surface is
    a ``stationary structure''
    if the net inter-atomic forces
    on each atom is acceptably close to zero.
    A ``minimum'' is a stationary structure
    for which a small distortion along any
    internal coordinate increases the energy
    (all curvatures are positive).
     A ``transition state structure''
    is a stationary structure for which
    a small distortion along one internal
    coordinate lowers the energy while distortions
    along any of the other internal coordinates
    increases the energy.
    The internal coordinate with the negative curvature
    is called the ``transition vector''.
    For the rotational transition state structures,
    the transition vector describes a rotation of
    the methyl group about the H$_3$C--C bond
    and serves to scramble the H atoms
    in the associated minimum structures
    (i.e., Naph1a can be realized with any
    one of the three methyl-Hs in the plane).
    }
Note that the $a$-conformation can be the minimum (i.e., Naph1a)
or the rotational transition state structure (i.e., Naph2a).

The molecules Tolu, Anth9 and Pyre2 are symmetric
and the $a$- and $b$-conformations are identical.
In these cases there exists an additional conformation type,
the $c$-conformation, in which one of the methyl-CH bonds is
almost perpendicular with respect to the arene plane.
For Tolu and Pery2, the $c$-conformation is the minimum
while the conformations with in-plane CH-bonds are
the rotational transition state structures.
In contrast, for Anth9 the $c$-conformation
serves as the transition state for interconversion
between the conformations with in-plane CH-bonds.

\begin{figure}
\vspace{-10cm}
\centerline
{
\includegraphics[width=14.8cm,angle=0]{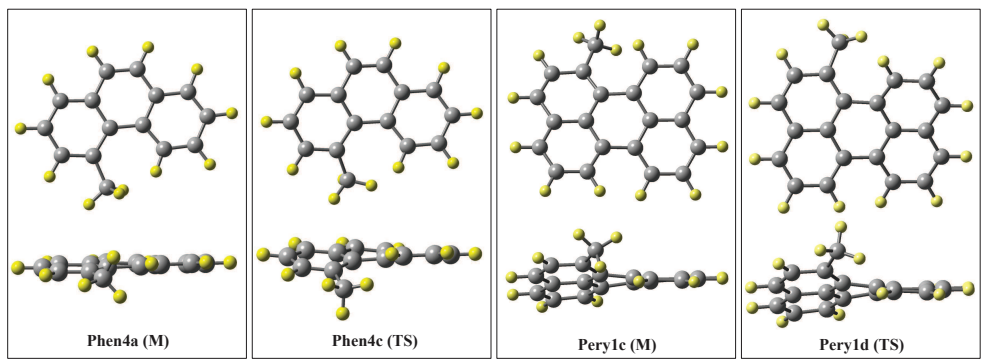}
}
\vspace{-7mm}
\caption{\footnotesize
         \label{fig:Phen4_Pery1}
         Optimized structures of 4-methylphenanthrene
         [Phen4a (M) and Phen4c (TS)]
         and 1-methylperylene [Pery1c (M) and Pery1d (TS)].
         H atoms are marked with
         color yellow and C atoms are color grey.
         M and TS respectively refer to
         the minimum (M) or the transition state (TS)
         structure for the methyl rotation.
         }
\end{figure}

The structures are generally unremarkable.
In most cases the $a$-, $b$- and $c$-conformations
all feature essentially planar arenes
and only Phen4 and Pery1 stand out
and their structures are shown
in Figure~\ref{fig:Phen4_Pery1}.
While there is no problem to accommodate
the in-plane CH-bond in Phen4a,
a Phen4b-type structure is prohibited
because of the extraordinary steric
interaction between the in-plane
methyl CH-bond and the C(5)H-bond.\footnote{%
    C(5) stands for the carbon atom numbered 5
    in the corresponding molecule
    as marked in Figure~\ref{fig:MonoMethylPAHs}.
    }
It is required that the entire methyl group is bent out of
the best arene plane to accomplish the methyl rotation
via the transition state structure Phen4c.
In the case of Pery1, the steric repulsion
between the methyl group and the C(12)H-bond
is so strong that both the minimum Pery1c
and the transition state structure Pery1d
contain twisted perylene moieties and,
in both structures, the methyl group is
moved out of the best plane of the arene
to which it is attached.
Note that one methyl-CH bond is nearly
perpendicular to the attached arene Pery1c and Pery1d.
The difference between Pery1c and Pery1d concerns
the orientation of this near-perpendicular CH bond
relative to to the other naphthalene moiety
(the one without the methyl group);
it points either to (in Pery1c)
or away (in Pery1d) from the best plane
of the other naphthalene.

The structure of Phen4a is no longer C$_{\rm s}$-symmetric:
the ``in-plane'' CH-bond is slightly out of the plane
($\angle$(H--C--C4--C3)\,=\,9.5$^\circ$),
the methyl-C no longer lies in the arene plane
($\angle$(H$_{3}$C--C4--C3--C2)\,=\,177.3$^\circ$),
and the phenanthrene frame is twisted significantly
($\angle$(C4--C4a--C4b--C5)\,=\,8.2$^\circ$).
In Phen4c, one CH-bond is almost perpendicular to
the attached benzene plane
($\angle$(H--C--C4--C3)\,=\,83.1$^\circ$),
the methyl-C is moved more out of the plane
of the attached benzene
($\angle$(H$_{3}$C--C4--C3--C2)\,=\,167.4$^\circ$),
and the phenanthrene frame twisting grows significantly
($\angle$(C4--C4a--C4b--C5)\,=\,20.6$^\circ$).\footnote{%
   C4(a) stands for the carbon atom numbered 4a
   in the corresponding molecule
   as marked in Figure~\ref{fig:MonoMethylPAHs}.
   }
As can be seen, in Pery1c the two naphthalene moieties
are twisted significantly
($\angle$(C6b--C6a--C12b--C12a)\,=\,13.9$^\circ$)
and the methyl group also is slightly pushed out of
the arene plane and away from the H atom at C12
($\angle$(H$_{3}$C--C1--C2--C3)\,=\,176.4$^\circ$).
In the transition state structure Pery1d
both of these deformations are enhanced with
$\angle$(C6b--C6a--C12b--C12a\,=\,13.9$^\circ$
and $\angle$(H$_{3}$C--C1--C2--C3)\,=\,170.5$^\circ$.

\section{Results\label{sec:results}}
All of the molecules are studied
in all conformations
at the {\rm B3LYP/6-31G$^{\ast}$} level
and the computed total energies
and the thermochemical parameters
are summarized
in Table~\ref{Table:Summary_min_631}
for the minima
and in Table~\ref{Table:Summary_TS_631}
for the transition states.
For reasons that will be given below,
the minima of all of the molecules are also
studied at the {\rm B3LYP/6-311+G$^{\ast\ast}$} level.
Table~\ref{Table:Summary_min_6311} lists
the computed total energies
and the thermochemical parameters
obtained at the {\rm B3LYP/6-311+G$^{\ast\ast}$} level.

%
%
%

The vibrational frequencies and intensities for
the aromatic and the methyl (aliphatic) C--H stretching
modes were computed. The standard scaling is applied
to the frequencies by employing the scale factors
listed in Table~\ref{Table:Freq_scale}.
The scaling for the intensities will be discussed
in \S\ref{sec:I_scale}.

\begin{figure*}
\begin{center}
\includegraphics[scale=0.75,clip]{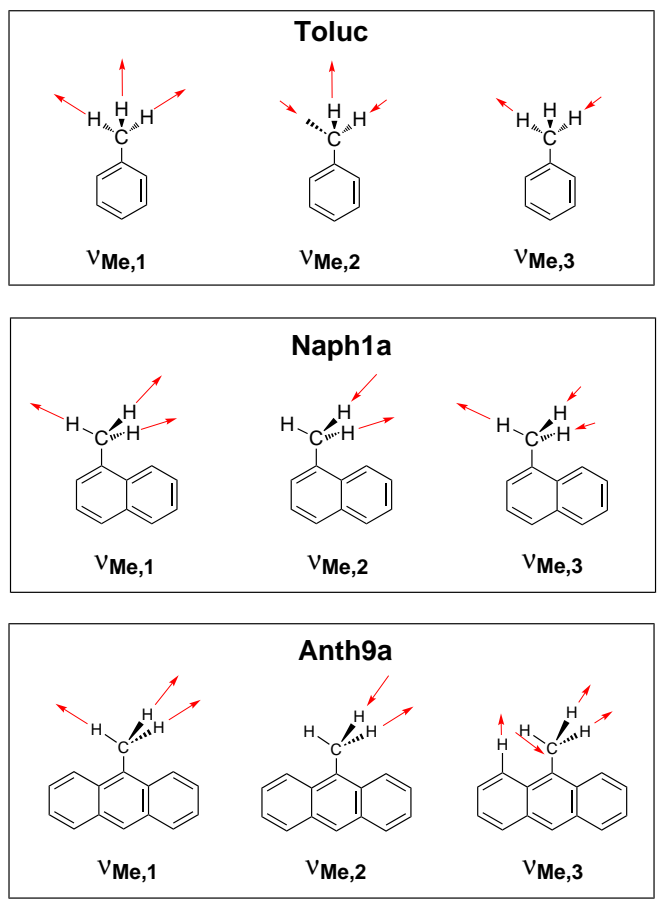}
\caption{\footnotesize
         \label{fig:methyl_schetch_modes}
         Schematic illustration of the C--H stretching modes
         of toluene (Toluc, M),
         naphthalene (Naph1a, M), and
         anthracene (Anth9a, M).
         Irrespective of the methyl conformation,
         $\nuMea$ refers to the one symmetric stretching mode
         in which three CH bonds lengthen/shorten
         at the same time, $\nuMeb$ refers to the asymmetric
         stretching mode in which one CH bond shortens
         while another lengthens, and $\nuMec$ refers to
         the asymmetric stretching modes in which two CH bonds
         change in phase and in opposite phase
         to the third CH bond.  These labels are used
         independently of the frequency of the three modes.
         }
\end{center}
\end{figure*}

The calculations always show three methyl
C--H stretches for all the methyl derivatives
of all the molecules, and we always describe
these three bands as $\nuMea$, $\nuMeb$, and $\nuMec$
as illustrated in Figure~\ref{fig:methyl_schetch_modes}.
For demonstration, we show in
Table~\ref{Table:Freq_int_tolu}
the frequencies and intensities
computed at some of the levels
for toluene along with
the gas-phase experimental data
of Wilmshurst \& Bernstein (1957)
and of the {\it National Institute
of Standards and Technology} (NIST).\footnote{%
   http://webbook.nist.gov
   }

\subsection{Frequencies and Intensities of Parent PAHs
         \label{sec:ParentPAHs}
         }
\subsubsection{Benzene}
Selected computed spectra of benzene are
shown in Figure~\ref{fig:ParentPAH_IR} together with
the experimental gas-phase spectrum taken from NIST.
The computed spectrum shows just one band which corresponds
to the fundamental mode $\nu_{12}$
in the Herzberg nomenclature (Herzberg 1945).
The experimental gas phase spectrum shows three bands
at 3045, 3073 and 3099$\cm^{-1}$
and only the last one of these is a pure C--H vibration.
The measured $\nu_{12}$ band appears
at $\simali$3099$\cm^{-1}$ and it is
about 22$\cm^{-1}$ above the scaled computed
frequency because of Fermi resonance
(FR; Herzberg 1945).
The pure C--H stretching mode $\nu_{12}$
(at $\simali$3076$\cm^{-1}$ without FR)
forms a resonance doublet with the combination band
($\nu_{13}$\,+\,$\nu_{16}$) which shows a similar
offset of $\simali$23.8$\cm^{-1}$ in the opposite
direction ($\nu_{13}$\,=\,1485$\cm^{-1}$;
$\nu_{16}$\,=\,1584.8$\cm^{-1}$).
%
The quantitative NIST spectrum\footnote{%
   The intensities for benzene are taken from
   the 3-term Blackman-Harris entries
   with a resolution of 0.125$\cm^{-1}$.
   }
gives an absorption intensity of $\simali$54.4$\km\mol^{-1}$
for the aromatic C--H stretches in benzene in gas phase,
and this value is in close agreement with
the intensity of $\simali$55$\km\mol^{-1}$
reported by Pavlyuchko et al.\ (2012).
On the other hand, Bertie \& Keefe (1994)
gave a significantly higher value of
$A_{\rm aro}(\nu_{12})\approx 73\pm{9}\km\mol^{-1}$
based on their integration over
the range of 3175--2925$\cm^{-1}$.
This region contains some intensity
from the (weak) combination bands
and all the experimental intensity data
are much lower than the value
$A_{\rm aro}(\nu_{12})\approx104\km\mol^{-1}$
computed at the {\rm B3LYP/6-31G$^{\ast}$} level.
Clearly, we will need to take a closer look
at the accuracy of the computed intensities.

\subsubsection{Important Parent PAHs}
\begin{figure*}
\begin{center}
\begin{minipage}[t]{0.45\textwidth}
\resizebox{7.5cm}{5cm}{\includegraphics[clip]{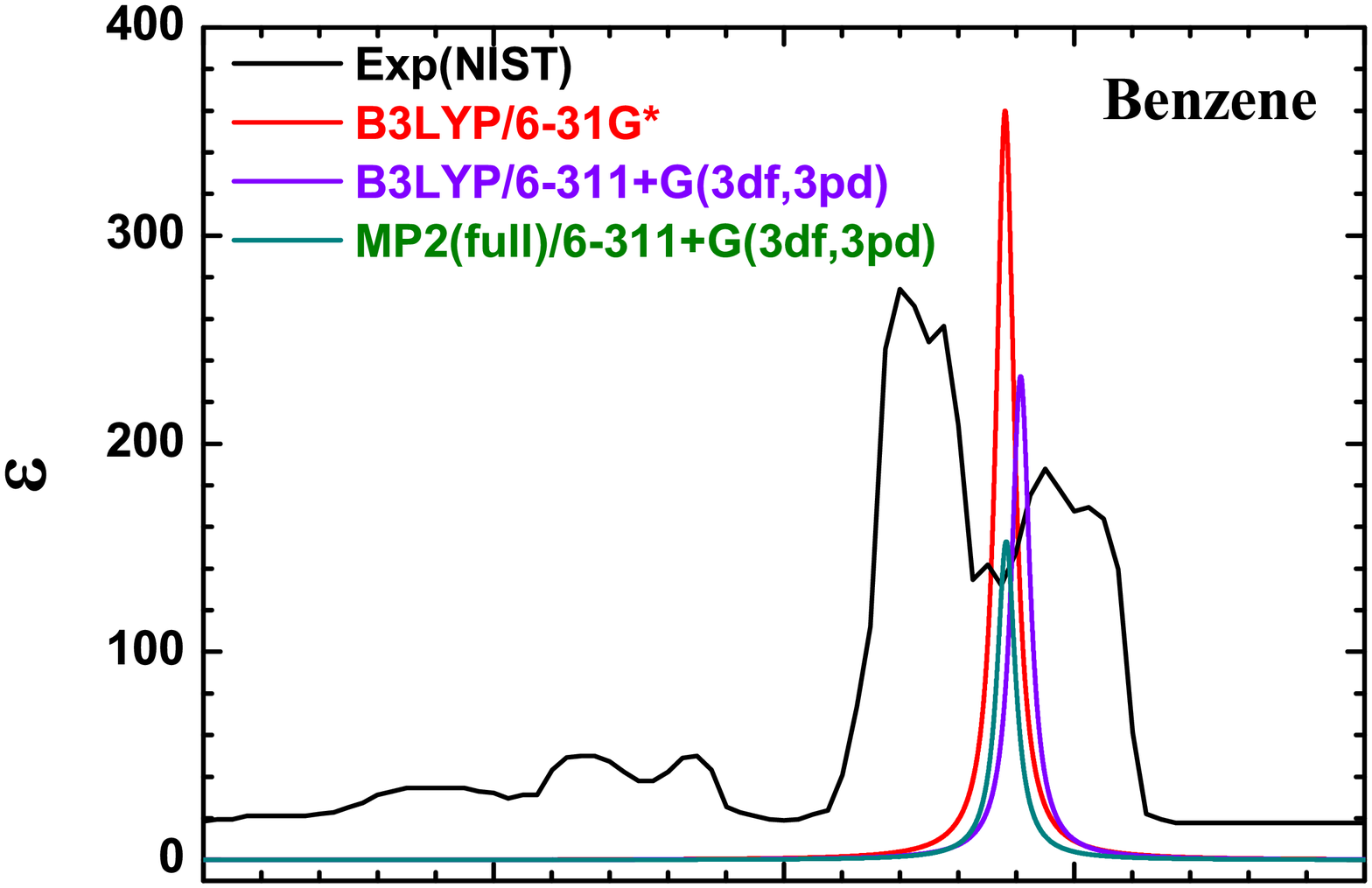}}\vspace{-1.2cm}
\resizebox{7.5cm}{5cm}{\includegraphics[clip]{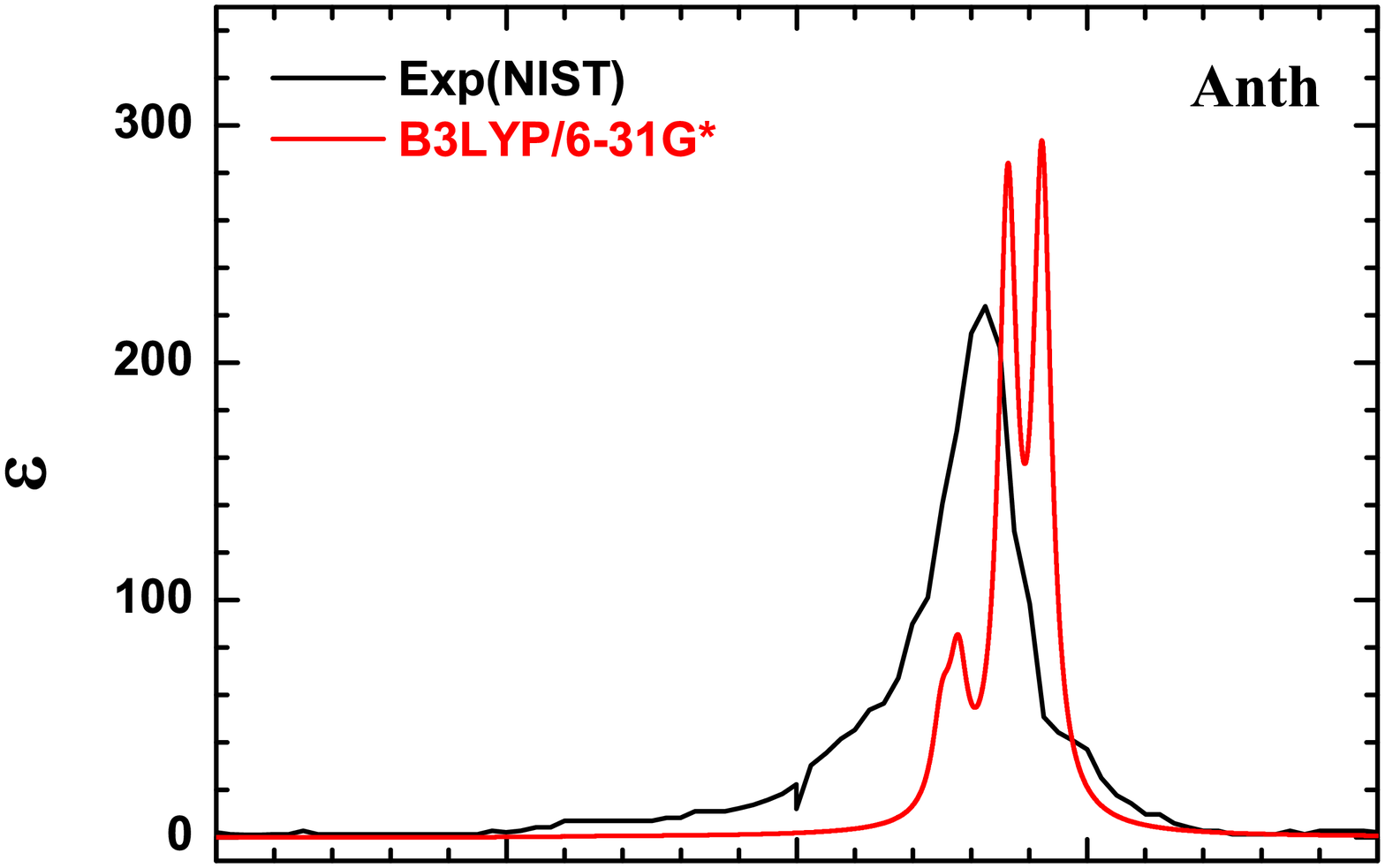}}\vspace{-1.2cm}
\resizebox{7.5cm}{5cm}{\includegraphics[clip]{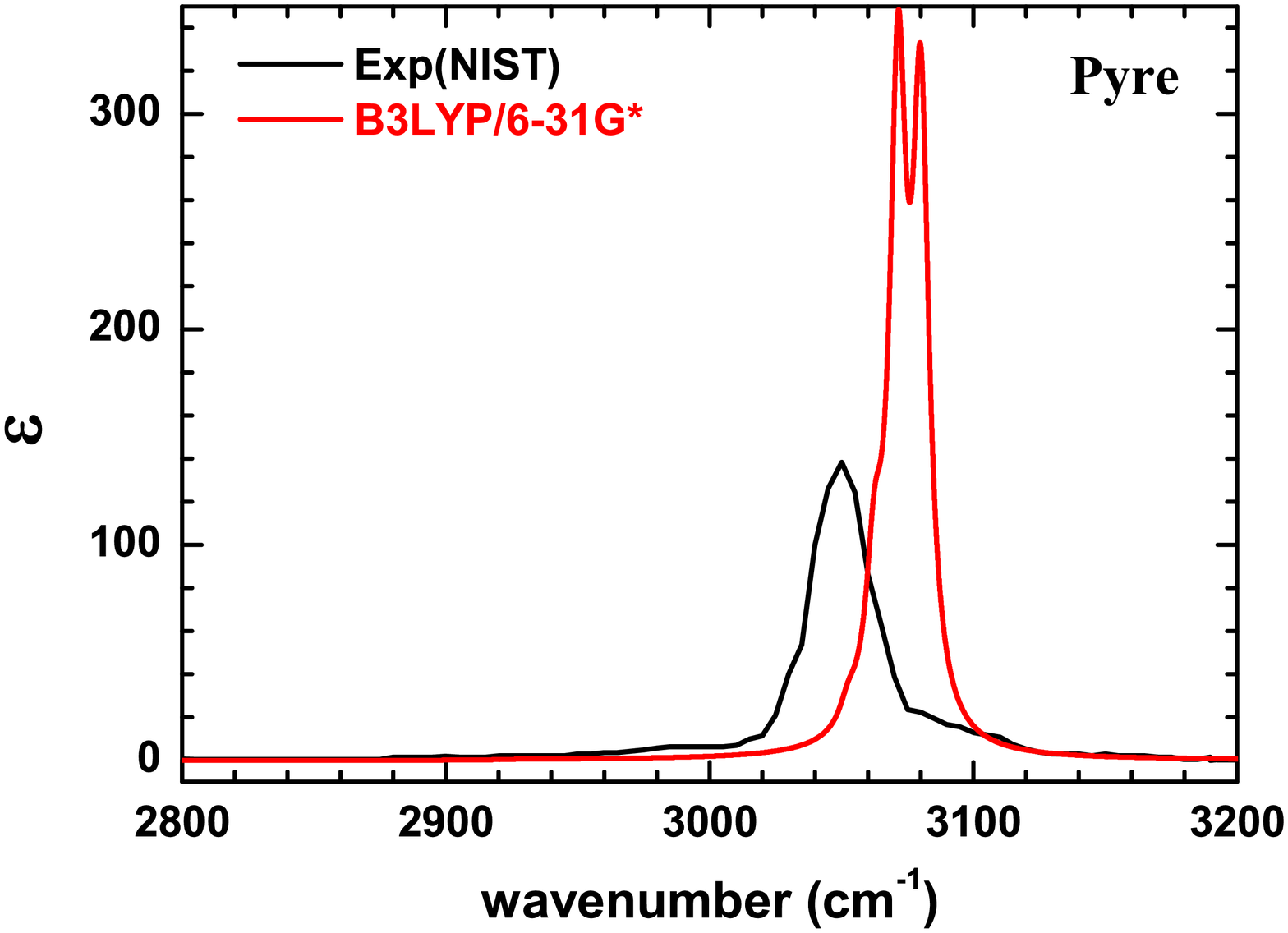}}
\end{minipage}\hspace{0.5cm}
\begin{minipage}[t]{0.33\textwidth}
\resizebox{7.5cm}{5cm}{\includegraphics[clip]{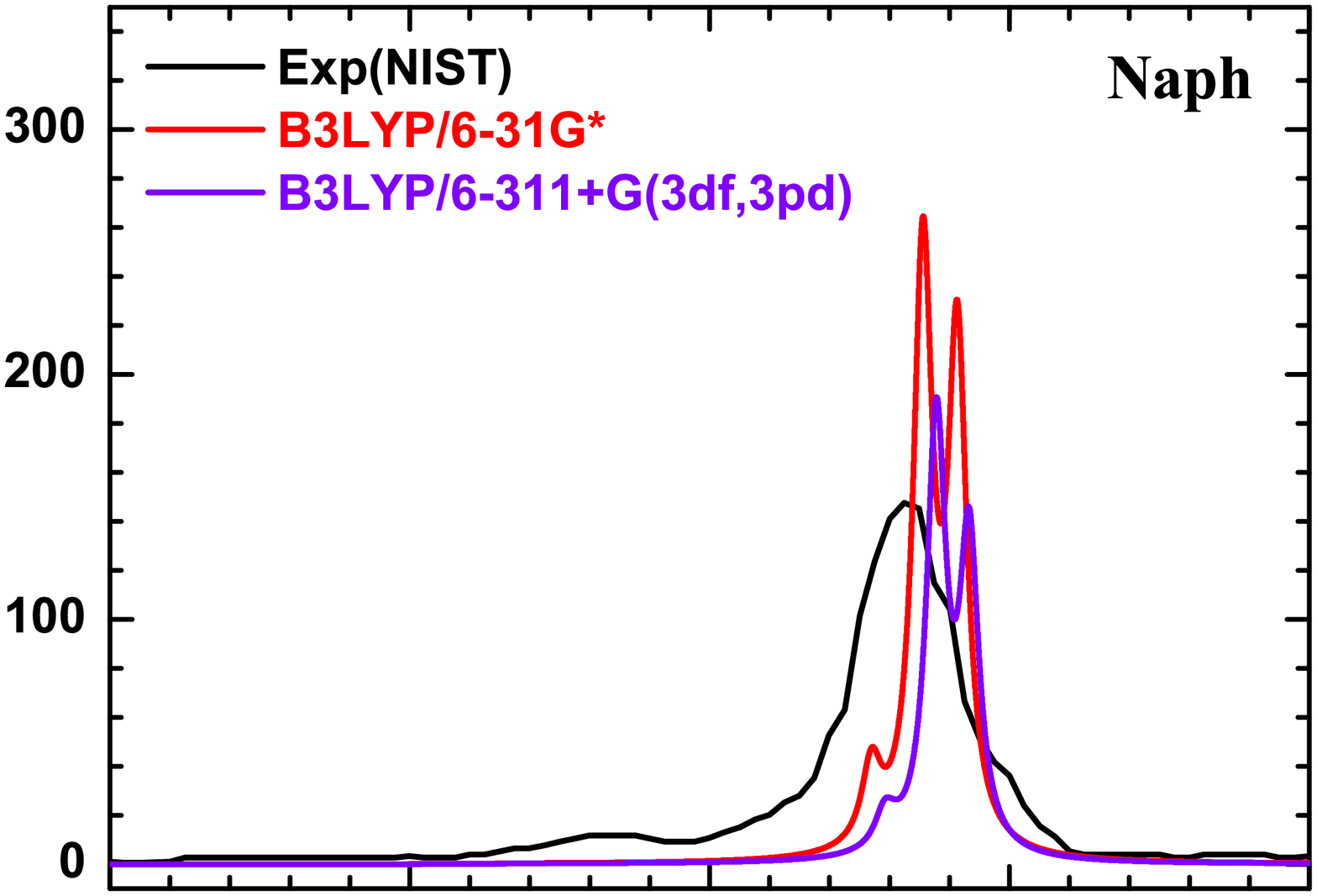}}\vspace{-1.2cm}
\resizebox{7.5cm}{5cm}{\includegraphics[clip]{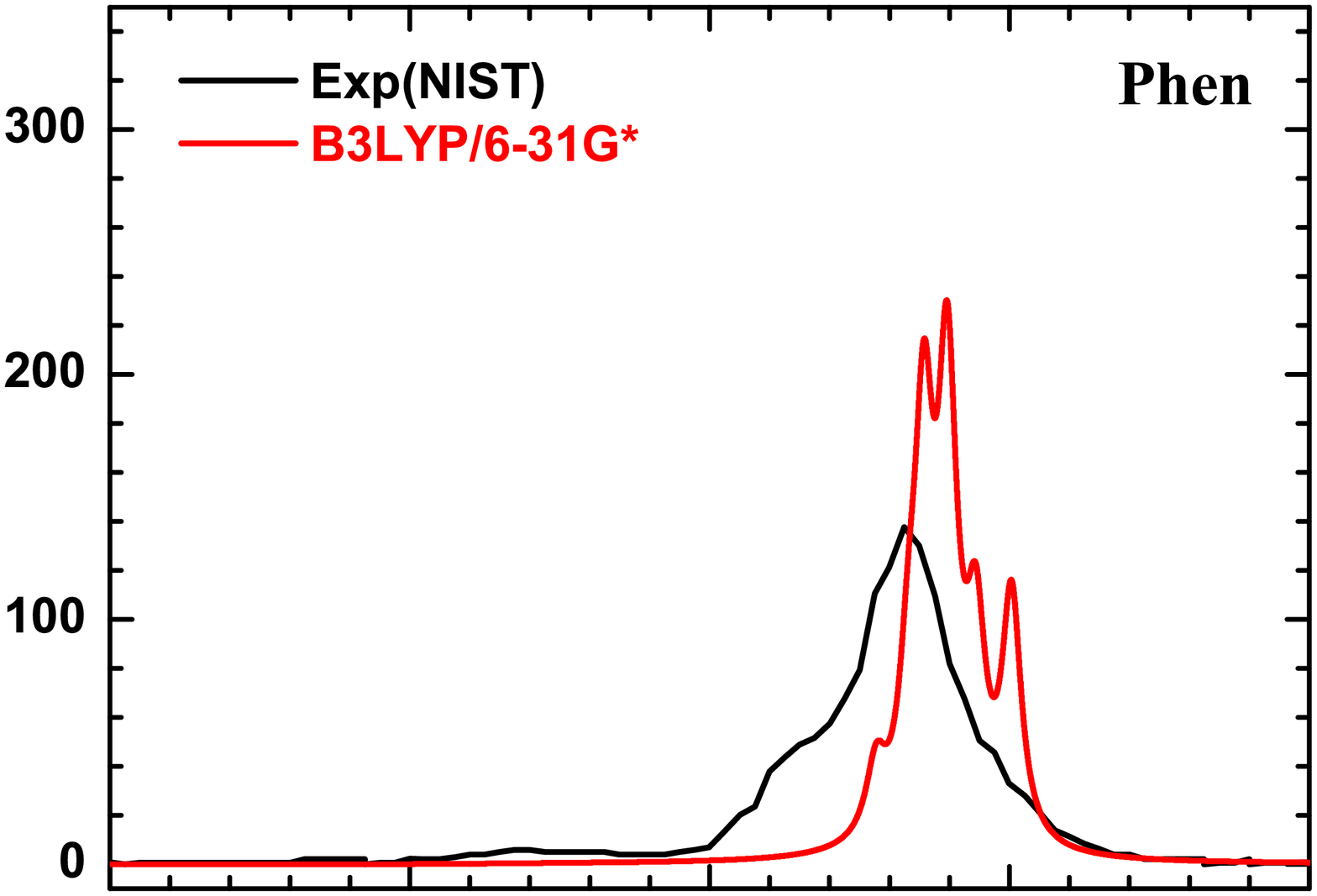}}\vspace{-1.2cm}
\resizebox{7.5cm}{5cm}{\includegraphics[clip]{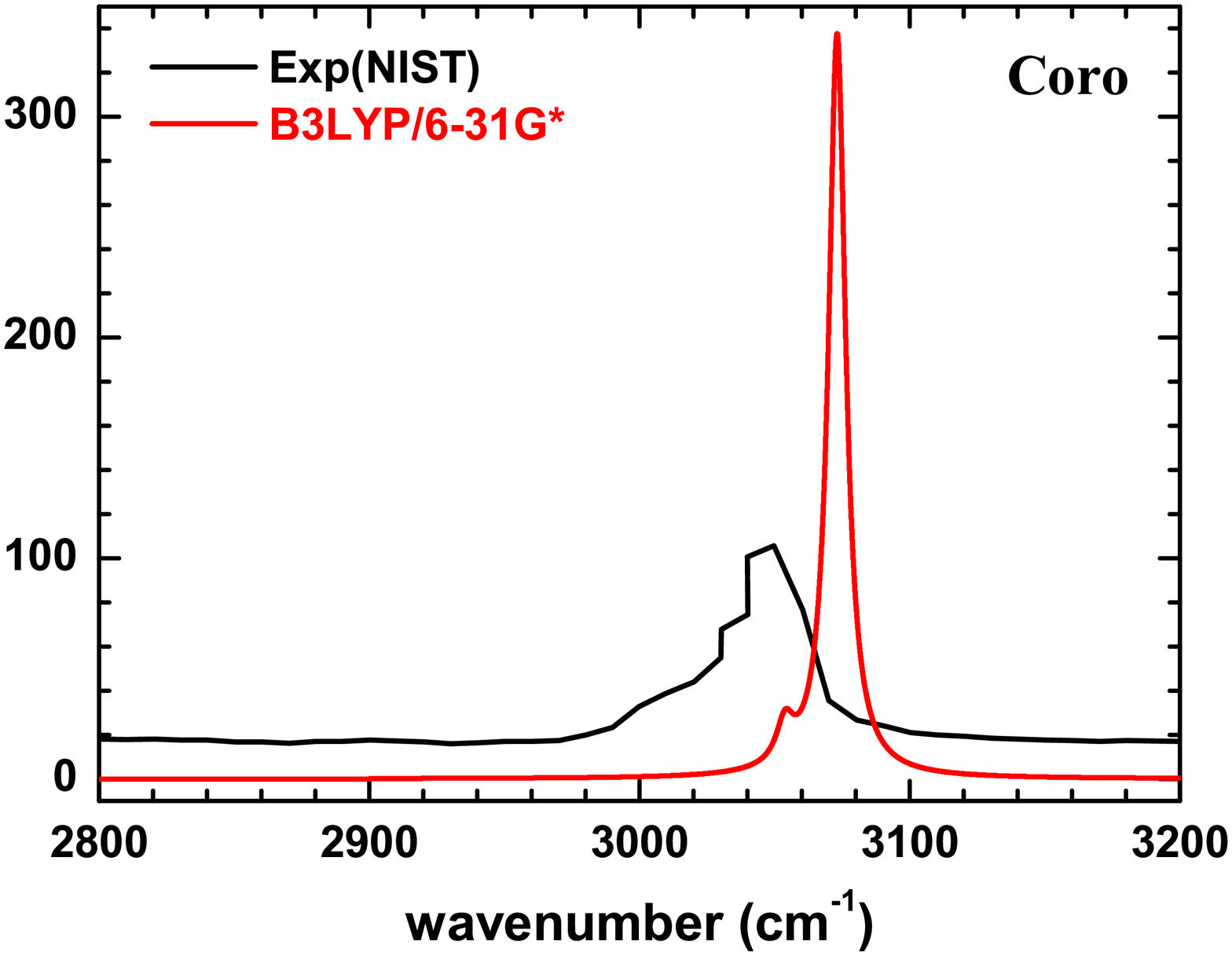}}
\end{minipage}\hspace{-1.5cm}
\end{center}
\caption{\footnotesize
         \label{fig:ParentPAH_IR}
         Comparison of the gas-phase absorption spectra
         experimentally measured by NIST
         (labelled with ``Exp\,(NIST)''; black lines)
         to the computed, frequency-scaled spectra
         (colored lines)
         of benzene, naphthalene (Naph),
         anthracene (Anth), phenanthrene (Phen),
         pyrene (Pyre), and coronene (Coro).
         The y-axis plots the molar absorptivity coefficient
         ($\varepsilon$) in units of $\mol^{-1}\cm^{-1}$.
         The $\varepsilon$ values for the NIST experimental data
         are scaled to be comparable to the computed values
         by multiplying the NIST absorbance
         with an artificial factor,
         as NIST only gives the absorbance and
         does not have information for the concentration to
         derive the absolute $\varepsilon$.
         }
\end{figure*}

The NIST experimental spectra of
the non-substituted PAHs
(naphthalene, anthracene, phenanthrene, and pyrene)
are shown in Figure~\ref{fig:ParentPAH_IR}
together with selected computed spectra.
The experimental spectra of these PAHs always
show one strong band at $\simali$3050$\cm^{-1}$
for the aromatic C--H stretches and several weak
bands in the region of 2800--3000$\cm^{-1}$.
These NIST spectra are not a part of
the NIST {\it Quantitative Infrared Database}
and therefore there is no intensity information
in the NIST database.
The gas-phase intensity measurements
of the aromatic C--H stretches
have been reported for naphthalene
($\simali$96$\km\mol^{-1}$;
Can\'{e} et al.\ 1996, Pauzat et al.\ 1999),
anthracene ($\simali$161$\km\mol^{-1}$;
Can\'{e} et al.\ 1997, Pauzat et al.\ 1999),
pyrene ($\simali$122$\km\mol^{-1}$;
Joblin et al.\ 1994, Pauzat et al. 1999),
and coronene ($\simali$161$\km\mol^{-1}$;
Joblin et al.\ 1994).
It appears that no gas phase IR intensities
have been published for phenanthrene and perylene.

Visual inspection of Figure~\ref{fig:ParentPAH_IR}
shows that good to excellent agreement
between the measured and computed C--H stretch
vibrational frequencies can be achieved in all cases
with the use of standard scale factors.
The computed spectra are usually drawn with the
line broadening set to be 4$\cm^{-1}$.
Note especially that the weak features
in the range of 2800--3000$\cm^{-1}$
(due to overtones and combinations,
Mitra \& Bernstein 1959) are much weaker for
the parent PAHs as compared to benzene itself
(vide supra).

Again, the experimental intensities
are much lower than our calculated
results for the aromatic C--H stretches
at {\rm B3LYP/6-31G$^{\ast}$} level.
With the calculated intensity
$A_{\rm aro}({\rm cal})$
and the experimental value
$A_{\rm aro}({\rm exp})$,
we find positive values for
$\Delta A = A_{\rm aro}({\rm cal})
- A_{\rm aro}({\rm exp})$:
naphthalene
[$A_{\rm aro}({\rm cal})$\,$\simali$139$\km\mol^{-1}$,
$\Delta A = A_{\rm aro}({\rm cal})
- A_{\rm aro}({\rm exp})\approx 43\km\mol^{-1}$],
anthracene
[$A_{\rm aro}({\rm cal})$\,$\simali$178$\km\mol^{-1}$,
$\Delta A \approx 17\km\mol^{-1}$],
pyrene
[$A_{\rm aro}({\rm cal})$\,$\simali$188$\km\mol^{-1}$,
$\Delta A \approx 66\km\mol^{-1}$],
and coronene
[$A_{\rm aro}({\rm cal})$\,$\simali$257$\km\mol^{-1}$,
$\Delta A \approx 96\km\mol^{-1}$].

\subsection{Frequencies and Intensities of Toluene
            and Methylated PAHs
           \label{sec:MethylPAHs}
           }
\subsubsection{Toluene}
\begin{figure*}
\centerline{
\includegraphics[scale=0.45,clip]{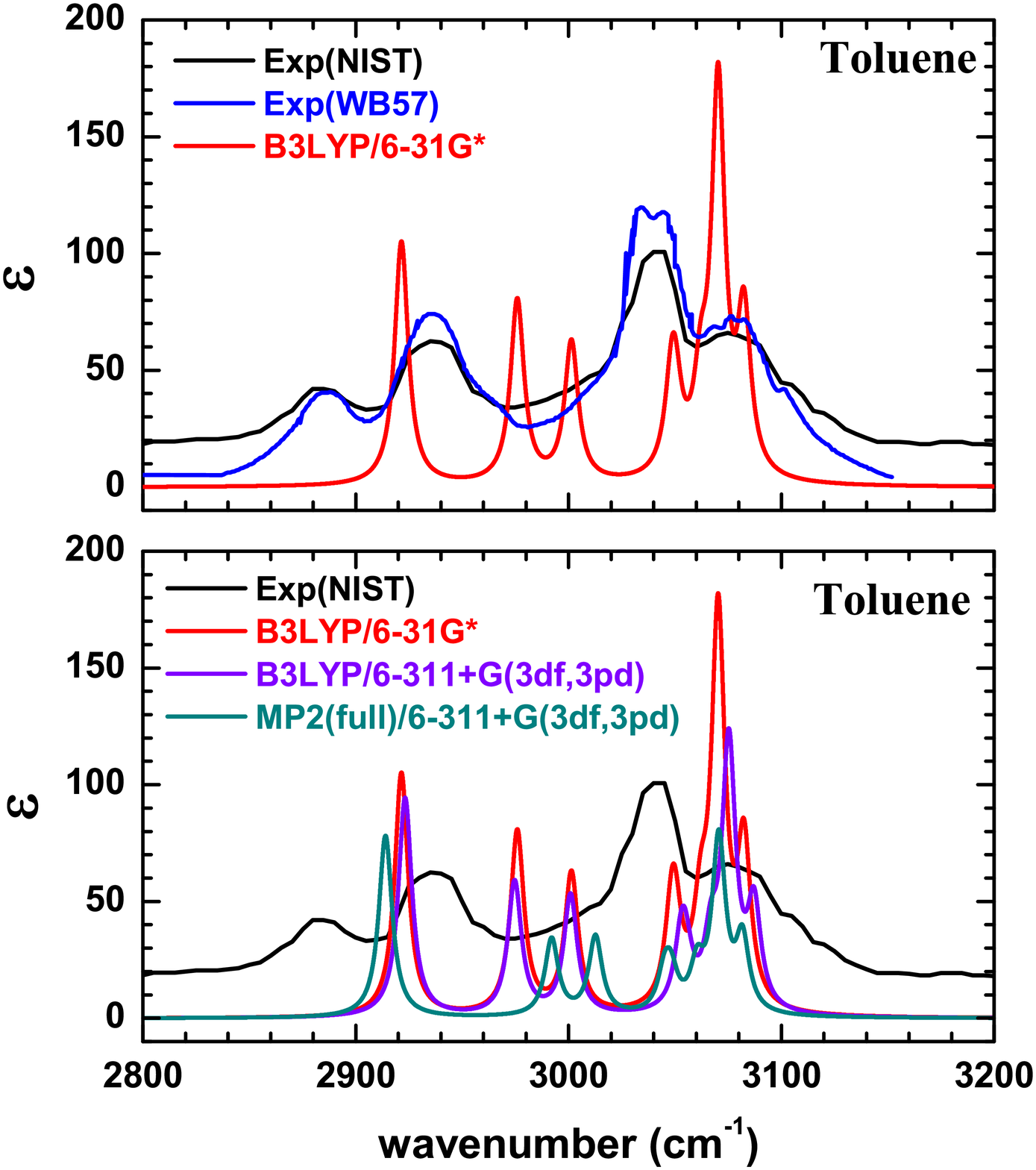}
}
\caption{\footnotesize
         \label{fig:toluene_IR}
         Comparison of the vibrational spectra of
         toluene computed at various levels
         with the experimental spectra
         of NIST [marked with ``Exp\,(NIST)'']
         and of Wilmshurst \&  Bernstein (1957)
         [marked with ``Exp\,(WB57)''].
         The NIST and WB57 experimental spectra
         are multiplied by a factor to be comparable
         with the computed spectra.
         }
\end{figure*}

Selected computed spectra of toluene
are shown in Figure~\ref{fig:toluene_IR}
together with the experimental spectra
taken from NIST
and from Wilmshurst \& Bernstein (1957).
The experimental spectra are similar,
and both contain just two bands
in the methyl region and also two bands
in the aromatic region.
The wavenumbers of those maxima are shown
in Table~\ref{Table:Freq_int_tolu}.

It is clear from Figure~\ref{fig:toluene_IR}
that the standard scaling works well for
the aromatic C--H stretches.
For the methyl (aliphatic) C--H stretches,
it is remarkable that the computations
(after normal scaling)
greatly overestimate the frequencies of
the asymmetric modes $\nuMeb$, $\nuMec$,
while $\nuMea$ is somewhat underestimated.
This remarkable discrepancy is a consequence
of the harmonic approximation and the free rotation
of the methyl group.
%
For our present purposes, it is important to recognize
that the experimental value of $A_{\rm ali}({\rm exp})$
-- the intensity of all the methyl (aliphatic)
C--H stretches -- does contain a substantial contribution
from overtones (about 19\%).

In the absence of absolute intensity data
for most of the molecules in our set,
we will discuss the ratio of the intensities
in the regions of the methyl (aliphatic)
and aromatic C--H stretches
and this ratio is determined as follows.
We digitize the NIST experimental spectra
and integrate over the selected regions.
We take 3000$\cm^{-1}$ as the wavenumber separation of
the methyl (aliphatic) and aromatic C--H stretching regions.
The integration over the range of 3000--3200$\cm^{-1}$
is taken as the total intensity of the aromatic
C--H stretches ($A_{\rm aro}$).
Similarly,
the integration in the range of 2800--3000$\cm^{-1}$
is taken as the total intensity of the aliphatic
C--H stretches ($A_{\rm ali}$).
The relative intensity of
the methyl (aliphatic) signal
to that of the aromatic band
is $\Arel \approx 0.79$.
A similar analysis of the spectrum
of Wilmshurst \& Bernstein (1957)
results in the experimental methyl
signal intensity of $\Arel\approx0.71$
relative to that of the aromatic C--H band.
Note that $A_{\rm aro}$ ($A_{\rm ali}$)
is the strength of all the aromatic (aliphatic)
C--H stretches while $\Aaro$ ($\Aali$) is the strength
of the aromatic (aliphatic) stretch per C--H bond.
For toluene,  $A_{\rm aro} = 5\Aaro$
and $A_{\rm ali} = 3\Aali$ and therefore
we have $\Aratio = \left(5/3\right)\,\Arel$.

Our integration of the NIST spectrum
gives a total intensity
of $\simali$97.2$\km\mol^{-1}$
for all the C--H stretches (both methyl and aromatic)
and is in excellent agreement with the value
of $\simali$95$\km\mol^{-1}$ reported by
Pavlyuchko et al.\ (2012)
and by Galabov et al.\ (1992).
According to our ratio of the measured intensities
for the methyl to aromatic regions
($\Arel \approx 0.79$),
this overall intensity corresponds to intensities
of $\simali$42.9$\km\mol^{-1}$ for the methyl bands
and of $\simali$54.3$\km\mol^{-1}$ for the aromatic bands.
The intensities computed
at the {\rm B3LYP/6-31G$^{\ast}$} level
for toluene are $\simali$165.3$\km\mol^{-1}$
for the entire region
and $\simali$70.4 and $\simali$94.9$\km\mol^{-1}$
for the methyl and aromatic sections, respectively.
Again, we see that the computed intensities
are much higher than the experimental values
from the gas phase measurements.

\subsubsection{Important Methylated PAHs}
\begin{figure*}
\centerline{
\includegraphics[scale=0.2,clip]{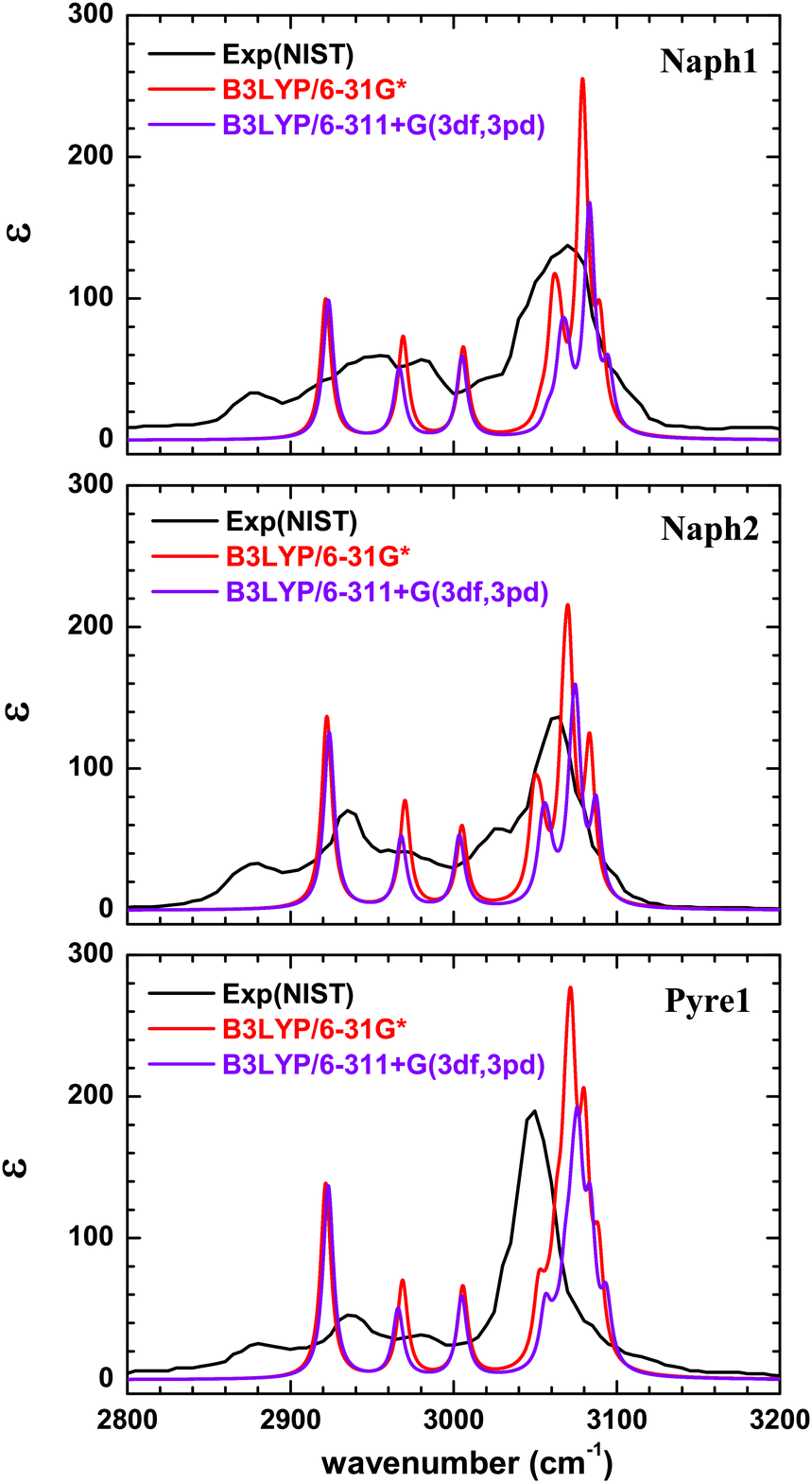}\hspace{-0.7cm}
\includegraphics[scale=0.2,clip]{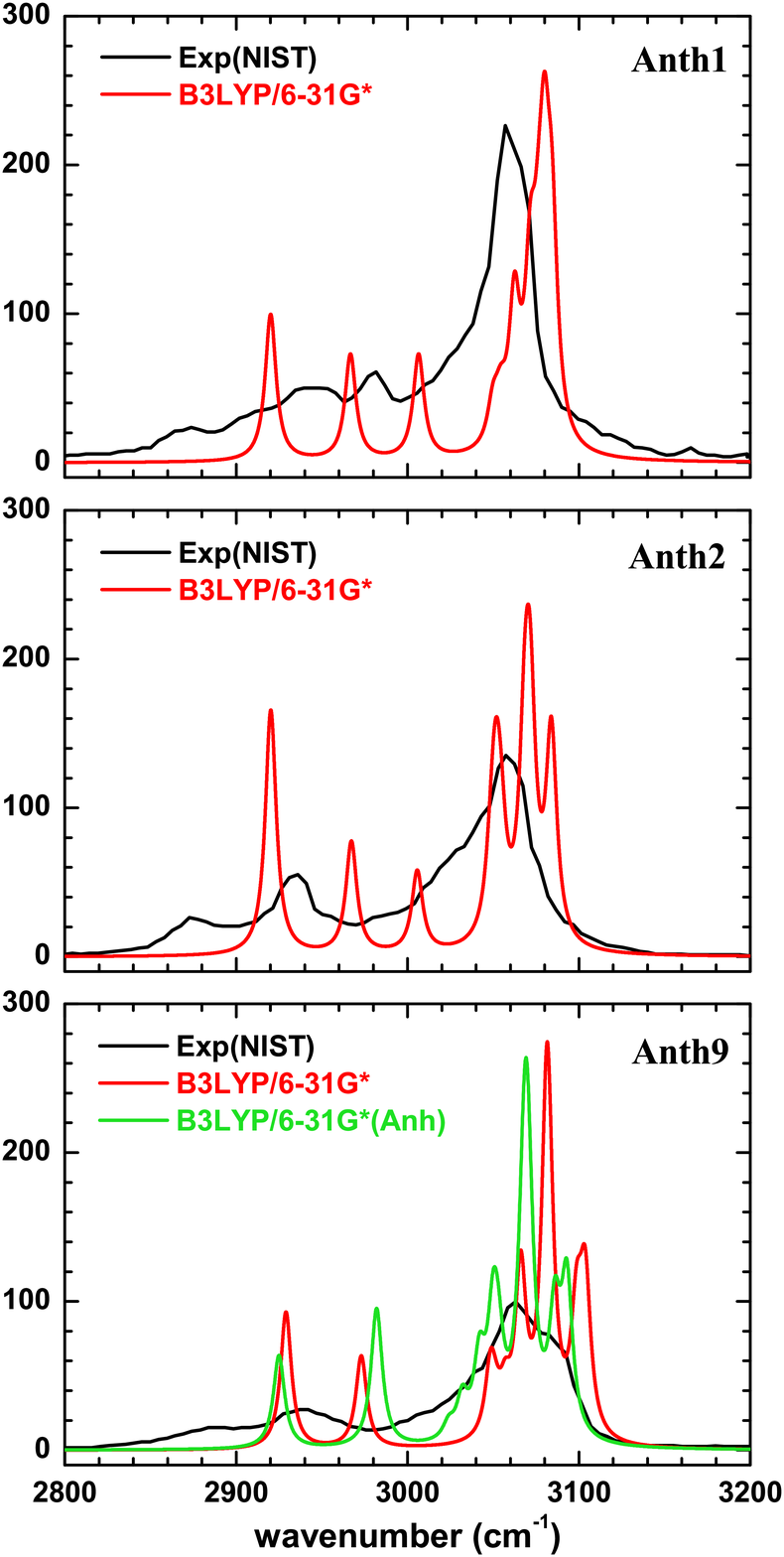}\hspace{-0.7cm}
\includegraphics[scale=0.2,clip]{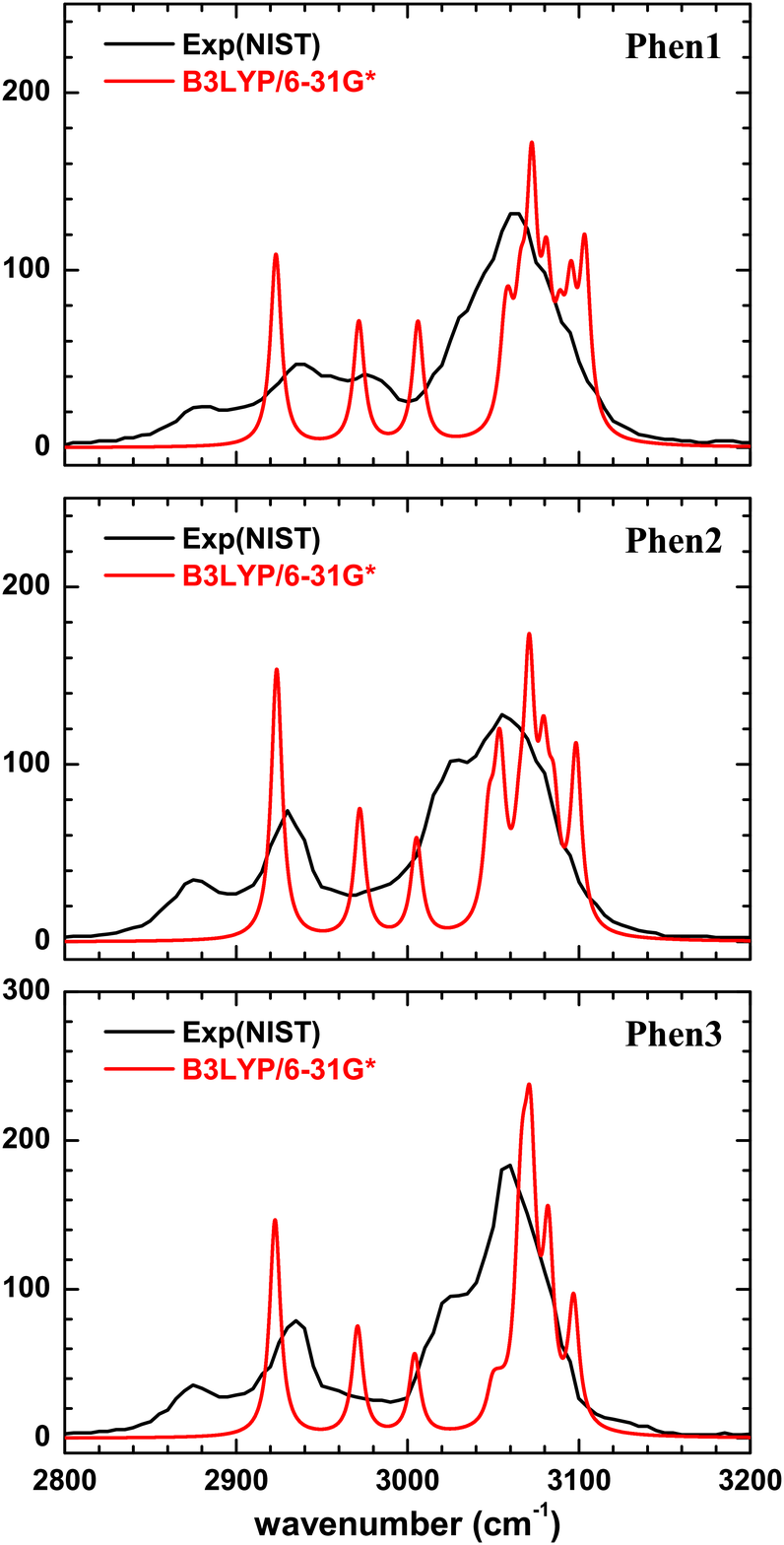}
}
\caption{\footnotesize
       \label{fig:MethylPAH_IR}
       Comparison of the computed spectra of methyl PAHs
       with their NIST experimental spectra
       [labelled with ``Exp\,(NIST)''].
       The NIST experimental spectra
       are multiplied by a factor in order
       to be comparable with the computed spectra.
       }
\end{figure*}

Selected computed spectra of
all the monomethyl-substituted PAHs
(naphthalene, 
anthracene, 
phenanthrene, 
and pyrene)
are shown in Figure~\ref{fig:MethylPAH_IR}
together with the corresponding gas-phase
experimental spectra taken from NIST (if available).
As expected, the NIST spectra of the methyl-PAHs
show several strong bands
in the region 3040--3070$\cm^{-1}$
due to the aromatic C--H stretching modes.
In all of these spectra, the computed spectra
with standard wavenumber scaling are shown with
a line width of 4$\cm^{-1}$.

The interesting aspect of Figure~\ref{fig:MethylPAH_IR}
concerns the bands due to
the methyl (aliphatic) C--H stretches
in the region 2800--3000$\cm^{-1}$.
As with toluene, the first band on
the low-wavenumber side
in the experimental spectra is due to overtones,
and the remaining band(s) in the aliphatic region
are due to the overlapping methyl C--H stretches.
The computed spectrum of Anth9 stands out in that
$\nuMec$ is extraordinarily high
and appears in the aromatic region
($\nuMec\approx3048\cm^{-1}$).
Pauzat et al.\ (1999) reported a similar
$\nuMec\approx3030\cm^{-1}$ for Anth9.
Hence, one must wonder whether part of the intensity
of the ``aromatic region'' is due to the aliphatic C--H
stretching mode $\nuMec$.

To address this question,
we also compute the anharmonic vibrations
at the {\rm B3LYP/6-31G$^{\ast}$} level for Anth9
(see Figure~\ref{fig:MethylPAH_IR}, the green line
in the Anth9 panel).
We again find that $\nuMeb$ and $\nuMec$ are
overestimated by $\simali$48$\cm^{-1}$
and $\simali$6$\cm^{-1}$ respectively,
and $\nuMea$ is underestimated by $\simali$53$\cm^{-1}$
in the harmonic approximation (even after scaling).
However, even with the anharmonic approximation,
$\nuMec$ is still in the aromatic region,
hence, we conclude that the aliphatic and aromatic regions
do overlap in the experimental spectrum of Anth9.

The unusual behavior of Anth9a results from
the fact that one methyl C--H bond lies in
the arene plane and stretches of this C--H bond
cannot occur independently of stretches of
the proximate C(1)--H bond,
i.e., the presence of an all-{\it cis}
H--C$_{\rm Me}$--(C$_{\rm aro}$)3--H moiety.
Instead, the stretching of the in-plane
methyl CH bond must be out of phase
with any stretching of the  C(1)--H bond
to avoid steric repulsion. This coupling is
illustrated schematically in the bottom row of
Figure~\ref{fig:methyl_schetch_modes}.
This structural feature of Anth9a is uncommon
in minima of other PAHs, but it is common in
transition state structures and this type of
vibrational coupling also occurs in Naph1b,
Anth1b, Phen1b, Phen9b, Pyre1b, Pyre4b,
Pery3b and Coro1b. The structural feature
is not present in the isomers Anth1a and Anth2b,
and the vibrational spectra of the minima of
the anthracene isomers therefore are not expected
and do not show the coupling of Anth9a.

Again, we integrate the aromatic C--H stretches
to obtain $A_{\rm aro}$
and integrate the methyl (aliphatic) C--H stretches
to obtain $A_{\rm ali}$.
The intensity of the methyl signals relative to
that of the aromatic stretches are
generally $\Arel$\,$\sim$0.53 at {\rm B3LYP/6-31G$^{\ast}$}.
The only exception is 9-methylanthracene (Anth9)
with its much smaller relative intensity of
$\Arel$\,$\simali$0.42.
Considering that the aliphatic C--H stretch ($\nuMec$)
contributes to the aromatic region,
this is not surprising.


\section{Theoretical Level Dependency of Computed IR Intensities
         \label{sec:I_scale}
         }
\subsection{B3LYP vs. MP2 IR Intensities of C--H Stretching Modes}
As we have seen above, the IR intensities calculated
at the {\rm B3LYP/6-31G$^{\ast}$} level are much higher
compared to the experimental results.
Using better basis sets in conjunction with
the B3LYP method, we found that the IR intensities
still differ by a factor of $\simali$30\%
compared to the experiment results
(see Table~\ref{Table:Average_Int_B3LYP_MP2}).
Pavlyuchko et al.\ (2012) recently reported that
the IR intensities calculated for benzene and toluene
at the level {\rm MP2/6-311G(3df,3pd)} would match
the experimental results very well.
Considering this report,
we try to reproduce their data for benzene
and toluene and we perform both MP2(fc)
and MP2(full) computations
with the 6-311G(3df,3pd) basis set.
The MP2(fc) results closely match the data
of Pavlyuchko et al.\ (2012;
see Table~\ref{Table:Average_Int_B3LYP_MP2}).
The IR intensities computed with and without
the frozen core (fc) approximation actually differ
noticeably for toluene and, hence, we then calculate
all other vibrational spectra with the MP2(full) method
in conjunction with the standard basis set {\rm 6-31G$^{\ast}$}
and the extended basis sets {\rm 6-311+G$^{\ast\ast}$}
and {\rm 6-311+G(3df,3pd)} for benzene, naphthalene
and their mono-methyl derivatives as test cases.
The most relevant results are given in
Table~\ref{Table:Average_Int_B3LYP_MP2}.

Note that the addition of sets of diffuse functions
in the large basis set 6-311+G(3df,3pd)
drastically changes the IR intensities for benzene
and toluene. This level is better than
the MP2(full)/6-311G(3df,3pd) level
and at this level the overall intensities are
significantly lower than the experimental data.
This finding shows that the seemingly excellent
agreement between experiment
and the MP2(fc)/6-311G(3df,3pd) data is fortuitous.
Second, this finding actually makes perfect sense
because the experimental IR intensities
in the C--H stretching regions always contain
significant contributions from overtones and
combinations (vide supra).
The overtones in the methyl region of toluene are
responsible for $\simali$19\% of the intensity.
Subtraction of 19\% of the intensity of
the methyl region results in the adjusted
experimental value of $\simali$89.0$\km\mol^{-1}$,
and this value is in excellent agreement
with the computed value of 86.5$\km\mol^{-1}$.

While the MP2(full)/6-311+G(3df,3pd) level data
reproduce the measured IR intensities reasonably well,
such calculations are far too expensive especially for
the larger molecules. The MP2(full) computations of the
naphthalene systems with the large basis sets
including the (3df,3pd) polarization functions
each requires several days of computer time on
eight processors. Considering that the absolute values
computed at all of the MP2 levels are better than
the respective values computed at the B3LYP levels,
one would be inclined to explore scaling approaches
of the MP2 data computed with modest basis sets.
However, we will show that scaling approaches that are
based on the B3LYP data can be just as successful
in spite of the fact that the absolute numbers
computed at the {\rm B3LYP/6-31G$^{\ast}$} level
differ much more from experiment than
do the {\rm MP2/6-31G$^{\ast}$} data.

Before we proceed, it is useful to clarify
the meaning of scaling approaches.
In the most typical approach to scaling,
it is attempted to reproduce a set
of experimental data with a set of data
obtained at a level $\Li$ such that
$p({\rm exp}) \approx f \cdot p(\Li)$,
that is, one scaling factor $f$ is applied
to all values in the data set and
this scale factor depends on the level, $f = f(\Li)$.
This kind of scaling is commonly employed for
vibrational frequencies (see Table~\ref{Table:Freq_scale}).
For intensities, however, we will see that approaches
of the type $p({\rm exp}) \approx f \cdot p(\Li)+C(\Li)$
are more successful, that is,
there will be a non-zero offset.

\subsection{Scaling Approaches for
            the Computed Total Intensities
            of C--H Stretching Modes}
\begin{figure*}
\vspace{-10cm}
\centerline
{
\includegraphics[scale=0.5,clip]{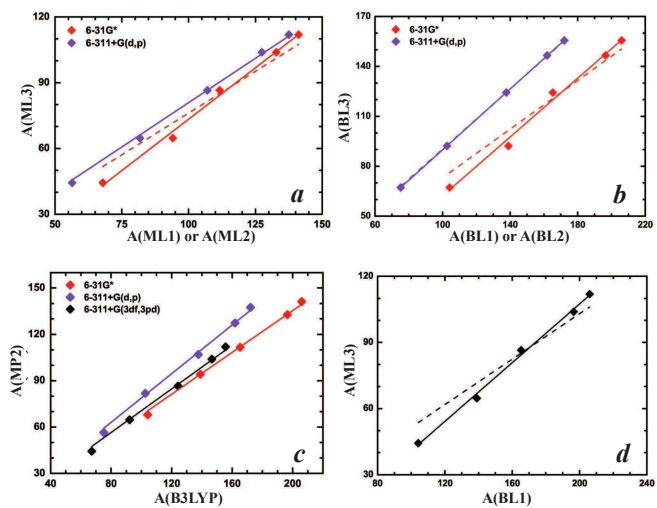}
}
\caption{\footnotesize
         \label{fig:Int_LevelDep}
         Level dependency of the total C--H stretch
         intensities (methyl plus aromatic)
         for benzene and naphthalene and for their methyl
         derivatives toluene and methylnaphthalene.
         Top left (a): Intensities calculated at MP2
         with small basis sets
         [i.e., 6-31G$^{\ast}$ (i.e., $ML1$),
          6-311+G(d,p) (i.e., $ML2$)]
         vs. that with a large basis set
         [6-311+G(3df,3pd) (i.e., $ML3$)].
         Dashed red line plots eq.\,\ref{eq:MP2_LS_SS_a},
         solid red line plots eq.\,\ref{eq:MP2_LS_SS_b},
         and solid blue line plots eq.\,\ref{eq:MP2_LS_MS}.
         Top right (b): Same as (a) but at B3LYP.
         Dashed red line plots eq.\,\ref{eq:B3LYP_LS_MS_a},
         solid red line plots eq.\,\ref{eq:B3LYP_LS_MS_b},
         dashed blue line plots eq.\,\ref{eq:B3LYP_LS_SS_a},
         and solid blue line plots eq.\,\ref{eq:B3LYP_LS_SS_b}.
         Bottom left (c): Intensities calculated
         at B3LYP vs. MP2 with the same basis set.
         Solid red line plots eq.\,\ref{eq:B3LYP_MP2_SS},
         solid blue line plots eq.\,\ref{eq:B3LYP_MP2_MS},
         and solid black line plots eq.\,\ref{eq:B3LYP_MP2_LS}.
         Bottom right (d): Intensities calculated
         at B3LYP/6-31G$^{\ast}$ (i.e., $BL1$)
         vs. MP2/6-311+G(3df,3pd) (i.e., $ML3$).
         Dashed black line plots eq.\,\ref{eq:B3LYP_MP2_a},
         and solid black line plots eq.\,\ref{eq:B3LYP_MP2_b}
         }
\end{figure*}

Let $ML1$, $ML2$ and $ML3$ respectively
represent the {\rm MP2(full)} computations
with the {\rm 6-31G$^{\ast}$}, 6-311+G(d,p),
and 6-311+G(3df,3pd) basis sets
[i.e.,  $ML1$~$\equiv$~{\rm MP2(full)/6-31G$^{\ast}$},
$ML2$~$\equiv$~{\rm MP2(full)/6-311+G(d,p)}, and
$ML3$~$\equiv$~{\rm MP2(full)/6-311+G(3df,3pd)}].
Let $BL1$, $BL2$ and $BL3$ respectively
represent the B3LYP computations
with the {\rm 6-31G$^{\ast}$}, 6-311+G(d,p),
and 6-311+G(3df,3pd) basis sets
(i.e.,  $BL1$~$\equiv$~{\rm B3LYP/6-31G$^{\ast}$},
$BL2$~$\equiv$~{\rm B3LYP/6-311+G(d,p)},
and $BL3$~$\equiv$~{\rm B3LYP/6-311+G(3df,3pd)}).
As can be seen from Figure~\ref{fig:Int_LevelDep} (top left),
the total intensities ($A$) computed at the MP2 level
but with different basis sets
[i.e., $A(ML1)$, $A(ML2)$, and $A(ML3)$]
are linearly related:
\begin{subequations}\label{eq:MP2_LS_SSMS}
\begin{align}
A(ML3)&\approx0.7615\,A(ML1)~~,~~(r^{2}\approx0.9575)
\label{eq:MP2_LS_SS_a} \\
A(ML3)&\approx0.9382\,A(ML1)-20.4880~~,~~(r^{2}\approx0.9949)
\label{eq:MP2_LS_SS_b} \\
A(ML3)&\approx0.8089\,A(ML2)~~,~~(r^{2}\approx0.9984)
\label{eq:MP2_LS_MS}
\end{align}
\end{subequations}
where $r^{2}$ is the linear-correlation coefficient.
While eq.\,\ref{eq:MP2_LS_MS} describes
an excellent linear correlation
between the intensities
computed with the $ML3$ method [$A(ML3)$]
and that with the $ML2$ method [$A(ML2)$]
without any need for an offset,
the analogous eq.\,\ref{eq:MP2_LS_SS_a}
is less successful and an excellent linear
correlation between $A(ML3)$ and $A(ML1)$
only is achieved when a non-zero offset
is allowed in eq.\,\ref{eq:MP2_LS_SS_b}.
The analogous relations also hold at the B3LYP level
(eq.\,\ref{eq:B3LYP_LS_SS})
and they are shown in
Figure~\ref{fig:Int_LevelDep} (top right),
where $A(BL1)$, $A(BL2)$, and $A(BL3)$ are
respectively the intensities computed at
the $BL1$, $BL2$ and $BL3$ levels.
%
%
%
%
\begin{subequations}\label{eq:B3LYP_LS_SS}
\begin{align}
A(BL3)&\approx0.7306\,A(BL1)~,~~(r^{2}\approx0.9610)
\label{eq:B3LYP_LS_MS_a} \\
A(BL3)&\approx0.8838\,A(BL1)-26.1670~,~~(r^{2}\approx0.9924)
\label{eq:B3LYP_LS_MS_b}\\
A(BL3)&\approx0.8089\,A(BL2)~,~~(r^{2}\approx0.9984)
\label{eq:B3LYP_LS_SS_a} \\
A(BL3)&\approx0.8395\,A(BL2)-3.3861~,~~(r^{2}\approx0.9998)
\label{eq:B3LYP_LS_SS_b}
\end{align}
\end{subequations}

Also shown in Figure~\ref{fig:Int_LevelDep} (bottom left)
are the nearly linear relations between the IR intensities
computed at the B3LYP and MP2(full) levels
with a common basis set.
The data are very well described by linear regression
and there is no need for a non-zero offset
in any of the following equations
(see eqs.\,\ref{eq:B3LYP_MP2_SS},
\ref{eq:B3LYP_MP2_MS}, and \ref{eq:B3LYP_MP2_LS}).
It is remarkable that these slopes are rather similar
for the various basis sets.
\begin{subequations} \label{eq:B3LYP_MP2_scale}
\begin{align}
A(ML1)&\approx0.6769\,A(BL1)~,~~(r^{2}\approx0.9971)
\label{eq:B3LYP_MP2_SS} \\
A(ML2)&\approx0.7877\,A(BL2)~,~~(r^{2}\approx0.9966)
\label{eq:B3LYP_MP2_MS} \\
A(ML3)&\approx0.7056\,A(BL3)~,~~(r^{2}\approx0.9949)
\label{eq:B3LYP_MP2_LS}
\end{align}
\end{subequations}

In light of these linear correlations,
it is clear that there must be a strong
linear correlation between the lowest DFT level,
our standard level {\rm B3LYP/6-31G${^{\ast}}$}
(i.e., $BL1$), and the best MP2 level,
the level MP2(full)/6-311+G(3df,3pd)
(i.e., $ML3$).
Eqs.\,\ref{eq:MP2_LS_SS_a}
and \ref{eq:B3LYP_MP2_SS}
suggest a proportionality constant of
$\approx 0.7615\times0.6769\approx 0.5154$
and the actual correlation coefficient of
eq.\,\ref{eq:B3LYP_MP2_a} is $\simali$0.5152
and it is essentially the same
(see Figure~\ref{fig:Int_LevelDep}, bottom right).
Considering the need for non-zero offset in
eq.\,\ref{eq:MP2_LS_SS_b},
we also explore eq.\,\ref{eq:B3LYP_MP2_b}
and achieve an excellent linear correlation:
\begin{subequations}\label{eq:B3LYP_MP2}
\begin{align}
A(ML3)&\approx0.5152\,A(BL1)~,~~(r^{2}\approx0.9428)
\label{eq:B3LYP_MP2_a}  \\
A(ML3)&\approx0.6655\,A(BL1)-25.6770~,~~(r^{2}\approx0.9964)
\label{eq:B3LYP_MP2_b}
\end{align}
\end{subequations}

We will demonstrate in the following that the offsets
come from the fact that the intensities of
methyl (aliphatic) and aromatic C--H stretches
do not scale alike (i.e., $\fali\neq \faro$).
Eqs.\,\ref{eq:I_L12_define_a}
and \ref{eq:I_L12_define_b}
show the total intensities of
the C--H stretching regions as a function
of the numbers of methyl ($n_{\Ali}$)
and aromatic ($n_{\Aro}$) C--H bonds
and the average IR intensities of a methyl ($\Aali$)
or of an aromatic ($\Aaro$) C--H stretching bond
for two theoretical levels $\Li$ and $\Lj$:
\begin{subequations}\label{eq:I_L12_define}
\begin{align}
A(\Li)&~=~n_{\Ali}\,A_{\Ali}(\Li) + n_{\Aro}\,A_{\Aro}(\Li)
\label{eq:I_L12_define_a} \\
A(\Lj)&~=~n_{\Ali}\,A_{\Ali}(\Lj) + n_{\Aro}\,A_{\Aro}(\Lj)
\label{eq:I_L12_define_b}
\end{align}
\end{subequations}
where $A_{\Ali}(\Li)$ and $A_{\Aro}(\Li)$
are respectively the strengths of
one aliphatic or one aromatic C--H bond
computed at the $\Li$ level,
and $A_{\Ali}(\Lj)$ and $A_{\Aro}(\Lj)$
are the same parameters but computed
at the $\Lj$ level.

Assuming that the intensities of the methyl (aliphatic)
and aromatic C--H stretches scale with factors
$\fali$ and $\faro$, respectively,
one can express the total intensity at level $\Lj$
as a function of the average IR intensities of
a methyl (aliphatic) or of an aromatic
C--H stretching bond at theoretical levels $\Li$
[i.e., $A_{\Ali}(\Li)$ and $A_{\Aro}(\Li)$;
see eq.\,\ref{eq:I_L12_scale_1a}].
By addition and subtraction of
the term $\faro\,n_{\Ali}\,A_{\Ali}(\Li)$,
it is possible to rewrite
eq.\,\ref{eq:I_L12_scale_1a}
such that $A(\Lj)$ is expressed as a function
of $A(\Li)$ and $A_{\Ali}(\Li)$
(see eq.\,\ref{eq:I_L12_scale_1d}).
Using instead the analogous term
$\fali\,n_{\Aro}\,A_{\Aro}(\Li)$ gives $A(\Lj)$
as a function of $A(\Li)$ and $A_{\Aro}(\Li)$
(see eq.\,\ref{eq:I_L12_scale_2d}).
\begin{subequations}\label{eq:I_L12_scale1}
\begin{align}
A(\Lj)&~=~\fali\,n_{\Ali}\,A_{\Ali}(\Li) + \faro\,n_{\Aro}\,A_{\Aro}(\Li)
\label{eq:I_L12_scale_1a} \\
     &~=~\fali\,n_{\Ali}\,A_{\Ali}(\Li) + \faro\,n_{\Aro}\,A_{\Aro}(\Li)
       + \faro\,n_{\Ali}\,A_{\Ali}(\Li) - \faro\,n_{\Ali}\,A_{\Ali}(\Li)
\label{eq:I_L12_scale_1b} \\
     &~=~\faro\,[n_{\Ali}\,A_{\Ali}(\Li) + n_{\Aro}\,A_{\Aro}(\Li)]
       + \fali\,n_{\Ali}\,A_{\Ali}(\Li) - \faro\,n_{\Ali}\,A_{\Ali}(\Li)
\label{eq:I_L12_scale_1c} \\
     &~=~\faro\,A(\Li) + \underline{(\fali - \faro)\,n_{\Ali}\,A_{\Ali}(\Li)}
\label{eq:I_L12_scale_1d}
\end{align}
\end{subequations}
or
\begin{subequations}\label{eq:I_L12_scale2}
\begin{align}
A(\Lj)&~=~\fali\,n_{\Ali}\,A_{\Ali}(\Li) + \faro\,n_{\Aro}\,A_{\Aro}(\Li)
\label{eq:I_L12_scale_2a} \\
     &~=~\fali\,n_{\Ali}\,A_{\Ali}(\Li) + \faro\,n_{\Aro}\,A_{\Aro}(\Li)
       + \fali\,n_{\Aro}\,A_{\Aro}(\Li) - \fali\,n_{\Aro}\,A_{\Aro}(\Li)
\label{eq:I_L12_scale_2b} \\
     &~=~\faro\,[n_{\Ali}\,A_{\Ali}(\Li) + n_{\Aro}\,A_{\Aro}(\Li)]
       + \faro\,n_{\Aro}\,A_{\Aro}(\Li) - \fali\,n_{\Aro}\,A_{\Aro}(\Li)
\label{eq:I_L12_scale_2c} \\
     &~=~\fali\,A(\Li) + \underline{(\faro - \fali)\,n_{\Aro}\,A_{\Aro}(\Li)}
\label{eq:I_L12_scale_2d}
\end{align}
\end{subequations}
where the underlined terms in
eqs.\,\ref{eq:I_L12_scale_1d}
and \ref{eq:I_L12_scale_2d}
are responsible for the offset
in the correlations between the total intensities
at levels $\Li$ and $\Lj$,
and these offsets vanish only when $\faro = \fali$.
We will show in \S\ref{subsubsec:BSEffect_B3LYP}
that this condition never holds and,
in addition, it also is not trivial to determine
at what level $\faro$ and $\fali$ converge.

\subsection{Theoretical Level Dependency of
            Intensity Scaling Factors $\faro$ and $\fali$:
            Basis Set Effects at the B3LYP Level
            \label{subsubsec:BSEffect_B3LYP}
            }
\begin{figure*}
\centerline{
\includegraphics[scale=0.38,clip]{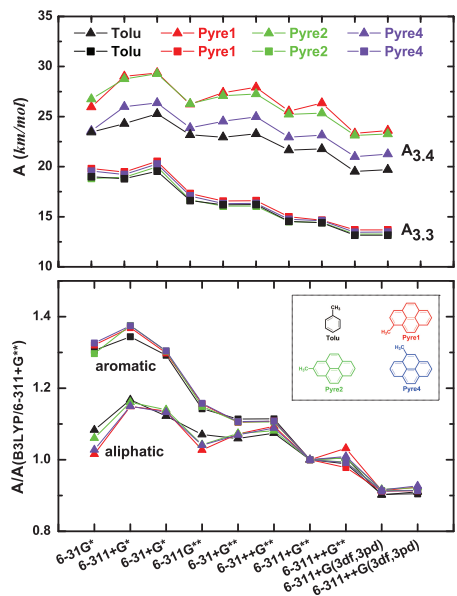}
}
\caption{\footnotesize
         \label{fig:IMeIAr}
         Top: Absolute intensities of the methyl (aliphatic)
         C--H stretch per chemical bond ($\Aali$; triangles)
         and of the aromatic C--H stretch ($\Aaro$; squares)
         for toluene and the three isomers of methyl pyrene
         computed at the B3LYP level with different basis sets.
         Bottom: Relative intensities of $\Aali$ and $\Aaro$
         computed at different basis sets
         with respect to those at B3LYP/6-311+G**
         (i.e, our standard level).
         }
\end{figure*}

\begin{figure}
\vspace{-10mm}
\centerline
{
\includegraphics[width=13.6cm,angle=0]{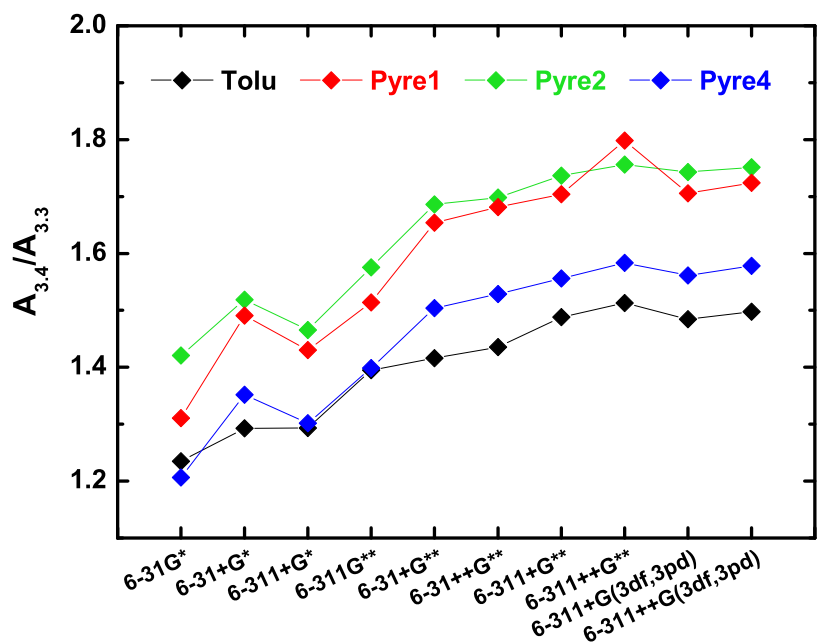}
}
\vspace{-4mm}
\caption{\footnotesize
         \label{fig:ARatio_BasisSet}
         Band-strength ratios ($\Aratio$) computed with
         different basis sets for toluene (i.e., methylbenzene)
         and the three isomers of methylpyrene.
         From left to right, the computations become
         increasingly more computer-time intensive
         and the results are expected to be more accurate.
         The results computed with the B3LYP method and
         in conjunction with the basis sets
         {\rm 6-311+G$^{\ast\ast}$},
         {\rm 6-311++G$^{\ast\ast}$},
         {\rm 6-311+G(3df,3pd)}, and
         6-311++G(3df,3pd)
         have essentially reached the convergence limit.
         For a compromise between accuracy
         and computational demand,
         the method of {\rm B3LYP/6-311+G$^{\ast\ast}$}
         is preferred.
         }
\end{figure}

The basis set effects were studied extensively
at the B3LYP level for toluene and the three isomers
of methylpyrene. The results are shown in
Table~\ref{Table:Average_Freq_BasisSet_2}
and illustrated in Figures~\ref{fig:IMeIAr}
and \ref{fig:ARatio_BasisSet}.

The first observation is that $A_{\Aro}$ is
almost invariant to the specific nature of the molecule.
The second observation is that there is a very large
basis set dependency in that $A_{\Aro}$ is greatly
reduced with the improvements of the basis set.
A typical $A_{\Aro}$ value
at the {\rm B3LYP/6-31G$^{\ast}$} level
is $\simali$18--20$\km\mol^{-1}$
and this value drops to $\simali$12.5--13.3$\km\mol^{-1}$,
i.e., a scaling factor of $\faro\approx0.7$.
In contrast, $A_{\Ali}$ greatly depends on
the specific isomer and the basis set dependency
of $A_{\Ali}$ is less than that of $A_{\Aro}$.
A typical $A_{\Ali}$ value
at the {\rm B3LYP/6-31G$^{\ast}$} level
is $\simali$23--27$\km\mol^{-1}$ and
this value drops to $\simali$19--24$\km\mol^{-1}$,
i.e., a scaling factor of $\fali\approx0.85$.

The plots in Figure~\ref{fig:IMeIAr} (left)
show that the absolute intensities are greatly
improved by the addition of at least single sets
of polarization functions on both C and H atoms
(and larger sets of polarization functions
provide only small additional benefits),
by the presence of single sets of diffuse functions
on carbons (while diffuse functions on H atoms
are less important), and by replacing
the split-valence basis set
with a triply-split valence basis set.
For our purposes the main question concerns
the convergence of the intensity ratio $\Aratio$
as a function of the theoretical level.
In this regard, the data suggest that an adequate
convergence value is obtained even with split-valence
basis sets and with single sets of diffuse function
on carbons and single sets of polarization functions
on all atoms (i.e., {\rm 6-31+G$^{\ast\ast}$},
{\rm 6-311+G$^{\ast\ast}$} or better).

The plots in Figure~\ref{fig:IMeIAr} (right)
show the scaling factor of the levels B3LYP/$\Lj$
relative to the intensity computed at the level
{\rm B3LYP/6-311+G$^{\ast\ast}$};
i.e., $A(\Lj)$/A({\rm B3LYP/6-311+G$^{\ast\ast}$}).
As can be seen, for all of the levels $\Lj$ equal to
or better than {\rm B3LYP/6-311+G$^{\ast\ast}$}
the scaling factors for the intensities
of the methyl and aromatic C--H stretches are very similar.
Figure~\ref{fig:ARatio_BasisSet} also shows that
the band-strength ratios $\Aratio$
computed with the basis sets
{\rm 6-311+G$^{\ast\ast}$},
{\rm 6-311++G$^{\ast\ast}$},
{\rm 6-311+G(3df,3pd)}, and
{\rm 6-311++G(3df,3pd)}
have essentially reached the convergence limit.
Meanwhile, as shown in
Table~\ref{Table:Average_Int_B3LYP_MP2},
the $A_{\Ali}/A_{\Aro}$ is less dependent on
the method than the basis sets.
This is expected since the overall intensity
calculated with different methods using a common basis set
are linearly correlated (cf. eqs.\,\ref{eq:B3LYP_MP2_scale}).
We therefore concluded that
the {\rm B3LYP/6-311+G$^{\ast\ast}$} method presents
an excellent compromise between accuracy
and computational demand.

Based on the insights derived from the above analysis,
we decide to determine the structures of and to perform
vibrational analyses for all parent systems
and all of their methyl-derivatives
at the {\rm B3LYP/6-311+G$^{\ast\ast}$} level
so as to ensure reliable $A_{\Ali}/A_{\Aro}$ values.
The energies and thermochemical parameters are
listed in Table~\ref{Table:Summary_min_6311}.
The most important results of
the vibrational analysis are summarized
in Table~\ref{Table:Average_Freq_BasisSet_2}.

\section{Recommended Band Intensities\label{sec:recomA}}
\begin{figure}
\centerline
{
\includegraphics[width=12.8cm,angle=0]{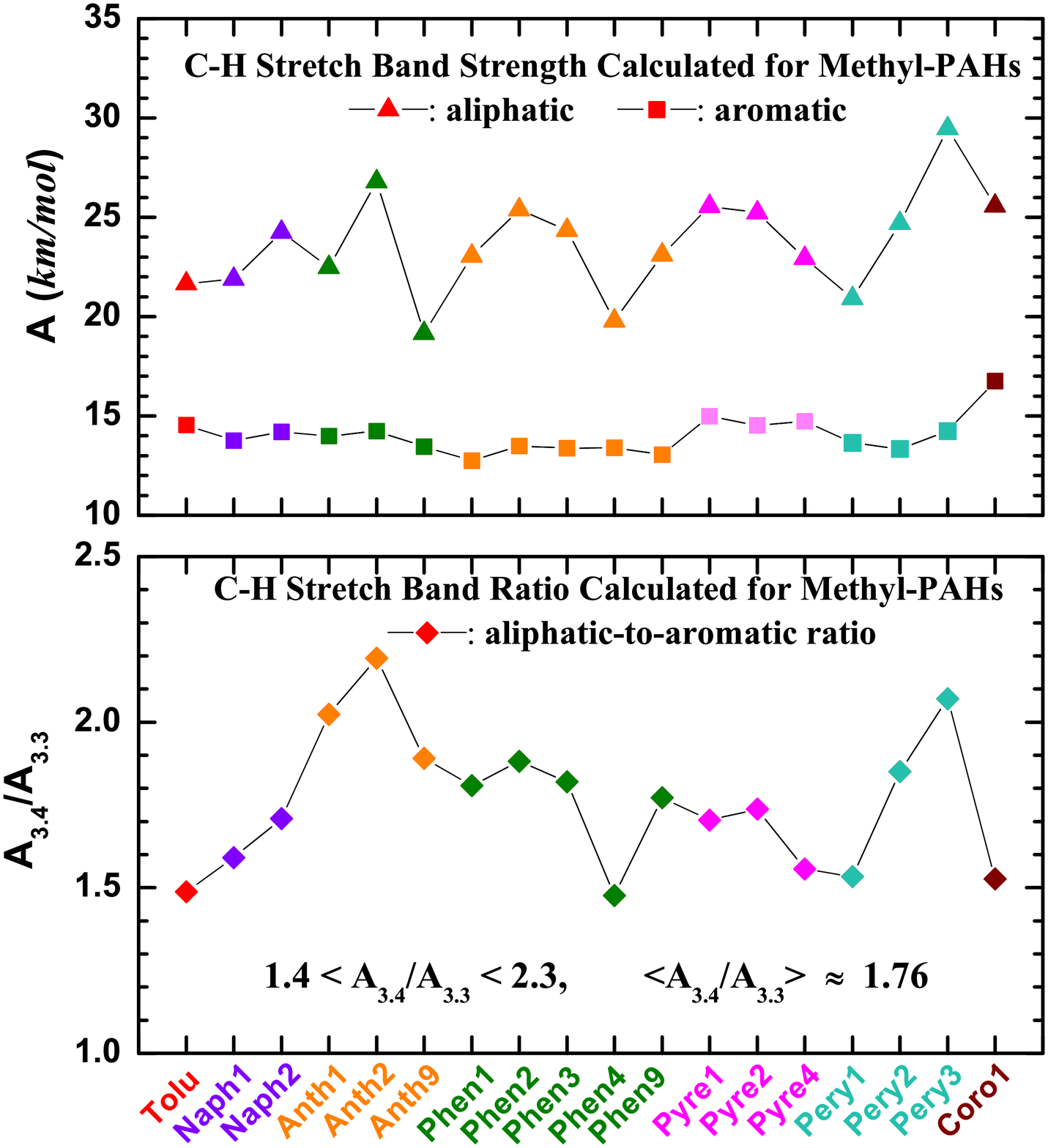}
}
\caption{\footnotesize
         \label{fig:A34A33_all}
         Band-strength as determined
         with the B3LYP/{\rm 6-311+G$^{\ast\ast}$} method
         for the mono-methyl derivatives
         of seven aromatic molecules and all of their isomers
         (benzene, naphthalene, anthracene,
         phenanthrene, pyrene, perylene, and coronene).
         }
\end{figure}

As shown in Figure~\ref{fig:A34A33_all} (top panel),
the aromatic C--H stretch band strength does not vary
significantly for different molecules.
It has an average value (per aromatic C--H bond) of
$\langle \Aaro\rangle \approx 14.03\km\mol^{-1}$,
with a standard deviation of
$\sigma(\Aaro)\approx 0.89\km\mol^{-1}$.
On the other hand, the aliphatic C--H stretch
band strength is more dependent on the nature
of the molecule and also on the specific isomer.
The average band strength (per aliphatic C--H bond)
is $\langle \Aali\rangle \approx 23.68\km\mol^{-1}$,
and the standard deviation is
$\sigma(\Aali)\approx 2.48\km\mol^{-1}$.

All of the above values are calculated at
the {\rm B3LYP/6-311+G$^{\ast\ast}$}
(i.e., $BL2$) level.
As discussed in \S\ref{sec:I_scale}, these values
need to be scaled.
By taking {\rm MP2(full)/6-311+G(3df,3pd)}
(i.e., $ML3$) to be the level
which gives the most reliable band strength,
the intensities need to be scaled with two formulae:
eqs.\,\ref{eq:MP2_LS_MS} and \ref{eq:B3LYP_MP2_MS}.
Thus, we recommend the value of
$\langle\Aaro\rangle \approx 14.03
\times 0.7877 \times 0.8089
\approx 8.94$$\km\mol^{-1}$
(i.e., $\simali$$1.49\times10^{-18}\cm$
per C--H bond), and
$\langle \Aali\rangle \approx 23.68
\times 0.7877 \times 0.8089
\approx 15.09$$\km\mol^{-1}$
(i.e., $\simali$$2.50\times10^{-18}\cm$
per C--H bond).

For the $\Aratio$ ratio, we have shown
in \S\ref{subsubsec:BSEffect_B3LYP} that
the {\rm B3LYP/6-311+G$^{\ast\ast}$} level
provides reliable values.
In Figure~\ref{fig:A34A33_all} (bottom panel)
we show the $\Aratio$ ratio calculated at this level.
We see that for all the molecules considered in this work
the $\Aratio$ values fall in the range
between $\simali$1.4 and $\simali$2.3,
with an average value of
$\langle\Aratio\rangle\approx1.76$.\footnote{%
   From the mean band strengths
   $\langle\Aaro\rangle \approx 8.94$$\km\mol^{-1}$
   and $\langle \Aali\rangle \approx 15.09$$\km\mol^{-1}$
   we obtain
   $\langle\Aali\rangle/\langle\Aaro\rangle \approx 1.69$.
   }
Considering that $\Aali$ depends significantly
on the specific molecule and isomers
while $\Aaro$ is relatively stable
for all the molecules,
it is not surprising that the $\Aratio$ values
show a high structure dependency.
For example, as shown in
Table~\ref{Table:Average_Freq_6311},
the methyl group of
the five phenanthrene isomers
give rise to $\Aratio$
between $\simali$1.4 and $\simali$1.9.
A similar isomer dependency is observed
for perylene with $\Aratio$
between $\simali$1.5 and $\simali$2.2.
It is important to fully realize this
high structure-dependency of
the $\Aratio$ ratio and this finding
stresses the need to study the formation
processes for methyl-substituted PAHs.

Finally, we also show in
Table~\ref{Table:Average_Freq_6311}
the experimental $\Aratio$ values obtained from
the NIST absorption spectra,
$\left(A_{\Ali}/A_{\Aro}\right)_{\rm NIST}$.

\section{Astrophysical Implications\label{sec:astro}}
In some HII regions, reflection nebulae
and planetary nebulae
(as well as extragalactic regions,
e.g., see Yamagishi et al.\ 2012,
Kondo et al.\ 2012, Kaneda et al.\ 2014),
the UIE band near 3$\mum$ exhibits a rich spectrum:
the dominant 3.3$\mum$ feature is usually accompanied
by a weaker feature at 3.4$\mum$
along with an underlying plateau
extending out to $\simali$3.6$\mum$
(see Figure~\ref{fig:astrospec}).
In some objects, a series of weaker features
at 3.46, 3.51, and 3.56$\mum$ are also seen superimposed
on the plateau, showing a tendency to decrease in strength
with increasing wavelength
(see Geballe et al.\ 1985, Jourdain de Muizon et al.\ 1986,
Joblin et al.\ 1996).
While the assignment of the 3.3$\mum$ emission feature to
the aromatic C--H stretch is widely accepted,
the precise identification of the 3.4$\mum$ feature
(and the accompanying weak features at 3.46, 3.51, and 3.56$\mum$
and the broad plateau) remains somewhat controversial.
By assigning the 3.4$\mum$ emission exclusively
to aliphatic C--H, one can place an upper limit
on the aliphatic fraction of the emitters
of the UIE features. 

Let $I_{3.4}$ and $I_{3.3}$ respectively
be the observed intensities of the 3.4$\mum$
and 3.3$\mum$ emission features.
Let $N_{\rm H,aliph}$ and $N_{\rm H,arom}$ respectively
be the numbers of aliphatic and aromatic C--H bonds
in the emitters of the 3.3$\mum$ UIE feature.
We obtain
$N_{\rm H,aliph}/N_{\rm H,arom}\approx\left(I_{3.4}/I_{3.3}\right)
\times\left(A_{3.3}/A_{3.4}\right)$.
We assume that one aliphatic C atom corresponds to
2.5 aliphatic C--H bonds (intermediate between methylene --CH$_2$
and methyl --CH$_3$) and one aromatic C atom corresponds to
0.75 aromatic C--H bond (intermediate between benzene C$_6$H$_6$
and coronene C$_{24}$H$_{12}$).
Therefore, in the UIE carriers the ratio of the number of C atoms
in aliphatic units to that in aromatic rings is
$N_{\rm C,aliph}/N_{\rm C,arom}\approx
\left(0.75/2.5\right)\,\times\,N_{\rm H,aliph}/N_{\rm H,arom}
= 0.3\times\,\left(I_{3.4}/I_{3.3}\right)
\times\,\left(A_{3.3}/A_{3.4}\right)$.
Yang et al.\ (2013) have compiled and analyzed
the UIE spectra of 35 sources
available in the literature
which exhibit both the 3.3$\mum$
and 3.4$\mum$ C--H features.
They derived a median ratio of
$\langle\Iratio\rangle\approx 0.12$,
with the majority (31/35) of
these sources having $\Iratio < 0.25$
(see Figure~\ref{fig:Iratio}).
With an average bond strength ratio
of $\Aali/\Aaro\approx 1.76$
(see \S\ref{sec:recomA}),
we obtain $N_{\rm C,aliph}/N_{\rm C,arom}\approx0.02$.
This suggests that the UIE emitters are predominantly
aromatic and the aliphatic component is only a very minor
part of the UIE emitters.



So far we have been focusing on the mono-methyl derivatives
of selected, relatively small PAHs. In reality, one would
assume that the PAH molecules in space cover a much larger
range of sizes, from a few tens of C atoms up to several
thousands, with a mean size of $\simali$100 C atoms
(see Li \& Draine 2001).
They may include defects,
substituents
(e.g., N in place of C;
see Hudgins et al.\ 2005),
partial dehydrogenation,
and sometimes superhydrogenation.
They could also include larger alkyl side chains
(ethyl, propyl, butyl, ...),
and several alkyl side chains might be present
in one PAH molecule.
Moreover, the alkyl side chains
and spacers might be unsaturated
(i.e., --CH=CH$_2$, --CH=CH--, C=CH$_2$, C=C--H).
Kwok \& Zhang (2013) argued that the 3.4$\mum$ interstellar
emission feature may not be the only manifestation of
the aliphatic structures of the UIE emitters.
They hypothesized that the clustering
of aromatic rings may break up the simple methyl- or
methylene-like side groups and hence the aliphatic
components may take many other forms.

\begin{figure*}
\centerline{
\includegraphics[scale=0.4,clip]{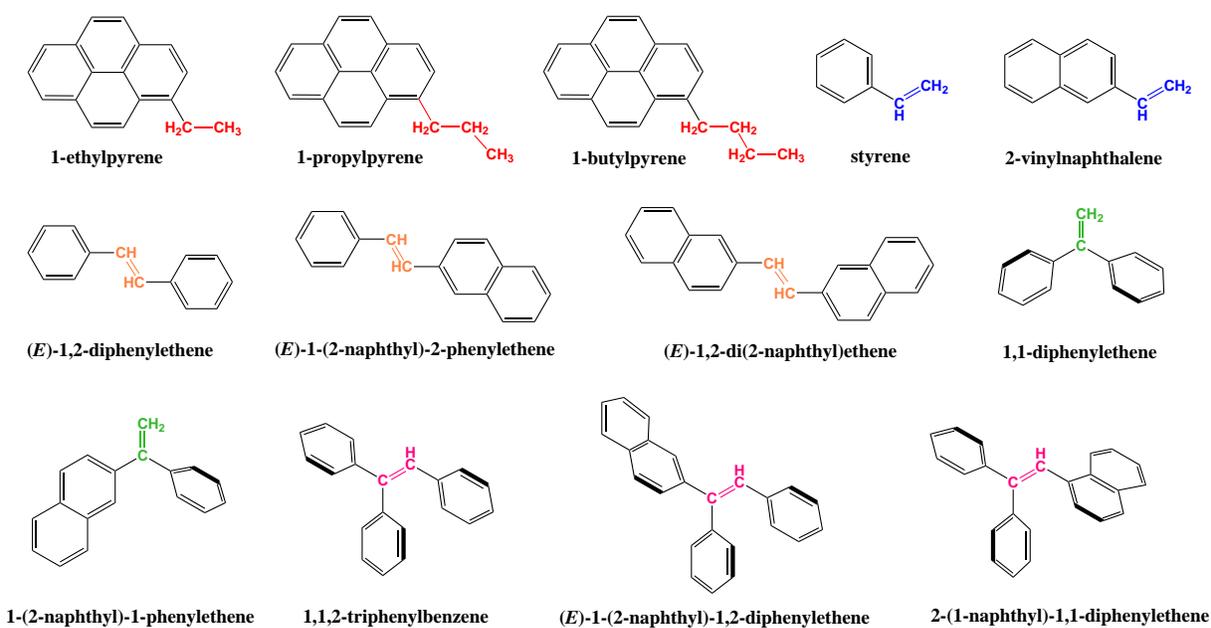}
}
\caption{\footnotesize
         Structures of PAHs attached with sidegroups
         other than methyl:
         ethyl (--CH$_2$--CH$_3$),
         propyl (--CH$_2$--CH$_2$--CH$_3$),
         butyl (--CH$_2$--CH$_2$--CH$_2$--CH$_3$),
         and unsaturated alkyl chains
         (--CH=CH$_2$, --CH=CH--, C=CH$_2$, C=C--H).
         \label{fig:other_sidegroup}
         }
\end{figure*}

Considering that many functional groups
other than methyl might also attach to
the aromatic rings and arise a feature at 3.4$\mum$,
we also included them in our computations
(see Yang et al.\ 2016a for details).
These computations were performed
at {\rm B3LYP/6-311+G$^{\ast\ast}$}, our standard level.
The structures considered are shown
in Figure~\ref{fig:other_sidegroup}
and they cover a wide range of sidegroups,
including ethyl (--CH$_2$--CH$_3$),
propyl (--CH$_2$--CH$_2$--CH$_3$),
butyl (--CH$_2$--CH$_2$--CH$_2$--CH$_3$),
and several unsaturated alkyl groups and spacers
(--CH=CH$_2$, --CH=CH--, C=CH$_2$, C=C--H).


From the top panel of
Figure~\ref{fig:A33A34_sidegroup_all},
we can see that the aliphatic C--H stretch
band strength varies within a wide range.
For ethyl, propyl and butyl,
the values ($\simali$25--30$\km\mol^{-1}$)
are generally consistent with methyl
(c.f. Figure~\ref{fig:A34A33_all} top panel),
while those for the unsaturated alkyl chains
(--CH=CH$_2$, --CH=CH--, C=CH$_2$, C=C--H)
are much lower ($\simali$5--15$\km\mol^{-1}$).
On the other hand, the aromatic C--H stretch
band strength stays stable for all the groups,
$\simali$10--15$\km\mol^{-1}$,
which is also consistent with
the corresponding value of methyl PAHs
(c.f. Figure~\ref{fig:A34A33_all} top panel).
%
%
Therefore, we conclude that the $\Aratio$ ratios
for PAHs with ethyl, propyl and butyl groups
are close to that of methyl PAHs.
The $\Aratio$ ratios for PAHs with unsaturated
alkyl chains could be lower by a factor of up to
$\simali$3 than that of methyl PAHs.
However, it is apparent that PAHs with unsaturated
alkyl chains are less stable than methyl PAHs
when subject to UV photons in the ISM.
We also note that PAHs with a large side chain
are not as stable against photolytic dissociation
as methyl-substituted PAHs.
If a large aliphatic chain (e.g., --CH$_2$--CH$_3$)
is attached to an aromatic structure, the most likely
photodissociation product is a benzyl radical
PAH-$\dot{\rm C}$H$_2$
(i.e., a --CH$_2$ group attached to a PAH molecule),
which, when subject to the reaction
PAH-$\dot{\rm C}$H$_2$ + H $\rightarrow$ PAH--CH$_3$
will rapidly lead to the product of a CH$_3$ group
at the periphery of an aromatic molecule
(Joblin et al.\ 1996; Hwang et al.\ 2002).
Therefore, neither PAHs with a large side chain
nor PAHs with unsaturated alkyl chains
are expected to be present in the ISM
in a large abundance.

Considering that several alkyl side chains
might be present in one PAH molecule,
we also consider the situation that there
are two methyl groups attached to a PAH molecule,
using pyrene as an example
(see Yang et al.\ 2016a for details).
We consider all possible isomers of
dimethyl-substituted pyrene
(see Figure ~\ref{fig:pyrene_dimethyl}).
For dimethyl pyrenes, the aliphatic C--H stretch
band strength varies within $\simali$18--27$\km\mol^{-1}$,
while these values for the aromatic C--H stretch
are generally $\simali$15$\km\mol^{-1}$
(c.f. Figure~\ref{fig:A34A33_dimethyl_all} top panel).
The $\Aratio$ ratios vary from
$\simali$1.25 (Pyre110) to $\simali$1.75 (Pyre27),
with an average ratio of
$\langle\Aratio\rangle\approx1.57$
(c.f. Figure~\ref{fig:A34A33_dimethyl_all} bottome panel),
which is only $\simali$11\% lower
than the mean ratio of
$\langle\Aratio\rangle\approx1.76$
computed from methyl PAHs
(see \S\ref{sec:recomA}).

The methyl groups are essentially independent of each other.
Noticeable effects on frequency and intensity only occur
when several alkyl groups are placed in direct proximity.
We note that for methyl PAHs the frequencies of
the aliphatic C--H stretch are always smaller than
$\simali$3000$\cm^{-1}$
and those for the aromatic C--H stretch
are larger than $\simali$3000$\cm^{-1}$.
The positions of the C--H stretches of
simple alkenes and dienes coincide with
the methyl signals of methyl-substituted PAHs.
However, for CH=CH$_2$ and C=CH$_2$,
one of the aliphatic C--H stretches
falls at $\simali$3120$\cm^{-1}$
(i.e, in the ``aromatic'' region).
For dimethyl pyrene Pyre45,
there is also one frequency of
the aliphatic C--H stretches that falls in
the ``aromatic'' region ($\simali$3070$\cm^{-1}$).

One may argue that the aliphatic chains and aromatic rings
in the MAON-, coal-, or kerogen-like UIE carriers may cluster
together and not every C atom is bonded to H atoms,
and therefore the C--H bands may not fully reveal
the aliphatic C.
We note that the clustering of aromatic rings
and aliphatic chains would be accompanied by
forming new C=C bonds and losing H atoms.
Laboratory measurements have also shown that the reduction
of H atoms leads to aromatization: with decreasing H/C and O/C
ratios, coal-like solid hydrocarbon materials
become more aromatic and exhibit weaker aliphatic
3.4$\mum$ features and stronger 3.3$\mum$ aromatic feature
(Papoular et al.\ 1989).

Finally, we note that,
in addition to the 3.4$\mum$ C--H stretching mode,
aliphatic hydrocarbon materials also have two C--H
deformation bands at 6.85$\mum$ and 7.25$\mum$.
These two bands have been observed in weak
{\it absorption} in the Galactic diffuse ISM
(Chiar et al.\ 2000).
They are also seen in {\it emission},
with the 6.85$\mum$ feature detected
both in the Milky Way
and in the Large and Small Magellanic Clouds
while the 7.25$\mum$ feature so far mostly seen
in the Magellanic Clouds
(e.g., see Sloan et al.\ 2014).
Their strengths (relative to the nearby 6.2 and 7.7$\mum$
C--C stretching bands) also allow an estimate of
the aliphatic fraction of the UIE carrier.
%
We have explored the aliphatic versus
aromatic content of the UIE carriers by examining
the ratio of the observed intensity of the 6.2$\mum$
aromatic C--C feature to that of the 6.85$\mum$ 
aliphatic C--H deformation features.
The fraction of C atoms in aliphatic form 
was derived to be at most $\simali$10\%,
confirming that the UIE emitters 
are predominantly aromatic
(see Yang et al.\ 2016b).

\section{Summary}\label{sec:summary}
The UIE carriers play an essential role in astrophysics
as an absorber of the UV starlight,
as an agent for photoelectrically
heating the interstellar gas,
and as a valid indicator of
the cosmic star-formation rates.
While the exact nature of the UIE carriers
remains unknown, the ratios of the observed intensities
of the 3.3$\mum$ aromatic C--H stretching emission feature
($\Iaro$) to that of the 3.4$\mum$ aliphatic C--H
emission feature ($\Iali$) could provide constraints
on the chemical structures of the UIE carriers,
i.e., are they mainly aromatic or largely aliphatic
with a mixed aromatic/aliphatic structure?
To this end, the knowledge of the intrinsic strengths
(per chemical bond)
of the 3.3$\mum$ aromatic C--H stretch ($\Aaro$)
and the 3.4$\mum$ aliphatic C--H stretch ($\Aali$)
is required. It is the purpose of this review to
present an overview on how $\Aratio$ is derived
from extensive computations of the vibrational 
frequencies and intensities of a range of methyl PAHs 
and PAHs with other side groups,
using density functional theory
and second-order perturbation theory
%
to compute their vibrational spectra.
The major results are:
\begin{enumerate}
\item The structures and excitation mechanisms of
      the major proposed carriers
      are examined in terms of two broad categories:
      free-flying PAH molecules
      or solid hydrocarbon materials with
      a mixed aromatic/aliphatic structure
      (HAC, QCC, soot, coal/kerogen, MAON).
\item The hybrid density functional
      theoretical method (B3LYP)
      in conjunction with a variety of basis sets
      [{\rm 6-31G$^{\ast}$},
       {\rm 6-31+G$^{\ast}$},
       {\rm 6-311+G$^{\ast}$},
       {\rm 6-311G$^{\ast\ast}$},
       {\rm 6-31+G$^{\ast\ast}$},
       {\rm 6-31++G$^{\ast\ast}$},
       {\rm 6-311+G$^{\ast\ast}$},
       {\rm 6-311++G$^{\ast\ast}$},
       {\rm 6-311+G(3df,3pd)}, and
       {\rm 6-311++G(3df,3pd)}
       ]
       are employed to calculate the vibrational
       spectra for a range of aromatic molecules
       (naphthalene, anthracene, phenanthrene,
        pyrene, perylene, and coronene)
       with a methyl side chain.
       M$\o$ller-Plesset perturbation theory (MP2)
       is also included with basis sets
       {\rm 6-311+G$^{\ast\ast}$}
       and {\rm 6-311++G(3df,3pd)}
       for some of the molecules.
\item With the use of the standard frequency scale factors,
      a good to excellent agreement between the measured
      and computed C--H stretch vibrational frequencies
      is achieved in all cases,
      for all molecules at all levels.
\item The band intensities calculated with {\rm B3LYP/6-31G$^{\ast}$}
      are much higher than the gas-phase experimental values.
      Using better basis sets in conjunction with
      the B3LYP method, the computed intensities
      are still considerably higher (by $\simali$30\%)
      compared to the experimental results.
\item The MP2(full) method with the basis set of
      6-311+G(3df,3pd) reproduces the measured
      intensities reasonably well.
      However, such calculations are far too expensive
      especially for large molecules.
      It is shown that intensity scaling approaches
      that are based on the B3LYP data
      can be just as successful.
\item By taking {\rm MP2(full)/6-311+G(3df,3pd)}
      to be the level
      which gives the most reliable band strengths,
      we determine and apply scaling factors
      to the intensities
      computed at the {\rm B3LYP/6-311+G$^{\ast\ast}$} level.
      The recommended band strengths (per chemical bond)
      are $\langle\Aaro\rangle \approx 8.94$$\km\mol^{-1}$
      (i.e., $\simali$$1.49\times10^{-18}\cm$
       per C--H bond), and
      $\langle \Aali\rangle \approx 15.09$$\km\mol^{-1}$
      (i.e., $\simali$$2.50\times10^{-18}\cm$ per C--H bond),
      where $\Aali$ depends significantly
      on the specific molecule and isomers
      while $\Aaro$ is relatively stable
      for all the molecules.
\item The band-strength ratios $\Aratio$
      computed with the basis sets
      {\rm 6-311+G$^{\ast\ast}$},
      {\rm 6-311++G$^{\ast\ast}$},
      {\rm 6-311+G(3df,3pd)}, and
      {\rm 6-311++G(3df,3pd)}
      essentially reach the convergence limit.
      The $A_{\Ali}/A_{\Aro}$ ratio is less
      dependent on the method than the basis sets.
      The {\rm B3LYP/6-311+G$^{\ast\ast}$} method
      presents an excellent compromise between accuracy
      and computational demand.
      Therefore, to ensure reliable
      $A_{\Ali}/A_{\Aro}$ values,
      we adopt this theoretical level to
      compute all of the molecules.
      For the molecules considered in this work
      the $\Aratio$ values fall in the range
      between $\simali$1.4 and $\simali$2.3,
      with an average value of
      $\langle\Aratio\rangle\approx1.76$.
\item By attributing the 3.4$\mum$ feature
      exclusively to aliphatic C--H stretch
      (i.e., neglecting anharmonicity and
       superhydrogenation),
      we derive the fraction of C atoms
      in aliphatic form from $\Iratio\approx 0.12$
      and $\Aali/\Aaro\approx 1.76$ to be $\sim$\,2\%,
      where $\Iratio$, the ratio of the power
      emitted from the 3.4$\mum$ feature to that from
      the 3.3$\mum$ feature, has a median ratio of
      $\langle\Iratio\rangle\approx 0.12$
      for 35 astronomical sources which exhibit
      both the 3.3$\mum$ and 3.4$\mum$ C--H features.
      We conclude that the UIE emitters are
      predominantly aromatic.
\item Dimethyl pyrene is studied in the context
      that several alkyl side chains
      might be present in one PAH molecule.
      The $\Aratio$ ratio averaged over
      all the isomers of dimethyl-substituted pyrene
      is $\sim 1.57$, which is only $\simali$11\%
      lower than that of mono-methyl PAHs.
\item A wide range of sidegroups
      (other than methyl and dimethyl)
      have also been considered,
      including ethyl (--CH$_2$--CH$_3$),
      propyl (--CH$_2$--CH$_2$--CH$_3$),
      butyl (--CH$_2$--CH$_2$--CH$_2$--CH$_3$)
      and several unsaturated alkyl chains
      (--CH=CH$_2$, --CH=CH--, C=CH$_2$, C=C--H).
      The corresponding $\Aratio$ ratios are close
      to that of mono-methyl PAHs,
      except PAHs with unsaturated alkyl chains
      (for which the $\Aratio$ ratios could be lower
      by a factor of up to $\simali$3).
      However, these molecules are photolytically
      less stable compared to methyl PAHs
      and are not expected to be present in the ISM
      in a large abundance.
      The aliphatic C--H stretches of PAHs
      with unsaturated alkyl chains often fall
      in the wavelength range even shortward of
      the aromatic C--H stretch
      which are not seen in the ISM.
\end{enumerate}

\section*{Acknowledgements}
We thank Dr. T.J.~Horscroft and 
Prof.~S.N. Zhang for inviting us 
to submit this review and also
for their support and patience
during the preparation of this review.
We thank Prof. B.T.~Draine, 
Dr. J.Y.~Seok, and the anonymous referee 
for very helpful suggestions. 
AL and XJY are supported in part by
NSFC\,11473023, NSFC\,11273022, 
NSF AST-1311804, NNX13AE63G,
Hunan Provincial NSF\,2015JJ3124,
and the University of Missouri Research Board.
RG is supported in part by NSF-PRISM grant
Mathematics and Life Sciences (0928053).
Computations were performed using the high-performance computer
resources of the University of Missouri Bioinformatics Consortium.



\clearpage
\vspace{-10cm}

\clearpage

\clearpage

\vspace{-10cm}

\clearpage

\vspace{-10cm}
\begin{figure*}
\vspace{-10cm}
\centerline
{
\includegraphics[width=15.6cm,angle=0]{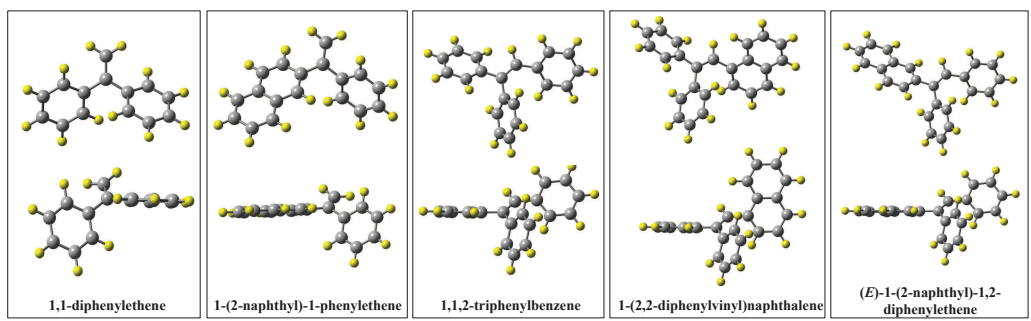}
}
\vspace{-10mm}
\caption{
         \label{fig:OptimStruct2}
         Optimized structures of phenyl- and
         naphthyl-substituted ethene.
         H atoms are shown in yellow and C atoms in grey.
         All structures are minima.
         }
\end{figure*}


\begin{figure}
\centerline
{
\includegraphics[width=12cm,angle=0]{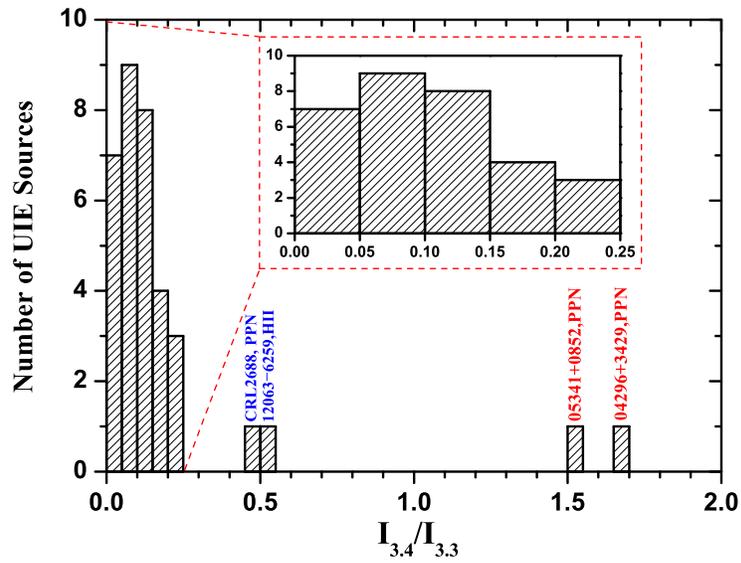}
}
\caption{\footnotesize
         \label{fig:Iratio}
         Histogram of the flux ratio ($\Iratio$) for 35 UIE sources.
         The median flux ratio is $\langle \Iratio\rangle\approx 0.12$.
         The insert panel enlarges the flux ratio distribution for the
         31 sources with $\Iratio\simlt 0.25$.
         }
\end{figure}


\begin{figure*}
\centerline{
\includegraphics[scale=0.48,clip]{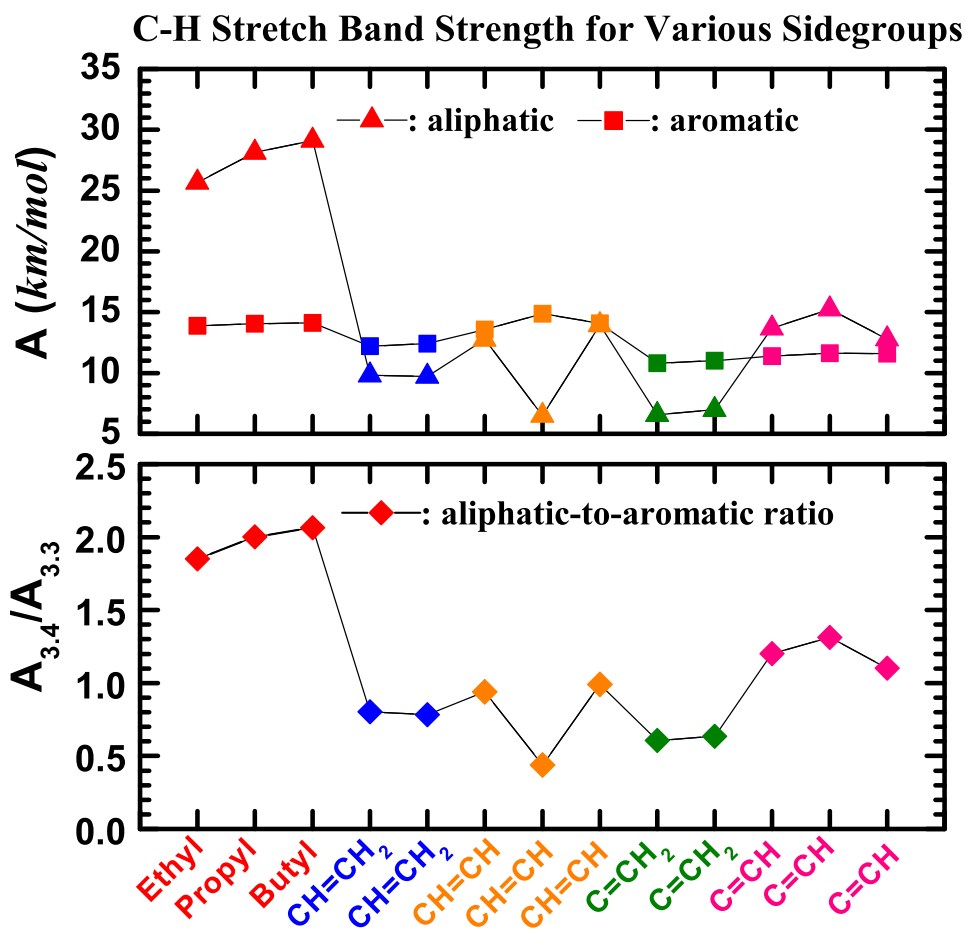}
}
\caption{\footnotesize
         Band-strengths as determined
         with the B3LYP/{\rm 6-311+G$^{\ast\ast}$} method
         for PAHs with sidegroups other than methyl:
         ethyl (--CH$_2$--CH$_3$),
         propyl (--CH$_2$--CH$_2$--CH$_3$),
         butyl (--CH$_2$--CH$_2$--CH$_2$--CH$_3$),
         and unsaturated alkyl chains
         (--CH=CH$_2$, --CH=CH--, C=CH$_2$, C=C--H).
         \label{fig:A33A34_sidegroup_all}
         }
\end{figure*}

\begin{figure*}
\centerline{
\includegraphics[scale=0.7,clip]{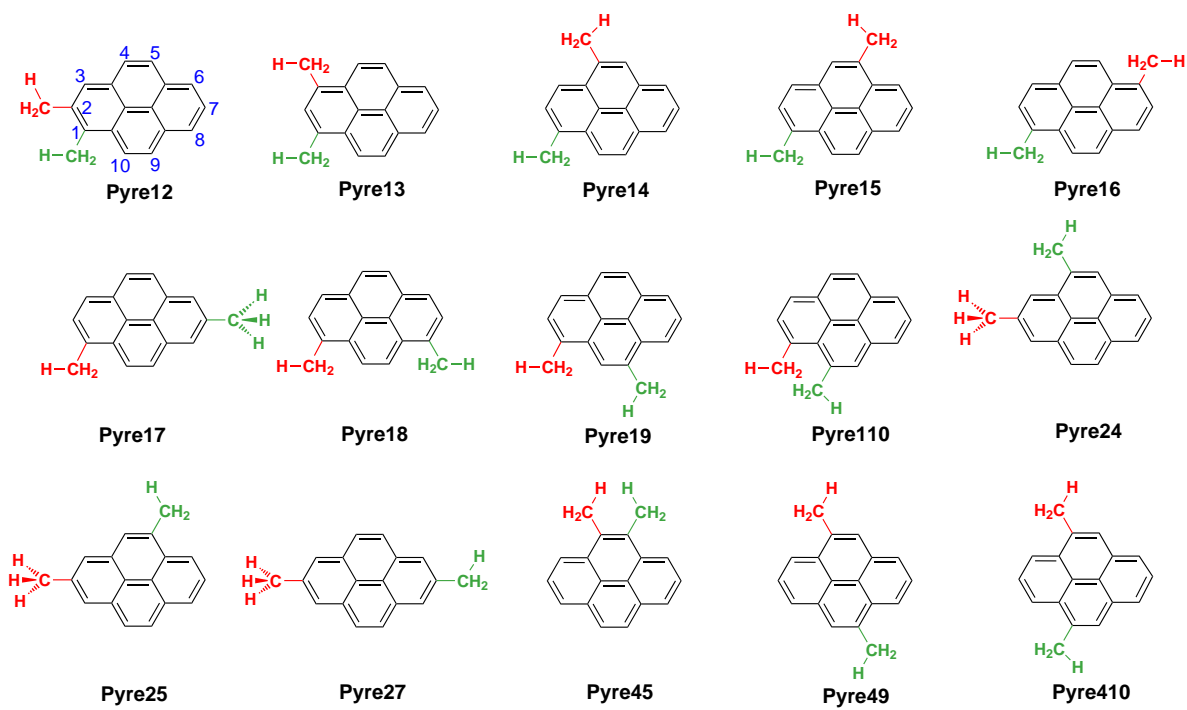}
}
\caption{\footnotesize
         Structures of all isomers of dimethylpyrene.
         The same naming method is used
         as in Figure~\ref{fig:MonoMethylPAHs}:
         ``Pyre'' stands for pyrene,
         and the digits specify the locations
         of the attached methyl groups
         (e.g., ``Pyre110'' means the two methyl
          groups are attached at positions 1 and 10).
          \label{fig:pyrene_dimethyl}
          }
\end{figure*}

\begin{figure*}
\centerline{
\includegraphics[scale=0.4,clip]{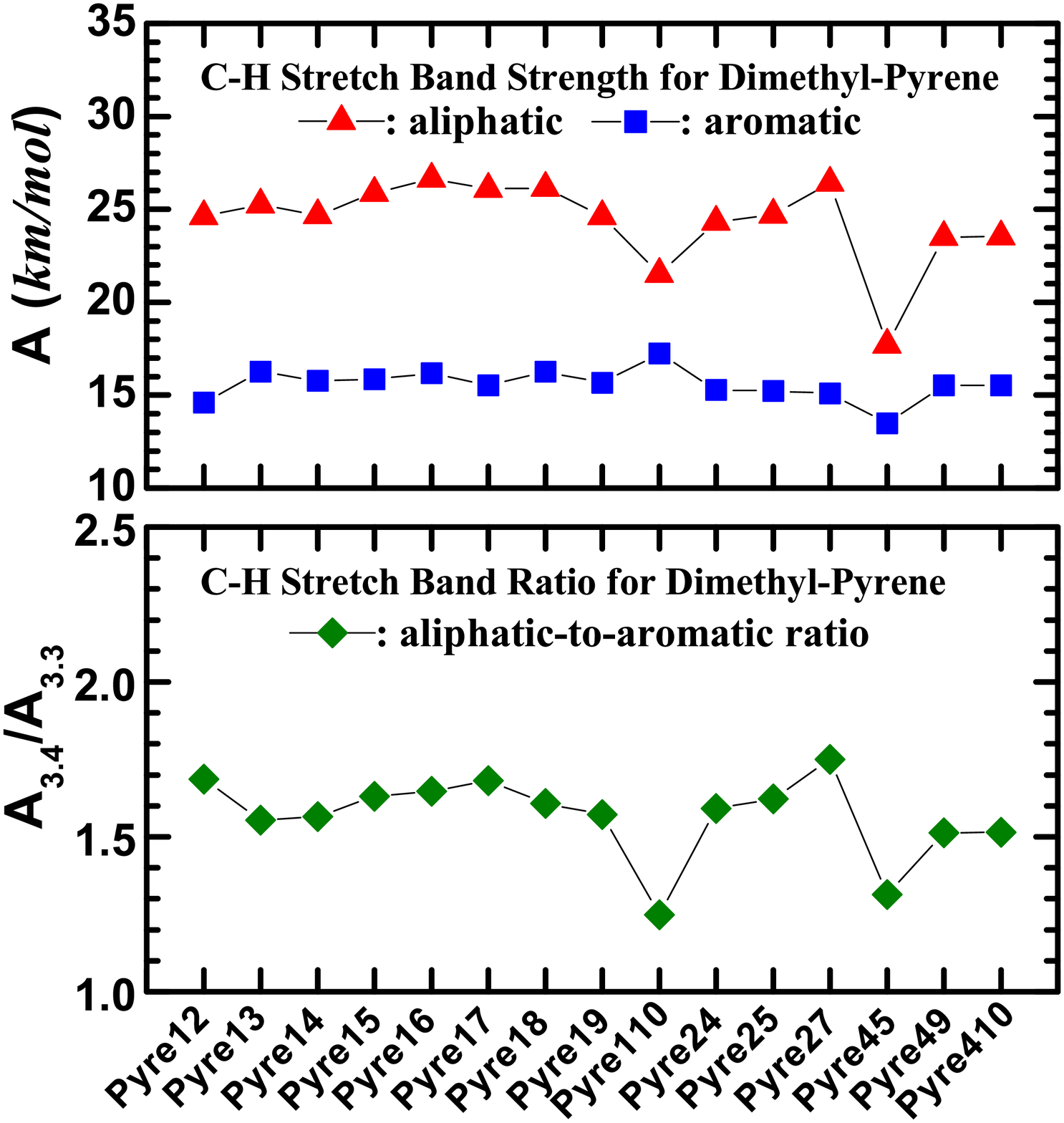}
}
\caption{\footnotesize
         \label{fig:A34A33_dimethyl_all}
         Band-strength as determined with
         the B3LYP/{\rm 6-311+G$^{\ast\ast}$} method
         for all the isomers of dimethyl pyrene.
         }
\end{figure*}
\clearpage


\begin{table*}[ht]
    \caption{Standard scale factors for computed vibrational frequencies.}
    \label{Table:Freq_scale}
     \begin{threeparttable}
    \centering
\begin{tabular}{lcc}
\noalign{\smallskip} \hline \hline \noalign{\smallskip}
Basis set     &      B3LYP   &       MP2            \\
\noalign{\smallskip} \hline \noalign{\smallskip}
6-31G$^\ast$       &     0.9613\tnote{a}   &\\
6-31+G$^\ast$      &     0.9640\tnote{b}   &\\
6-311+G$^{\ast}$   & 0.9678\tnote{c}   &\\
6-311G$^{\ast\ast}$  & 0.9670\tnote{b}   &\\
6-31+G$^{\ast\ast}$  & 0.9640\tnote{b}   &\\
6-31++G$^{\ast\ast}$ & 0.9640\tnote{b}   &\\
6-311+G$^{\ast\ast}$ & 0.9688\tnote{a} & 0.9523\tnote{a}\\
6-311++G$^{\ast\ast}$ & 0.9679\tnote{c} & \\
6-311+G(3df,3pd)  & 0.9672\tnote{d} & 0.9540\tnote{a}\\
6-311++G(3df,3pd) & 0.9673\tnote{d} & \\
\hline \noalign{\smallskip}
\noalign{\smallskip} \noalign{\smallskip}
\end{tabular}
        \begin{tablenotes}
\item[a] Borowski (2012)
\item[b] Precomputed vibrational scaling factors
                  taken from the {\it Computational Chemistry
                  Comparison and Benchmark Database} (CCVCBC).
                  Online at http://cccbdb.nist.gov/vibscalejust.asp
                  (accessed 06/01/13)
\item[c] Andersson et al.\ 2005
\item[d] Merrick et al.\ 2007
        \end{tablenotes}
     \end{threeparttable}
    \end{table*}


\begin{table*}[ht]
    \caption{Computed total energies and thermochemical parameters
                  for the minima of all the methyl PAHs
                  shown in Figure~\ref{fig:MonoMethylPAHs}
                  at {\rm B3LYP/6-31G$^\ast$}.
                  }
    \label{Table:Summary_min_631}
     \begin{threeparttable}
    \centering
\begin{tabular}{lccccccc}
\noalign{\smallskip} \hline \hline \noalign{\smallskip}
Compound	&	$\Etot$\tnote{a}	
                &	VZPE\tnote{b}	
                &	TE\tnote{c}	
                &	$S$\tnote{d}	
                &	$\nu_1$\tnote{e}	
                &	$\nu_2$\tnote{e}	
                &	$\mu$\tnote{f}\\	
\noalign{\smallskip} \hline \noalign{\smallskip}
Toluc	&	-271.566650 	&	80.53 	&	84.42 	&	79.27 	&	42.95 	&	211.45 	&	0.3196 	\\
Naph1a	&	-425.209691 	&	110.36 	&	115.62 	&	89.08 	&	135.46 	&	170.33 	&	0.2956 	\\
Naph2b	&	-425.211193 	&	110.11 	&	115.49 	&	90.50 	&	95.66 	&	125.11 	&	0.4170 	\\
Anth1a	&	-578.847859 	&	139.67 	&	146.59 	&	102.01 	&	91.96 	&	93.37 	&	0.2897 	\\
Anth2b	&	-578.849214 	&	139.39 	&	146.43 	&	103.32 	&	71.42 	&	111.02 	&	0.4971 	\\
Anth9a	&	-578.842107 	&	139.61 	&	146.67 	&	104.52 	&	37.81 	&	106.76 	&	0.3224 	\\
Phen1a	&	-578.854850 	&	139.86 	&	146.78 	&	102.32 	&	69.98 	&	98.69 	&	0.2931 	\\
Phen2b	&	-578.857120 	&	139.56 	&	146.64 	&	104.24 	&	69.47 	&	80.52 	&	0.4800 	\\
Phen3b	&	-578.856977 	&	139.59 	&	146.66 	&	104.21 	&	71.92 	&	79.29 	&	0.4748 	\\
Phen4a	&	-578.845615 	&	140.23 	&	147.03 	&	102.47 	&	34.77 	&	91.70 	&	0.2896 	\\
Phen9a	&	-578.855626 	&	139.94 	&	146.84 	&	102.02 	&	85.30 	&	99.09 	&	0.3250 	\\
Pyre1a	&	-655.089741 	&	147.89 	&	155.21 	&	104.58 	&	86.96 	&	117.44 	&	0.4176 	\\
Pyre2c	&	-655.091084 	&	147.55 	&	155.06 	&	107.98 	&	32.65 	&	76.60 	&	0.4572 	\\
Pyre4a	&	-655.090587 	&	147.93 	&	155.22 	&	104.23 	&	88.52 	&	126.68 	&	0.3563 	\\
Pery1c	&	-808.713736 	&	177.50 	&	186.40 	&	116.34 	&	60.05 	&	81.63 	&	0.4009 	\\
Pery2a	&	-808.724609 	&	176.97 	&	186.23 	&	121.10 	&	25.21 	&	82.46 	&	0.4250 	\\
Pery3a	&	-808.722947 	&	177.13 	&	186.31 	&	122.62 	&	6.21 	&	72.41 	&	0.4337 	\\
Coro1a	&	-961.214987 	&	193.45 	&	203.31 	&	122.27 	&	71.73 	&	89.44 	&	0.4294 	\\
\hline
\noalign{\smallskip}
\noalign{\smallskip}
\end{tabular}
        \begin{tablenotes}
\item[a] Total energies in atomic units.
\item[b] Vibrational zero-point energies (VZPE)
                  in $\kcal\mol^{-1}$.
\item[c] Thermal energies (TE) in $\kcal\mol^{-1}$.
\item[d] Molecular entropies ($S$) in $\cals\mol^{-1}\K^{-1}$.
\item[e] The lowest vibrational modes $\nu_1$ and $\nu_2$
                  in $\cm^{-1}$.
\item[f] Dipole moment in Debye.
        \end{tablenotes}
     \end{threeparttable}
    \end{table*}

\begin{table*}
\footnotesize
\begin{center}
\caption[]{Computed total energies and thermochemical parameters
                for the transition states of the methyl PAHs
                shown in Figure~\ref{fig:MonoMethylPAHs}
                at {\rm B3LYP/6-31G$^\ast$}.
                Note that the transition state has 
                an imaginary frequency (i.e., $\nu_1$)
                which is sometimes negative.
                }
\label{Table:Summary_TS_631}
\begin{tabular}{lccccccc}
\noalign{\smallskip} \hline \hline \noalign{\smallskip}
Compound	&	$\Etot$
                &	VZPE
                &	TE
                &	$S$
                &	$\nu_1$
                &	$\nu_2$
                &	$\mu$
\\\noalign{\smallskip} \hline \noalign{\smallskip}
Tolua	&	-271.566621 	&	80.46 	&	83.82 	&	74.15 	&	-33.24 	    &	212.77 	&	0.3218 	\\
Naph1b	&	-425.206459 	&	110.10 	&	115.00 	&	86.92 	&	-184.14 	&	122.52 	&	0.3865 	\\
Naph2a	&	-425.210243 	&	109.96 	&	114.88 	&	87.05 	&	-105.36 	&	118.80 	&	0.4218 	\\
Anth1b	&	-578.844198 	&	139.39 	&	145.96 	&	99.97 	&	-205.78 	&	85.65 	&	0.4125 	\\
Anth2a	&	-578.847870 	&	139.24 	&	145.83 	&	100.10 	&	-117.02 	&	72.64 	&	0.5099 	\\
Anth9c	&	-578.842064 	&	139.60 	&	146.13 	&	99.50 	&	-56.97 	    &	84.56 	&	0.3059 	\\
Phen1b	&	-578.851600 	&	139.56 	&	146.15 	&	100.48 	&	-188.17 	&	61.85 	&	0.4120 	\\
Phen2a	&	-578.856498 	&	139.44 	&	146.03 	&	100.28 	&	-87.20 	    &	74.80 	&	0.4752 	\\
Phen3a	&	-578.856363 	&	139.47 	&	146.05 	&	100.23 	&	-80.21 	    &	77.31 	&	0.4686 	\\
Phen4c	&	-578.838725 	&	139.68 	&	146.20 	&	99.60 	&	-205.94 	&	79.23 	&	0.2660 	\\
Phen9b	&	-578.851402 	&	139.60 	&	146.18 	&	100.34 	&	-212.72 	&	80.36 	&	0.4080 	\\
Pyre1b	&	-655.087304 	&	147.66 	&	154.60 	&	102.20 	&	-157.10 	&	83.70 	&	0.5462 	\\
Pyre2a	&	-655.091051 	&	147.49 	&	154.45 	&	102.31 	&	-16.07 	    &	77.15 	&	0.4598 	\\
Pyre4b	&	-655.086624 	&	147.68 	&	154.62 	&	102.24 	&	-208.36 	&	81.97 	&	0.4647 	\\
Pery1d	&	-808.707553 	&	177.09 	&	185.72 	&	114.66 	&	-220.94 	&	66.36 	&	0.4084 	\\
Pery2b	&	-808.723876 	&	176.86 	&	185.63 	&	117.28 	&	-89.40 	    &	25.41 	&	0.4541 	\\
Pery3b	&	-808.719443 	&	176.89 	&	185.69 	&	118.48 	&	-193.39 	&	17.37 	&	0.6114 	\\
Coro1b	&	-961.211602 	&	193.22 	&	202.72 	&	120.08 	&	-194.54 	&	68.28 	&	0.5411 	\\ \hline
\noalign{\smallskip} \noalign{\smallskip}
\end{tabular}
\end{center}
\normalsize
\end{table*}
\clearpage


\begin{table*}
\footnotesize
\caption[]{
           Same as Table~\ref{Table:Summary_min_631}
           but at the {\rm B3LYP/6-311+G$^{\ast\ast}$} level.
           }
\label{Table:Summary_min_6311}
\begin{tabular}{lccccccc}
\noalign{\smallskip} \hline \hline \noalign{\smallskip}
Compound	&	$\Etot$
                &	VZPE
                &	TE
                &	$S$
                &	$\nu_1$
                &	$\nu_2$
                &	$\mu$
\\ \noalign{\smallskip} \hline \noalign{\smallskip}
Benzene	&	-232.311242 	&	62.84 	&	65.61 	&	64.13 	&	409.45 	&	409.45 	&	0.0000 	\\
Toluc	&	-271.638814 	&	79.90 	&	83.80 	&	79.24 	&	44.80 	&	208.95 	&	0.4056 	\\\hline
Naph	&	-385.988871 	&	92.17 	&	96.46 	&	79.31 	&	173.26 	&	185.09 	&	0.0000 	\\
Naph1a	&	-425.315358 	&	109.49 	&	114.79 	&	89.37 	&	131.97 	&	164.79 	&	0.3704 	\\
Naph2b	&	-425.316894 	&	109.27 	&	114.67 	&	90.66 	&	93.95 	&	123.71 	&	0.5180 	\\\hline
Anth	&	-539.660248 	&	121.27 	&	127.22 	&	95.16 	&	91.25 	&	120.32 	&	0.0000 	\\
Anth1a	&	-578.987081 	&	138.58 	&	145.54 	&	102.39 	&	90.23 	&	90.87 	&	0.3688 	\\
Anth2b	&	-578.988540 	&	138.33 	&	145.40 	&	103.61 	&	69.69 	&	108.40 	&	0.6126 	\\
Anth9a	&	-578.981699 	&	138.56 	&	145.67 	&	105.09 	&	35.51 	&	99.39 	&	0.3669 	\\\hline
Phen	&	-539.668406 	&	121.51 	&	127.46 	&	94.01 	&	95.52 	&	99.88 	&	0.0154 	\\
Phen1a	&	-578.994150 	&	138.82 	&	145.80 	&	102.68 	&	68.76 	&	97.16 	&	0.3927 	\\
Phen2b	&	-578.996368 	&	138.53 	&	145.65 	&	104.51 	&	68.39 	&	79.96 	&	0.5846 	\\
Phen3b	&	-578.996254 	&	138.57 	&	145.67 	&	104.39 	&	72.49 	&	79.37 	&	0.5647 	\\
Phen4a	&	-578.984766 	&	139.13 	&	145.98 	&	102.64 	&	38.29 	&	90.02 	&	0.3458 	\\
Phen9a	&	-578.994871 	&	138.88 	&	145.84 	&	102.41 	&	83.88 	&	96.66 	&	0.4309 	\\\hline
Pyre	&	-615.915984 	&	129.36 	&	135.70 	&	94.90 	&	97.80 	&	151.01 	&	0.0000 	\\
Pyre1a	&	-655.242211 	&	146.67 	&	154.06 	&	105.09 	&	84.94 	&	115.06 	&	0.5126 	\\
Pyre2c	&	-655.243498 	&	146.32 	&	153.89 	&	108.34 	&	34.24 	&	74.34 	&	0.5614 	\\
Pyre4a	&	-655.243011 	&	146.69 	&	154.06 	&	104.76 	&	86.08 	&	123.96 	&	0.4367 	\\\hline
Pery	&	-769.582201 	&	158.55 	&	166.73 	&	111.71 	&	24.99 	&	94.75 	&	0.0000 	\\
Pery1c	&	-808.899758 	&	176.07 	&	185.06 	&	116.99 	&	59.11 	&	80.06 	&	0.5012 	\\
Pery2a	&	-808.910315 	&	175.56 	&	184.89 	&	121.35 	&	27.21 	&	85.03 	&	0.5198 	\\
Pery3a	&	-808.908572 	&	175.70 	&	184.96 	&	121.79 	&	13.26 	&	70.40 	&	0.5330 	\\\hline
Coro	&	-922.100621 	&	174.27 	&	183.26 	&	116.26 	&	86.41 	&	86.81 	&	0.0007 	\\
Coro1a	&	-961.427339 	&	191.61 	&	201.64 	&	123.33 	&	70.24 	&	87.85 	&	0.5207 	\\\hline
\noalign{\smallskip} \noalign{\smallskip}
\end{tabular}

\end{table*}
\clearpage

\begin{table*}
\footnotesize
\caption[]{Calculated frequencies and intensities
                for methyl (aliphatic) C--H and aromatic
                C--H stretches of toluene at the most
                pertinent levels and comparison to
                the NIST gas-phase experimental values.
                }
\label{Table:Freq_int_tolu}
     \begin{threeparttable}
         \centering
\begin{tabular}{lccccccccc}
\noalign{\smallskip} \hline \hline \noalign{\smallskip}
            & \multicolumn{3}{c}{B3LYP/6-31G$^{\ast}$}
            & \multicolumn{3}{c}{NIST}
            &  \multicolumn{2}{c}{WD57\tnote{g}}
\\ \noalign{\smallskip} \hline \noalign{\smallskip}
     	    &   $\nu({\rm cal})$\tnote{a}
            &   $\nu$\tnote{b}
            &	$A({\rm cal})$\tnote{c}
            &   $\nu({\rm exp})$\tnote{d}
            &	$\Arel$\tnote{e}
            &	$A({\rm exp})$\tnote{f}
            &   $\nu({\rm exp})$\tnote{d}
            &	$\Arel$\tnote{e}
            &	$A({\rm exp})$\tnote{f}\\
\noalign{\smallskip} \hline \noalign{\smallskip}
$    \nu_{\Me,1}$\tnote{h}    &	3039.1 	&	2921.5 	&	30.3 	& (2884.2)& \multirow{2}{*}{0.79} & \multirow{2}{*}{42.9} & (2885.8) & \multirow{2}{*}{0.71}& \multirow{2}{*}{--} \\	
     $\nu_{\Me,2}$\tnote{h}    &	3095.8 	&	2975.9 	&	22.7 	& (2935.4) 	&	                   &	 	              & (2936.2) &&	\\
     $\nu_{\Me,3}$\tnote{h}    &	3122.2 	&	3001.4 	&	17.4 	&		&		&		&		&		&		\\ \hline
\multirow{5}{*}{$\nu_{\rm aro}$\tnote{i}}	&	3171.3 	&	3048.6 	&	10.2 	&		&		&		&		&		 &		 \\
                                    &	3173.1 	&	3050.3 	&	6.9 	&		&		&		&		&		&		\\
                                    &	3185.4 	&	3062.2 	&	9.8 	 &(3040.7)&\multirow{2}{*}{}&\multirow{2}{*}{54.3}&(3039.4)&\multirow{2}{*}{}& \multirow{2}{*}{--}	\\	
                                    &	3194.0 	&	3070.3 	&	48.4 	&(3076.9)&		              &		                &(3079.6)&&	\\
                                    &	3206.6 	&	3082.5 	&	19.7 	&		&		&		&		&		&		\\
\hline
\noalign{\smallskip} \hline \noalign{\smallskip}
            & \multicolumn{3}{c}{B3LYP/6-311+G**}           & \multicolumn{3}{c}{B3LYP/6-311+G(3df,3pd)}    &  \multicolumn{3}{c}{MP2(full)/6-311+G**} \\
\noalign{\smallskip} \hline \noalign{\smallskip}
     	    &   $\nu({\rm cal})$\tnote{a}
            &   $\nu$\tnote{b}
            &	$A({\rm cal})$\tnote{c}
     	    &   $\nu({\rm cal})$\tnote{a}
            &   $\nu$\tnote{b}
            &	$A({\rm cal})$\tnote{c}
     	    &   $\nu({\rm cal})$\tnote{a}
            &   $\nu$\tnote{b}
            &	$A({\rm cal})$\tnote{c}\\
\noalign{\smallskip} \hline \noalign{\smallskip}
     $\nu_{\Me,1}$\tnote{h}            &	3019.8 	&	2925.5 	&	28.9 	&	3022.4	&	2923.2	&	27.2	&	 3072.7	&	 2926.1	&	25.8	\\
     $\nu_{\Me,2}$\tnote{h}            &	3073.5 	&	2977.7 	&	19.4 	&	3075.6	&	2974.7	&	16.6	&	 3151.3	&	 3001.0	&	13.2	\\
     $\nu_{\Me,3}$\tnote{h}            &	3099.3 	&	3002.6 	&	16.7 	&	3102.7	&	3001.0	&	14.8	&	 3170.3	&	 3019.1	&	12.1	\\ \hline
\multirow{5}{*}{$\nu_{\rm aro}$\tnote{i}}	&	3151.8 	&	3053.5 	&	9.0 	&	3156.6	&	3053.0	&	7.6	&	 3195.3	&	 3042.9	&	7.2	\\
                                    &	3153.4 	&	3055.0 	&	6.0 	&	3158.7	&	3055.1	&	5.2	&	3197.1	&	3044.6	 &	6.2	\\
                                    &	3166.1 	&	3067.3 	&	7.2 	&	3171.0	&	3067.0	&	7.4	&	3211	&	3057.8	 &	4.0	\\
                                    &	3174.4 	&	3075.3 	&	36.4 	&	3179.7	&	3075.4	&	32.9	&	3219.8	&	 3066.2	&	26.8	\\
                                    &	3186.9 	&	3087.5 	&	14.2 	&	3191.9	&	3087.2	&	12.6	&	3232.3	&	 3078.1	&	11.9	\\
\hline\noalign{\smallskip} \noalign{\smallskip}
\end{tabular}
        \begin{tablenotes}
\item[a] Computed frequency in $\cm^{-1}$.
\item[b] Frequency scaled with the corresponding
                  scaling factors listed in
                  Table~\ref{Table:Freq_scale}.

\item[c] Computed intensity in $\km\mol^{-1}$.
\item[d] Experimental frequency in $\cm^{-1}$
                  (they are given in parentheses as they
                   may not necessarily correspond to
                   $\nu_{\Me,x}$ or $\nu_{\rm aro}$).

\item[e] Ratio of the total methyl (aliphatic)
                  C--H stretch intensity
                  to the combined intensity of
                  all aromatic C--H stretches
                  as described in \S\ref{sec:MethylPAHs}.

\item[f] Experimental intensity in $\km\mol^{-1}$.

\item[g] Wilmshurst \&  Bernstein (1957).
\item[h] Methyl (aliphatic) C--H stretch frequencies
                  as described in the beginning
                  of \S\ref{sec:results}.

\item[i] Aromatic C--H stretch frequencies.
        \end{tablenotes}
     \end{threeparttable}
\end{table*}
\clearpage



\begin{table*}
\tiny
\caption[]{\footnotesize
           IR intensities ($\km\mol^{-1}$)
           computed with the B3LYP and MP2 methods
           with different basis sets
           for benzene, toluene, and
           1- and 2-methyl naphthalene
           [ML0: MP2/6-311G(3df,3pd);
            BL1: B3LYP/6-31G$^{\ast}$,
            BL2: B3LYP/6-311+G$^{\ast\ast}$,
            BL3: B3LYP/6-311+G(3df,3pd);
            ML1: MP2/6-31G$^{\ast}$,
            ML2: MP2/6-311+G$^{\ast\ast}$,
            ML3: MP2/6-311+G(3df,3pd)].
           }
\label{Table:Average_Int_B3LYP_MP2}
     \begin{threeparttable}
    \centering
\begin{tabular}{lccccccccccc}
\noalign{\smallskip} \hline \hline \noalign{\smallskip}
	&		&	Exp\tnote{a}	&	ML0\tnote{b}	&	ML0(fc)	&	ML0(full)	&	BL1	&	BL2	&	BL3	&	ML1	 &	ML2	&	ML3	 
\\ \noalign{\smallskip} \hline \noalign{\smallskip}
Benzene	&	$A$	&	55 	&	53 	&	53.83 	&	52.38 	&	104.00 	&	75.16 	&	67.23 	&	67.88 	&	56.40 	&	44.27 	 \\ \hline
\multirow{4}{*}{Tolu}	&	$A$\tnote{c}	&	95 	&	98 	&	97.08 	&	94.73 	&	165.30 	&	137.70 	&	124.32 	 &	 111.55 	&	107.04 	&	86.53 	 \\
	&	$A_{\Ali}$\tnote{d}	&		&		&	44.25 	&	43.19 	&	70.40 	&	64.96 	&	58.57 	&	47.17 	&	 51.04 	 &	41.83 	\\
	&	$A_{\Aro}$\tnote{e}	&		&		&	52.83 	&	51.53 	&	94.90 	&	72.74 	&	65.75 	&	64.38 	&	 56.00 	 &	44.69 	\\
	&	$A_{\Ali}/A_{\Aro}$	&		&		&		&		&	1.24 	&	1.49 	&	1.48 	&	1.22 	&	1.52 	&	 1.56 	\\ \hline
\multirow{4}{*}{Naph1}	&	$A$\tnote{c}	&		&		&		&		&	196.50 	&	161.92 	&	146.64 	&	 132.77 	 &	127.39 	&	103.87 	\\
	&	$A_{\Ali}$\tnote{d}	&		&		&		&		&	67.70 	&	65.66 	&	59.48 	&	43.44 	&	49.90 	 &	 41.94 	\\
	&	$A_{\Aro}$\tnote{e}	&		&		&		&		&	128.80 	&	96.27 	&	87.16 	&	89.33 	&	77.49 	 &	 61.93 	\\
	&	$A_{\Ali}/A_{\Aro}$	&		&		&		&		&	1.23 	&	1.59 	&	1.59 	&	1.13 	&	1.50 	&	 1.58 	\\ \hline
\multirow{4}{*}{Naph2}	&	$A$\tnote{c}	&		&		&		&		&	206.00 	&	172.08 	&	155.60 	&	 141.23 	 &	137.45 	&	111.88 	\\
	&	$A_{\Ali}$\tnote{d}	&		&		&		&		&	77.60 	&	72.75 	&	65.58 	&	44.34 	&	55.29 	 &	 46.45 	\\
	&	$A_{\Aro}$\tnote{e}	&		&		&		&		&	128.40 	&	99.32 	&	90.02 	&	96.89 	&	82.16 	 &	 65.43 	\\
	&	$A_{\Ali}/A_{\Aro}$	&		&		&		&		&	1.41 	&	1.71 	&	1.70 	&	1.07 	&	1.57 	&	 1.66 	\\
\hline \noalign{\smallskip} \noalign{\smallskip}
\end{tabular}
\footnotesize
        \begin{tablenotes}
\item[a] Experimental values
                  listed in Pavlyuchko et al.\ (2012).
                  
\item[b] Computed values of Pavlyuchko et al.\ (2012)
                  at the MP2/6-311G(3df,3pd) (i.e., ML0) level.
                  
\item[c] $A = A_{\Ali} + A_{\Aro}$.
\item[d] Intensity of the methyl (aliphatic)
                  C--H stretch (per chemical bond).
                  
\item[e] Intensity of the aromatic
                  C--H stretch (per chemical bond).
      \end{tablenotes}
     \end{threeparttable}        
\end{table*}

\clearpage


\begin{table*}
\tiny
\caption[]{\footnotesize
                Averages of scaled, characteristic frequencies
               and intensities of the aliphatic and aromatic
               C--H stretches for toluene and all the isomers
               of methyl pyrene computed at B3LYP
               with different basis sets.
               }
\label{Table:Average_Freq_BasisSet_2}
     \begin{threeparttable}
    \centering
\begin{tabular}{lllccccccccc}
\noalign{\smallskip} \hline \hline \noalign{\smallskip}
 &  Basis Set
         & $\bar \nu_{\Ali}$\tnote{a}
         & $\sigma(\nu_{\Ali})$\tnote{b}
         & $\bar \nu_{\Aro}$\tnote{a}
         & $\sigma(\nu_{\Aro})$\tnote{b}
         & $A_{\Ali}$\tnote{c}
         & $\sigma(A_{\Ali})$\tnote{d}
         & $A_{\Aro}$\tnote{c}
         & $\sigma(A_{\Aro})$\tnote{d}
         & $A_{\Ali}/A_{\Aro}$\\
\noalign{\smallskip} \hline \noalign{\smallskip}
Tolu	&	6-31G*	&	2966.3 	&	40.8 	&	3062.8 	&	14.2 	&	23.5 	&	6.5 	&	19.0 	&	17.1 	&	1.23 	 \\
	&	6-31+G*	&	2968.9 	&	39.2 	&	3068.7 	&	14.4 	&	24.3 	&	8.0 	&	18.8 	&	16.4 	&	1.29 	\\
	&	6-31+G**	&	2968.3 	&	41.7 	&	3068.2 	&	14.2 	&	22.9 	&	8.2 	&	16.2 	&	13.9 	&	1.42 	 \\
	&	6-31++G**	&	2968.0 	&	41.9 	&	3068.0 	&	14.2 	&	23.3 	&	8.4 	&	16.2 	&	14.1 	&	1.44 	 \\
	&	6-311G**	&	2965.5 	&	39.6 	&	3063.2 	&	14.3 	&	23.2 	&	5.6 	&	16.6 	&	14.7 	&	1.39 	 \\
	&	6-311+G*	&	2967.7 	&	38.0 	&	3062.7 	&	14.5 	&	25.3 	&	6.8 	&	19.6 	&	17.5 	&	1.29 	 \\
	&	6-311+G**	&	2968.6 	&	39.3 	&	3067.7 	&	14.3 	&	21.7 	&	6.4 	&	14.5 	&	12.6 	&	1.49 	 \\
	&	6-311++G**	&	2965.7 	&	39.4 	&	3064.6 	&	14.2 	&	21.8 	&	6.6 	&	14.4 	&	12.5 	&	1.51 	 \\
	&	6-311+G(3df,3pd)	&	2966.3 	&	39.5 	&	3067.5 	&	14.3 	&	19.5 	&	6.7 	&	13.1 	&	11.4 	&	 1.48 	\\
	&	6-311++G(3df,3pd)	&	2966.5 	&	39.6 	&	3068.0 	&	14.2 	&	19.7 	&	6.6 	&	13.2 	&	11.4 	&	 1.50 	\\\hline
Pyre1	&	6-31G*	&	2965.3 	&	42.1 	&	3065.7 	&	12.8 	&	25.9 	&	12.1 	&	19.8 	&	17.6 	&	1.31 	 \\
	&	6-31+G*	&	2966.2 	&	40.4 	&	3071.6 	&	13.1 	&	29.0 	&	16.8 	&	19.5 	&	16.6 	&	1.49 	\\
	&	6-31+G**	&	2965.3 	&	43.1 	&	3071.4 	&	12.9 	&	27.4 	&	16.7 	&	16.6 	&	13.9 	&	1.65 	 \\
	&	6-31++G**	&	2966.5 	&	43.3 	&	3070.9 	&	12.8 	&	27.9 	&	16.3 	&	16.6 	&	14.1 	&	1.68 	 \\
	&	6-311G**	&	2964.4 	&	41.3 	&	3066.4 	&	12.7 	&	26.2 	&	12.1 	&	17.3 	&	15.0 	&	1.51 	 \\
	&	6-311+G*	&	2966.6 	&	39.2 	&	3066.0 	&	13.0 	&	29.4 	&	15.0 	&	20.5 	&	18.1 	&	1.43 	 \\
	&	6-311+G**	&	2967.3 	&	40.9 	&	3070.5 	&	12.6 	&	25.5 	&	14.0 	&	15.0 	&	12.7 	&	1.70 	 \\
	&	6-311++G**	&	2964.4 	&	40.9 	&	3067.4 	&	12.6 	&	26.4 	&	14.7 	&	14.7 	&	12.4 	&	1.80 	 \\
	&	6-311+G(3df,3pd)	&	2964.7 	&	40.9 	&	3070.2 	&	12.8 	&	23.3 	&	14.0 	&	13.7 	&	11.5 	&	 1.71 	\\
	&	6-311++G(3df,3pd)	&	2964.9 	&	41.1 	&	3070.6 	&	12.8 	&	23.6 	&	14.1 	&	13.7 	&	11.5 	&	 1.72 	\\\hline
Pyre2	&	6-31G*	&	2969.2 	&	41.2 	&	3060.7 	&	11.2 	&	26.8 	&	14.7 	&	18.8 	&	26.2 	&	1.42 	 \\
	&	6-31+G*	&	2970.1 	&	39.3 	&	3066.1 	&	11.6 	&	28.8 	&	17.4 	&	18.9 	&	25.4 	&	1.52 	\\
	&	6-31+G**	&	2969.5 	&	41.9 	&	3066.0 	&	11.4 	&	27.1 	&	17.2 	&	16.1 	&	21.8 	&	1.69 	 \\
	&	6-31++G**	&	2970.6 	&	42.2 	&	3065.8 	&	11.2 	&	27.3 	&	17.3 	&	16.1 	&	22.0 	&	1.70 	 \\
	&	6-311G**	&	2968.1 	&	40.0 	&	3061.6 	&	11.0 	&	26.3 	&	13.2 	&	16.7 	&	23.6 	&	1.58 	 \\
	&	6-311+G*	&	2970.1 	&	38.4 	&	3061.0 	&	11.2 	&	29.3 	&	15.7 	&	20.0 	&	27.6 	&	1.47 	 \\
	&	6-311+G**	&	2970.9 	&	39.7 	&	3065.7 	&	11.0 	&	25.2 	&	14.2 	&	14.5 	&	20.4 	&	1.74 	 \\
	&	6-311++G**	&	2968.1 	&	39.7 	&	3062.6 	&	11.0 	&	25.3 	&	14.4 	&	14.4 	&	20.3 	&	1.76 	 \\
	&	6-311+G(3df,3pd)	&	2968.3 	&	39.8 	&	3065.4 	&	10.8 	&	23.1 	&	13.9 	&	13.3 	&	18.5 	&	 1.74 	\\
	&	6-311++G(3df,3pd)	&	2968.5 	&	39.9 	&	3065.9 	&	10.8 	&	23.3 	&	13.9 	&	13.3 	&	18.6 	&	 1.75 	\\\hline
Pyre4	&	6-31G*	&	2963.8 	&	43.2 	&	3066.6 	&	14.5 	&	23.6 	&	9.8 	&	19.6 	&	18.1 	&	1.21 	 \\
	&	6-31+G*	&	2966.1 	&	41.4 	&	3072.2 	&	14.8 	&	26.0 	&	13.6 	&	19.2 	&	17.0 	&	1.35 	\\
	&	6-31+G**	&	2965.2 	&	44.0 	&	3072.0 	&	14.6 	&	24.5 	&	13.4 	&	16.3 	&	14.5 	&	1.50 	 \\
	&	6-31++G**	&	2965.0 	&	44.1 	&	3071.9 	&	14.6 	&	25.0 	&	13.4 	&	16.3 	&	14.7 	&	1.53 	 \\
	&	6-311G**	&	2962.8 	&	42.2 	&	3067.6 	&	14.4 	&	23.9 	&	9.8 	&	17.1 	&	16.3 	&	1.40 	 \\
	&	6-311+G*	&	2965.1 	&	40.0 	&	3067.0 	&	14.7 	&	26.4 	&	12.2 	&	20.3 	&	19.1 	&	1.30 	 \\
	&	6-311+G**	&	2965.8 	&	41.7 	&	3071.7 	&	14.4 	&	22.9 	&	11.4 	&	14.7 	&	13.8 	&	1.56 	 \\
	&	6-311++G**	&	2962.8 	&	41.6 	&	3068.6 	&	14.4 	&	23.2 	&	11.7 	&	14.6 	&	13.8 	&	1.58 	 \\
	&	6-311+G(3df,3pd)	&	2963.1 	&	41.7 	&	3071.5 	&	14.6 	&	21.0 	&	11.5 	&	13.5 	&	12.5 	&	 1.56 	\\
	&	6-311++G(3df,3pd)	&	2963.3 	&	41.8 	&	3072.0 	&	14.6 	&	21.3 	&	11.5 	&	13.5 	&	12.5 	&	 1.58 	\\
\hline \noalign{\smallskip} \noalign{\smallskip}
\end{tabular}
\normalsize
\footnotesize
        \begin{tablenotes}
\item[a] Average scaled frequency
                  for the methyl (aliphatic)
                  or aromatic C--H stretch
                  in $\cm^{-1}$

\item[b] Standard deviation of frequency
                  for the methyl or aromatic C--H stretch
                  in $\cm^{-1}$
                  
\item[c] Intensity of each C--H bond
                  for the methyl or aromatic C--H stretch
                  in $\km\mol^{-1}$
                  
\item[d] Standard deviation of intensity of
                  each C--H bond for the methyl or aromatic
                  C--H stretches in $\km\mol^{-1}$

      \end{tablenotes}
     \end{threeparttable}
\end{table*}
\clearpage


\begin{table*}
\footnotesize
\caption[]{\footnotesize
           Averages of scaled, characteristic frequencies
           and intensities of the aliphatic and aromatic
           C--H stretches of PAHs and their methyl derivatives
           computed at {\rm B3LYP/6-311+G$^{\ast\ast}$}
           (with the same units as that in
            Table~\ref{Table:Average_Freq_BasisSet_2})
           }
\label{Table:Average_Freq_6311}
\begin{tabular}{lcccccccccc}
\noalign{\smallskip} \hline \hline \noalign{\smallskip}
&	$\bar{\nu}_{\Ali}$	
        & $\sigma(\nu_{\Ali})$	
        & $\bar{\nu}_{\Aro}$	
        & $\sigma(\nu_{\Aro})$	
        & $A_{\Ali}$	
        & $\sigma(A_{\Ali})$	
        & $A_{\Aro}$	
        & $\sigma(A_{\Aro})$	
        & $A_{\Ali}/A_{\Aro}$		
        & $\left(A_{\Ali}/A_{\Aro}\right)_{\rm NIST}$\\
\noalign{\smallskip} \hline \noalign{\smallskip}
Benzene	&		&		&	3077.3 	&	12.8 	&		&		&	12.5 	&	19.4 	&		&		\\
Toluc	&	2968.6 	&	39.3 	&	3067.7 	&	14.3 	&	21.7 	&	6.4 	&	14.5 	&	12.6 	&	1.49 	&	1.32 	 \\ \hline
Naph	&		&		&	3071.0 	&	12.2 	&		&		&	12.8 	&	22.3 	&		&		\\
Naph1a	&	2967.6 	&	40.9 	&	3073.7 	&	13.8 	&	21.9 	&	7.1 	&	13.8 	&	10.0 	&	1.59 	&	1.66 	 \\
Naph2b	&	2967.8 	&	39.9 	&	3066.5 	&	12.6 	&	24.3 	&	12.4 	&	14.2 	&	11.6 	&	1.71 	&	1.89 	 \\ \hline
Anth	&		&		&	3070.0 	&	12.4 	&		&		&	13.2 	&	22.9 	&		&		\\
Anth1a	&	2966.5 	&	41.9 	&	3071.0 	&	12.1 	&	22.5 	&	6.8 	&	14.0 	&	11.1 	&	2.02 	&	1.77 	 \\
Anth2b	&	2966.1 	&	41.2 	&	3064.8 	&	11.9 	&	26.8 	&	17.1 	&	14.2 	&	12.2 	&	2.19 	&	1.62 	 \\
Anth9a	&	2985.0 	&	59.4 	&	3078.8 	&	18.6 	&	19.2 	&	8.6 	&	13.5 	&	10.1 	&	1.89 	&	1.26 	 \\ \hline
Phen	&		&		&	3077.0 	&	15.9 	&		&		&	12.4 	&	14.6 	&		&		\\
Phen1a	&	2969.2 	&	40.4 	&	3080.5 	&	17.7 	&	23.1 	&	8.5 	&	12.7 	&	8.3 	&	1.81 	&	1.50 	 \\
Phen2b	&	2969.0 	&	39.4 	&	3072.8 	&	17.4 	&	25.4 	&	15.3 	&	13.5 	&	9.0 	&	1.88 	&	1.62 	 \\
Phen3b	&	2967.8 	&	39.2 	&	3073.4 	&	15.2 	&	24.3 	&	14.3 	&	13.4 	&	11.9 	&	1.82 	&	1.53 	 \\
Phen4a	&	2973.7 	&	39.2 	&	3084.9 	&	40.8 	&	19.8 	&	3.4 	&	13.4 	&	13.4 	&	1.48 	&		\\
Phen9a	&	2967.9 	&	41.5 	&	3079.8 	&	17.7 	&	23.1 	&	10.8 	&	13.0 	&	7.7 	&	1.77 	&		\\ \hline
Pyre	&		&		&	3069.6 	&	10.4 	&		&		&	14.0 	&	25.3 	&		&		\\
Pyre1a	&	2967.3 	&	40.9 	&	3070.5 	&	12.6 	&	25.5 	&	14.0 	&	15.0 	&	12.7 	&	1.70 	&	1.59 	 \\
Pyre2c	&	2970.9 	&	39.7 	&	3065.7 	&	11.0 	&	25.2 	&	14.2 	&	14.5 	&	20.4 	&	1.74 	&		\\
Pyre4a	&	2965.8 	&	41.7 	&	3071.7 	&	14.4 	&	22.9 	&	11.4 	&	14.7 	&	13.8 	&	1.56 	&		\\ \hline
Pery	&		&		&	3084.1 	&	17.3 	&		&		&	13.2 	&	26.8 	&		&		\\
Pery1c	&	2970.3 	&	45.6 	&	3083.0 	&	25.1 	&	20.9 	&	9.9 	&	13.6 	&	15.1 	&	1.53 	&		\\
Pery2a	&	2968.8 	&	39.5 	&	3079.8 	&	17.2 	&	24.7 	&	14.4 	&	13.3 	&	13.9 	&	1.85 	&		\\
Pery3a	&	2966.4 	&	41.4 	&	3084.1 	&	17.7 	&	29.5 	&	20.5 	&	14.2 	&	13.5 	&	2.07 	&		\\ \hline
Coro	&		&		&	3068.7 	&	9.1 	&		&		&	16.2 	&	35.2 	&		&		\\
Coro1a	&	2968.1 	&	41.4 	&	3068.9 	&	11.8 	&	25.6 	&	15.5 	&	16.8 	&	21.8 	&	1.53 	&		\\ \hline \noalign{\smallskip} \noalign{\smallskip}
\end{tabular}

\end{table*}

\end{document}